\documentclass[12pt, draftcls, onecolumn]{IEEEtran}

\usepackage{color}
\usepackage{graphicx}
\usepackage{amsmath, amsfonts, amssymb, psfrag, epsfig, epstopdf}
\usepackage{caption}

\newcommand\scalemath[2]{\scalebox{#1}{\mbox{\ensuremath{\displaystyle #2}}}}

\newcommand{\markov}{\mathrel{\multimap}\joinrel\mathrel{-}%
\joinrel\mathrel{\mkern-6mu}\joinrel\mathrel{-}}

\newtheorem{proposition}{{Proposition}}
\newtheorem{definition}{{Definition}}
\newtheorem{theorem}{{Theorem}}
\newtheorem{lemma}{{Lemma}}
\newtheorem{example}{{Example}}
\newtheorem{corollary}{{Corollary}}
\newtheorem{remark}{{Remark}}

\DeclareMathAlphabet{\mathpzc}{OT1}{pzc}{m}{it}
\DeclareMathOperator*{\argmax}{arg\,max}

\ifCLASSOPTIONcompsoc
\usepackage[caption=false,font=normalsize,labelfont=sf,textfont=sf]{subfig}
\else
\usepackage[caption=false,font=footnotesize]{subfig}
\fi

\makeatletter

\newcommand{\xleftrightarrow}[2][]{\ext@arrow 3359\leftrightarrowfill@{#1}{#2}}
\newcommand{\xdashleftrightarrow}[2][]{\ext@arrow 3359\leftrightarrowfill@@{#1}{#2}}
\def\rightarrowfill@@{\arrowfill@@\relax\relbar\rightarrow}
\def\leftarrowfill@@{\arrowfill@@\leftarrow\relbar\relax}
\def\leftrightarrowfill@@{\arrowfill@@\leftarrow\relbar\rightarrow}
\def\arrowfill@@#1#2#3#4{%
  $\m@th\thickmuskip0mu\medmuskip\thickmuskip\thinmuskip\thickmuskip
   \relax#4#1
   \xleaders\hbox{$#4#2$}\hfill
   #3$%
}

\makeatother
\newcounter{parentnumber}
\allowdisplaybreaks

\begin{document}

\title{Capacity of Two-Way Channels\\ with Symmetry Properties}

\author{
\IEEEauthorblockN{Jian-Jia Weng\IEEEauthorrefmark{2},~\IEEEmembership{Student Member,~IEEE}, Lin Song\IEEEauthorrefmark{3}, Fady Alajaji\IEEEauthorrefmark{2},~\IEEEmembership{Senior Member,~IEEE}, and Tam\'as Linder\IEEEauthorrefmark{2},~\IEEEmembership{Fellow,~IEEE}}
\thanks{%
    \IEEEauthorrefmark{2}The authors are with the Department of Mathematics and Statistics, Queen's University, Kingston, ON K7L 3N6, Canada (Emails: jian-jia.weng@queensu.ca, \{fady, linder\}@mast.queensu.ca).}
\thanks{%
    \IEEEauthorrefmark{3} L.~Song was with the Department of Mathematics and Statistics, Queen's University, Kingston, ON K7L 3N6, Canada. She is now with Contextere Ltd., Ottawa, ON K1Y 2C5, Canada  (Email: lin@contextere.com).}
\thanks{
This work was supported in part by NSERC of Canada. Parts of this work were presented at the 2016 IEEE International Symposium on Information Theory \cite{Song:2016} and the 2018 IEEE International Symposium on Information Theory \cite{JJW:2018}.}
}

\maketitle

\begin{abstract}
In this paper, we make use of channel symmetry properties to determine the capacity region of three types of two-way networks: (a) two-user memoryless two-way channels (TWCs), (b) two-user TWCs with memory, and (c) three-user multiaccess/degraded broadcast (MA/DB) TWCs. 
For each network, symmetry conditions under which a Shannon-type random coding inner bound (under independent non-adaptive inputs) is tight are given. 
For two-user memoryless TWCs, prior results are substantially generalized by viewing a TWC as two interacting state-dependent one-way channels. 
The capacity of symmetric TWCs with memory, whose outputs are functions of the inputs and independent stationary and ergodic noise processes, is also obtained. 
Moreover, various channel symmetry properties under which the Shannon-type inner bound is tight are identified for three-user MA/DB TWCs. 
The results not only enlarge the class of symmetric TWCs whose capacity region can be exactly determined but also imply that interactive adaptive coding, not improving capacity, is unnecessary for such channels.
\end{abstract}

\begin{IEEEkeywords}
Network information theory, two-way channels, capacity region, inner and outer bounds, channel symmetry, multiple access and broadcast channels, channels with memory, adaptive coding. 
\end{IEEEkeywords}

\section{Introduction}\label{sec:introduction}
Shannon's two-way channel (TWC) \cite{Shannon:1961}, which allows two users to exchange data streams in a full-duplex manner, is a basic component of communication systems. 
To mitigate the interference incurred from two-way simultaneous transmission, TWCs are often used in conjunction with orthogonal multiplexing \cite{Jama:2005}. 
With increasing demands for fast data transmission, many industrial standards have enabled the use of non-orthogonal multiplexing to accommodate more users \cite{Yuan:2016}, \cite{Ding:2017}. 
From an information-theoretic viewpoint, the challenge is how each user can effectively maximize its individual transmission rate over the shared channel and concurrently provide sufficient feedback to help the other users' transmissions.
These competing objectives impose on each user the challenging task of optimally adapting their channel inputs to the previously received signals of the other users. 
As finding such an optimal coding procedure is still elusive, the exact characterization of the capacity region of general TWCs remains open \cite{Meulen:1977}, \cite[Section~17.5]{Gamal:2012}. 

\begin{figure*}[!t]
\centering
\subfloat[]{\includegraphics[draft=false, scale=0.65]{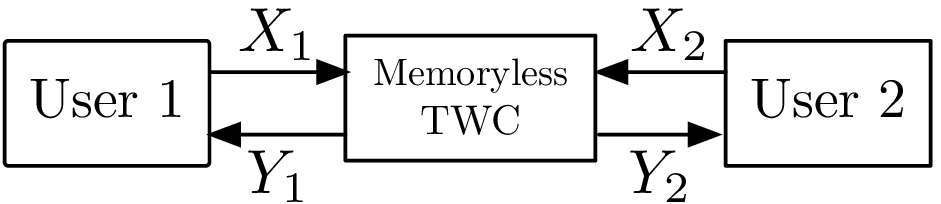}
\label{model1}}
\qquad
\subfloat[]{\includegraphics[draft=false, scale=0.65]{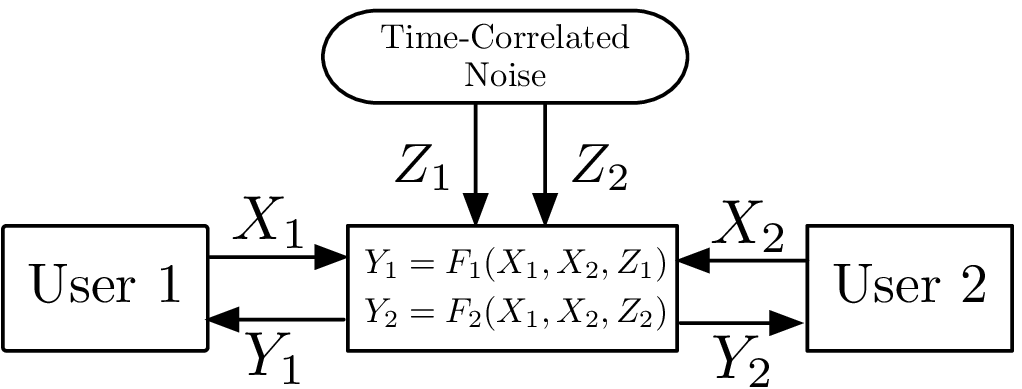}
\label{model2}}
\\
\subfloat[]{\includegraphics[draft=false, scale=0.52]{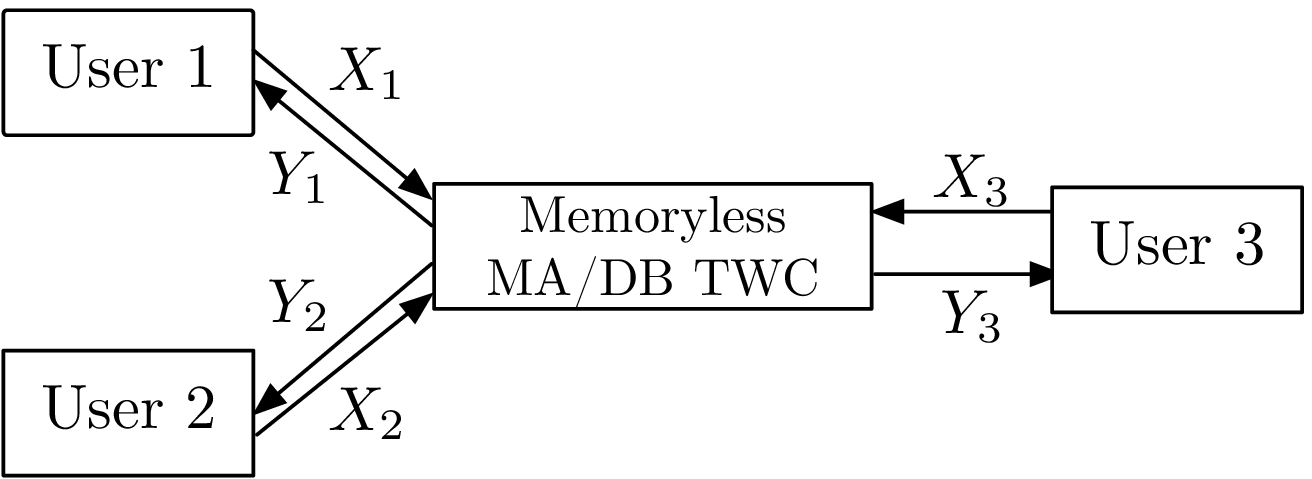}
\label{model3}}
\caption{Block diagrams of the two-way networks considered: (a) point-to-point memoryless TWC with two channel inputs $X_1$ and $X_2$ and two channel outputs $Y_1$ and $Y_2$; (b) point-to-point TWC with memory, where $F_1$ and $F_2$ are deterministic functions and $(Z_1, Z_2)$ is a time-correlated channel noise pair generated from a joint stationary and ergodic process; (c) three-user memoryless MA/DB TWC, where $X_i$ and $Y_i$ respectively denote channel input and output at user~$j$ for $j=1, 2, 3$.}
\label{TWCmodels}
\end{figure*}

This paper revisits this open problem by finding larger classes of TWCs whose capacity region can be exactly obtained. 
Our approach is to identify channel symmetry properties under which a Shannon-type random coding inner bound (under independent non-adaptive inputs) is tight, thus directly determining the capacity region. 
As a result, we identify TWCs for which interactive adaptive coding is useless in terms of improving the users' transmission rates. 
In particular, we focus on three two-way networks which we depict in Fig.~\ref{TWCmodels}. 
The two-user (point-to-point) memoryless TWC in Fig.~\ref{model1} models device-to-device communication \cite{Asadi:2014}. 
The simplified TWC with memory in Fig.~\ref{model2}, which is a generalization of additive-noise TWC in \cite{Song:2016}, can capture the effect of time-correlated channel noise which commonly arises in wireless communications.  
The three-user memoryless multiaccess/degraded broadcast (MA/DB) TWC \cite{Devroye:2014} in Fig.~\ref{model3} models the communication between two mobile users and one base station, where the shared channel in the users-to-base-station (uplink) direction acts as a multiple-access channel (MAC) while the reverse (downlink) direction acts as a degraded broadcast channel (DBC).   
For these networks, we derive conditions under which the Shannon-type inner bound is optimal in terms of achieving channel capacity.
Such a result also has a practical significance since communication without adaptive coding simplifies system design.

\subsection{Capacity Bounds for TWCs}
We briefly review some general results on the capacity of TWCs.  
In \cite{Shannon:1961}, Shannon derived inner and outer capacity bounds in the form of a single-letter expression for two-user memoryless TWCs.
The inner bound is obtained via random coding where the users' channel inputs are independent (and non-adaptive), while the inputs are allowed to have arbitrary correlation in the outer bound. 
In general, the two bounds do not coincide.  
Follow-up work in \cite{Dueck:1979}-\cite{Han:1984} was devoted to improving Shannon's inner bound by using adaptive coding.
Two novel outer bounds \cite{Zhang:1986}, \cite{Hekstra:1989}, which restrict the dependency among channel inputs, were proposed to refine Shannon's result. 
Moreover, methods to efficiently utilize TWCs were investigated by studying the role of feedback \cite{Mee:1998}. 
In \cite{Krama:1998}, directed mutual information \cite{Massey:1990}, which is widely used in the study of one-way channels with feedback \cite{Tatikonda:2000}-\cite{Permuter2:2009}, was used to characterize the capacity of TWCs, but the obtained multi-letter expressions are often not computable. 
Recently, the Shannon-type random coding scheme was shown to be optimal in several deterministic multi-user TWC settings \cite{Devroye:2014} such as MA/BC, Z, and interference TWCs, hence finding the channel capacity in these cases. 
The channel capacity for a variant of these multi-user TWCs, called three-way channels, was also investigated in different network setups such as three-way multi-cast finite-field or phase-fading Gaussian channels \cite{Ong:2012} and three-way Gaussian channels with multiple unicast sessions \cite{Chaaban:2016}. 
An additional capacity result for deterministic interference TWCs was derived in \cite{Tse:2018}. 
For TWCs with memory, Shannon provided a multi-letter capacity characterization in \cite[Section~16]{Shannon:1961} which in general is incalculable. 

\subsection{Related Work}
Channel symmetry properties, which are extensively investigated to simplify the computation of the capacity of one-way channels, play a key role in determining the capacity region for TWCs. 
The first channel symmetry property for TWCs was proposed by Shannon \cite[Section~12]{Shannon:1961}. Let $[P_{Y_1, Y_2|X_1, X_2}(\cdot, \cdot|\cdot, \cdot)]$ denote
the channel transition matrix of a two-user discrete memoryless TWC, where $X_j$ and $Y_j$ denote the channel input and output at user $j$, respectively.
Shannon gave two permutation invariance conditions on $[P_{Y_1, Y_2|X_1, X_2}(\cdot, \cdot|\cdot, \cdot)]$ which guarantee the equality of his inner and outer bounds (see Propositions~1 and~2 in Section~II for details). 
A recent work \cite{Chaaban:2017} by Chaaban, Varshney, and Alouini (CVA) presented another tightness condition, where the channel symmetry property is given in terms of conditional entropies for the marginal channel distribution $[P_{Y_j|X_1, X_2}(\cdot|\cdot, \cdot)]$ (see Proposition~3).  

The above conditions delineate classes of two-user memoryless TWCs for which Shannon's capacity inner bound is tight, hence exactly yielding their capacity region. 
Examples include Gaussian TWCs \cite{Han:1984}, $q$-ary additive-noise TWCs \cite{Song:2016}, and more general channel models such as injective semi-deterministic TWCs (ISD-TWCs) \cite{Chaaban:2017}, Cauchy \cite{Chaaban:2017} and exponential family type TWCs \cite{Varshney:2013}.
It is worth mentioning that Hekstra and Willems \cite{Hekstra:1989} also presented a condition under which Shannon's inner bound is tight. 
However, their result is only valid for single-output memoryless TWCs.

For three-user MA/BC memoryless TWCs, Cheng and Devroye \cite{Devroye:2014} investigated a class of symmetric TWCs. 
In particular, they considered deterministic, invertible, and alphabet-restricted MA/BC TWCs, proving that the Shannon-type inner bound is tight for that class of channels. 
However, to the best of our knowledge, symmetry properties for TWCs beyond these have not been investigated.
It is also important to point out that two-user TWCs with memory are not well understood either. 

\subsection{A Motivational Example and Proposed Approach}
Consider a point-to-point binary-input and binary-output memoryless TWC with transition probability matrix (see Section~\ref{subsec:p2pmemoryless} for the formal description of the channel model)
\begin{IEEEeqnarray}{l}
[P_{Y_1,Y_2|X_1,X_2}(\cdot, \cdot|\cdot, \cdot)] =\scalemath{1}{\bordermatrix{
& 00& 01 & 10 & 11\cr
00& 0.783 & 0.087 & 0.117 & 0.013\cr
01& 0.0417 & 0.3753 & 0.0583 & 0.5247\cr
10& 0.261 & 0.609 & 0.039 & 0.091 \cr
11& 0.2919 & 0.1251 & 0.4081 & 0.1749}},\nonumber
\end{IEEEeqnarray}
where the rows and columns are indexed by the channel inputs and outputs, respectively. 
The corresponding marginal channel transition matrices are 
\begin{IEEEeqnarray}{l}
\scalemath{1}{{[P_{Y_2|X_1,X_2}(\cdot|\cdot,0)]= 
\begin{pmatrix}
0.9 & 0.1\\
0.3 & 0.7
\end{pmatrix},}}\ 
\scalemath{1}{{[P_{Y_2|X_1,X_2}(\cdot|\cdot,1)]= 
 \begin{pmatrix}
0.1 & 0.9\\
0.7 &  0.3
\end{pmatrix},}}\nonumber
\end{IEEEeqnarray}
and
\begin{IEEEeqnarray}{l}
\scalemath{1}{{[P_{Y_1|X_1,X_2}(\cdot|0,\cdot)]= [P_{Y_1|X_1,X_2}(\cdot|1,\cdot)]= 
\begin{pmatrix}
0.87 & 0.13\\
0.417 & 0.583
\end{pmatrix}}}.\nonumber
\end{IEEEeqnarray}
A thorough examination reveals that for this TWC Shannon's inner bound is actually exact due to the symmetric structures of the channel's marginal transition matrices. However, none of the previously proposed symmetry conditions in the literature are satisfied.

We address this problem by viewing a TWC as two state-dependent one-way channels \cite{Shannon:1961}, \cite{Jelinek:1964}. 
Taking the two-user setting as an example, the state-dependent one-way channel from users~1 to~2 has input $X_1$, output $Y_2$, state $X_2$, and transition matrix given by $[P_{Y_2|X_1, X_2}(\cdot|\cdot, \cdot)]$; similarly, the one-way channel $[P_{Y_1|X_1, X_2}(\cdot|\cdot, \cdot)]$  in the reverse direction has input $X_2$, output $Y_1$, and channel state $X_1$. 
Note that this viewpoint\footnote{Another viewpoint for two-user TWCs is based on compound MACs, see \cite[Problem 14.11]{Csiszar:2011} and \cite{Gunduz:2009}.} may also be useful for all previously mentioned two-way networks. 
Another useful tool is the rich set of symmetry concepts for single-user one-way channels.\footnote{Channel symmetry properties for single-user one-way memoryless channels can be roughly classified into two types. One type focuses on the structure of the channel transition probability such as Gallager symmetric channels \cite{Gallager:1968}, weakly symmetric and symmetric channels \cite{Cover:2006}, and quasi-symmetric channels \cite{Alajaji}. The other type aims at the invariance of information quantities including $T$-symmetric channels \cite{Grant:2004} and channels with input-invariance symmetry \cite{Wesel:2008}.} 
From this perspective, the two one-way channels now interact with each other through the channel states. 
Clearly, this interaction could improve bi-directional transmission rates by making use of adaptive coding.

Our approach is to study symmetry properties for state-dependent one-way channels that imply that the capacity cannot be increased with the availability of channel state information at the transmitter (in addition to the receiver). 
Such properties can potentially render interactive adaptive coding useless in terms of enlarging TWC capacity. 
In the two-user memoryless setting, we develop the following two important channel symmetry notions. 
The {\it common optimal input distribution} condition identifies a state-dependent one-way channel that has an identical capacity-achieving input distribution for all channel states. 
The {\it invariance of input-output mutual information} condition then identifies a state-dependent one-way channel that produces the same input-output mutual information for all channel states under any fixed input distribution. 
If a TWC satisfies both conditions, one for each direction of the two-way transmission, the optimal transmission scheme of one user is irrelevant to the other user's transmission scheme, implying that the interaction between the users does not increase their transmission rates and hence channel capacity. 
In fact, the preceding motivational example illustrates this. 
More formally, we can prove that under certain symmetry properties (identified by the derived conditions), any rate pair inside Shannon's outer bound region is always contained in the inner bound region, implying that the latter bound is tight. 

Furthermore, it should be expected that validating generalized channel symmetry properties can be a very complex procedure. 
However, we show that such a verification can be greatly simplified  for some TWCs. 
For instance, the channel transition matrices $[P_{Y_1|X_1, X_2}(\cdot|\cdot, 0)]$ and $[P_{Y_1|X_1, X_2}(\cdot|\cdot, 1)]$ in the above example are column permutations of each other and the matrices $[P_{Y_1|X_1, X_2}(\cdot|0, \cdot)]$ and $[P_{Y_1|X_1, X_2}(\cdot|1, \cdot)]$ are identical. 
It turns out (as we will see later) that these two symmetry properties imply that Shannon's inner bound is tight.
Therefore, we not only seek general conditions but also look for conditions which are simple to verify. 

\subsection{Summary of Contributions}
\begin{figure}[tb]
\begin{centering}
\includegraphics[scale=0.7, draft=false]{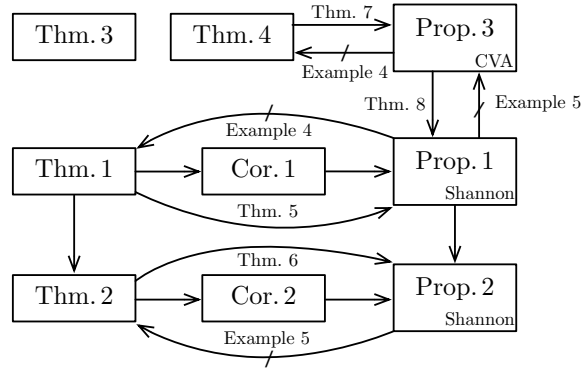}
\caption{The relationships between the results yielding the equality of Shannon's capacity bounds in point-to-point memoryless TWCs. Here, $A\rightarrow B$ indicates that result $A$ subsumes result $B$, and $B\nrightarrow A$ indicates that result $B$ does not subsume result $A$. For example, Prop.~3 $\rightarrow$ Prop.~1 and Prop.~1 $\nrightarrow$ Prop.~3 mean that the CVA result in Prop.~3 is more general than the Shannon result in Prop.~1.}
\label{fig:impDia1}
\end{centering}
\end{figure}

Most of the conditions that we establish in this paper comprise two parts, one for each direction of the two-way transmission. 
Our contributions are summarized as follows.\\
\noindent$\bullet$ \textbf{Point-to-Point Memoryless TWCs:} six sufficient conditions (Theorems~\ref{thm:23}-\ref{thm:24} and Corollaries~\ref{cor:nc1}-\ref{cor:77}) guaranteeing that Shannon's inner and outer bounds coincide are derived. 
Three of these are shown to be substantial generalizations of the Shannon and CVA conditions (in Theorems~\ref{thm:23toSC}-\ref{thm:CVAto24}); our simplest condition can be verified by only observing the channel marginal distributions.
Moreover, the capacity region of $q$-ary additive-noise TWCs with erasures, which subsume several classical TWCs, is fully characterized by our conditions. 
Several examples illustrating the difference between these conditions are  provided. 
We also refine Shannon's result to show that the CVA condition is a strict generalization of the Shannon condition (Theorem~\ref{thm:SCtoCVA}), thus answering a question raised in \cite{Chaaban:2017}. Implications among our results (and prior results) are depicted in Fig.~\ref{fig:impDia1}.

\begin{figure*}[tb]
\begin{centering}
\includegraphics[scale=0.7, draft=false]{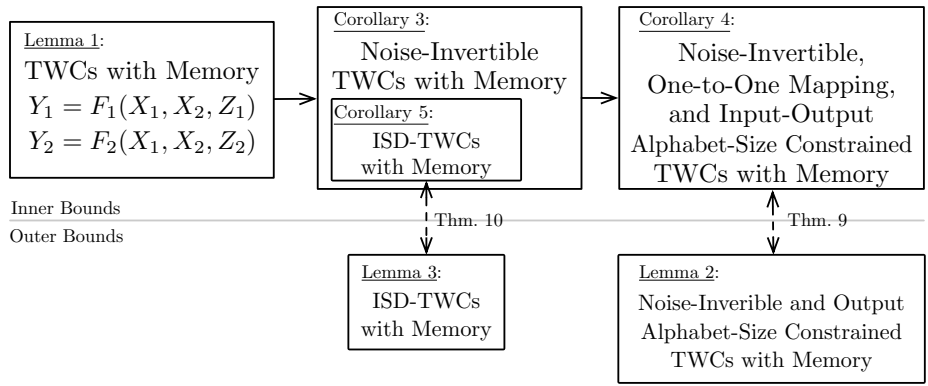}
\caption{The relationships between the results for point-to-point TWCs with memory. Here, $A\xdashleftrightarrow{\ \text{Thm.~C\ }} B$ indicates that results $A$ and $B$ are combined in Theorem~$C$ to determine the capacity region.}
\label{fig:impDia2}
\end{centering}
\end{figure*}

\noindent$\bullet$ \textbf{Point-to-Point TWCs with Memory:} 
a Shannon-type inner bound and an outer bound for the capacity of TWCs with memory under certain invertibility, one-to-one mapping, and alphabet size constraints are derived (Lemmas~\ref{thm:memin1}\mbox{-}\ref{lem:outer} and Corollaries~\ref{cor:memin1}-\ref{cor:inner2}). 
Two sufficient conditions for the tightness of the bounds are given (Theorems~\ref{thm:mem1} and~\ref{thm:ISDCap}). 
The first condition is derived for TWCs with strict invertibility and alphabet size constraints, characterizing channel capacity in single-letter form. 
The other condition is specialized for injective semi-deterministic TWCs with memory.\footnote{ISD-TWC model with memoryless noise were introduced in \cite{Chaaban:2017}. Here, we merely extend this setting by allowing noise processes with memory.} 
The obtained results are related as shown in Fig.~\ref{fig:impDia2}.
We also illustrate via a simple example that when the channel's memory is strong, the Shannon-type random coding scheme does not achieve capacity and adaptive coding is useful. 

\noindent$\bullet$ \textbf{Three-User Memoryless MA/DB TWCs}: we establish a Shannon-type inner bound and an outer bound for the capacity region of MA/DB TWCs (Theorems~\ref{the:3inner} and~\ref{the:3outer}) where both bounds admit a common rate expression but have different input distribution requirements. 
Three sufficient conditions (based on different techniques) for these bounds to coincide are established (Theorems~\ref{thm:exMain}-\ref{thm:exSC}). 
The first condition involves the existence of independent inputs that can achieve the outer bound (similar to the CVA approach).  
The second condition is derived from the viewpoint of two interacting state-dependent one-way channels. 
The last one focuses on the permutation invariance structure of the channel transition matrix (mirroring the Shannon symmetry method). 
The obtained results extend the results in \cite{Devroye:2014} and readily provide the capacity region for a larger class of MA/DB TWCs. 
While the channel model here is admittedly simplified, we note that our intention is to illustrate a potential methodology for determining the capacity regions of multi-user two-way channels and to motivate future work in this area.

The rest of the paper is organized as follows. In Section~\ref{sec:p2p}, point-to-point memoryless TWCs are investigated. 
TWCs with memory are studied in Section \ref{sec:memory}, and memoryless MA/DB TWCs are examined in Section \ref{sec:MADBC}. 
Concluding remarks are given in Section \ref{sec:conclusion}.

\section{Point-to-Point Memoryless TWCs}\label{sec:p2p}
In this section, we study two-user memoryless two-way networks. 
We first formally describe the general model for point-to-point TWCs (not necessarily memoryless) in Section~\ref{subsec:p2pmodel}, and then review the prior results for the memoryless case in Section~\ref{subsec:p2pmemoryless}.
New symmetry conditions are derived in Section~\ref{subsec:newcondset}, and we demonstrate how to apply these conditions to finding the channel capacity in Section~\ref{subsec:example}. 
Comparisons between prior results and our conditions are presented in Section~\ref{subsec:comp}, and the relationship between Shannon's condition and the CVA condition is examined in Section~\ref{subsec:summary}. 

\subsection{General Channel Model}\label{subsec:p2pmodel}
\begin{figure}[!t]
\begin{centering}
\includegraphics[scale=0.5]{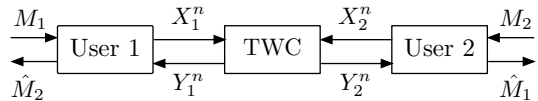}
\caption{The information flow of point-to-point two-way transmission.\label{fig:TWModel}}
\end{centering}
\end{figure}

In point-to-point two-way communication as shown in Fig.~\ref{fig:TWModel}, two users exchange messages $M_1$ and $M_2$ via $n$ channel uses. 
Here, $M_1$ and $M_2$ are assumed to be independent and uniformly distributed on the finite sets $\mathcal{M}_1 \triangleq \{ 1,2,..., 2^{nR_1} \} $ and $\mathcal{M}_2 \triangleq \{ 1,2,...,2^{nR_2} \}$, respectively, for some $R_1, R_2\ge 0$. 
Let $\mathcal{X}_j$ and $\mathcal{Y}_j$ be the channel input and output alphabets, respectively for $j=1, 2$. 
For $i=1, 2, \dots, n$, let $X_{j,i} \in \mathcal{X}_j$ and $Y_{j,i}\in \mathcal{Y}_j$ denote the channel input and output of user $j$ at time $i$, respectively.  
The joint probability distribution of all random variables for the entire transmission period is given by
\begin{IEEEeqnarray}{l}
P_{M_1, M_2, X_1^n, X_2^n, Y_1^n, Y_2^n}=P_{M_1}\cdot P_{M_2}\cdot\left(\prod\limits_{i=1}^n P_{X_{1,i}|M_1, Y_1^{i-1}}\right)\nonumber\\
\qquad\qquad\qquad\qquad\qquad\cdot\left(\prod\limits_{i=1}^n P_{X_{2, i}|M_2, Y_2^{i-1}}\right)\cdot\left(\prod\limits_{i=1}^n P_{Y_{1,i}, Y_{2, i}|X_1^i, X_2^i, Y_1^{i-1}, Y_2^{i-1}}\right),\nonumber
\end{IEEEeqnarray}
where $X_j^i\triangleq (X_{j, 1}, X_{j, 2}, \dots, X_{j, i})$ and $Y_j^i\triangleq (Y_{j, 1}, Y_{j, 2}, \dots, Y_{j, i})$ for $j=1, 2$. 
The $n$ transmissions over a point-to-point TWC can be then described by the sequence of conditional probabilities $\{ P_{Y_{1,i},Y_{2,i}|X_{1}^i,X_{2}^i,Y_{1}^{i-1},Y_{2}^{i-1} }\}_{i=1}^n$.

\begin{definition}{\label{def:CCforTWC}}
An $(n, R_1, R_2)$ code for a TWC consists of two message sets $\mathcal{M}_1=\{1, 2, \dots, \allowbreak 2^{nR_1}\}$ and $\mathcal{M}_2=\{1, 2, \dots, 2^{nR_2}\}$, two sequences of encoding functions $f_1^n\triangleq (f_{1,1}, f_{1,2}, \dots, f_{1,n})$ and $f_2^n\triangleq (f_{2,1}, f_{2,2}, \dots, f_{2,n})$ such that
\[
\begin{array}{ll}
X_{1,1}= f_{1, 1}(M_1),& X_{1,i}=f_{1, i}(M_1, Y_1^{i-1}),\\
X_{2,1}=f_{2, 1}(M_2),& X_{2,i}=f_{2, i}(M_2, Y_2^{i-1}),
\end{array}
\]
for $i=2, 3, \dots, n$, and two decoding functions $g_1$ and $g_2$ such that $\hat{M}_2=g_1(M_1, Y_1^n)$ and $\hat{M}_1=g_2(M_2, Y_2^n)$.
\end{definition}

When messages $M_1$ and $M_2$ are encoded via an $(n, R_1, R_2)$ channel code, the probability of decoding error is defined as $P^{(n)}_{\text{e}}(f_1^n, f_2^n, g_1, g_2)=\text{Pr}\{\hat{M}_1 \neq M_1\ \text{or}\ \hat{M}_2 \neq M_2\}.$

\begin{definition}{\label{def:AchRforTWC}}
A rate pair $(R_1,R_2)$ is said to be achievable if there exists a sequence of $(n, R_1, R_2)$ codes with $\lim_{n \to \infty} P^{(n)}_{\text{e}} =0.$
\end{definition}

\begin{definition}{\label{def:CapforTWC}}
The capacity region $\mathcal{C}$ of a point-to-point TWC is defined as the closure of the convex hull of all achievable rate pairs. 
\end{definition}

\subsection{Prior Results for Memoryless TWCs}\label{subsec:p2pmemoryless}
A point-to-point TWC is said to be \emph{memoryless} if its transition probabilities satisfy
\begin{align}
P_{Y_{1,i},Y_{2,i}|X_{1}^i,X_{2}^i,Y_{1}^{i-1},Y_{2}^{i-1} }  = P_{Y_{1},Y_{2}|X_{1},X_{2} }\nonumber
\end{align}
for some $P_{Y_{1},Y_{2}|X_{1},X_{2}}$ and all $i \ge 1$.
For a memoryless TWC with transition probability $P_{Y_1, Y_2|X_1, X_2}$ and input distribution $P_{X_1,X_2}$, let $\mathcal{R}(P_{X_1,X_2},P_{Y_1,Y_2|X_1,X_2})$ denote the set of all rate pairs $(R_1,R_2)$ constrained by 
\begin{align}
R_1 \le I(X_1;Y_2|X_2)\ \text{and}\ R_2 \le I(X_2;Y_1|X_1). \label{eq:rateR}
\end{align} 
In \cite{Shannon:1961}, Shannon showed that the capacity region of a discrete memoryless point-to-point TWC is inner bounded by
\begin{align*}
\mathcal{C}_{\text{I}} (P_{Y_1,Y_2|X_1,X_2})\triangleq \scalemath{1}{\overline{\text{co}}\left({\bigcup_{P_{X_1}, P_{X_2}}}\mathcal{R}(P_{X_1}{\cdot}P_{X_2}, P_{Y_1,Y_2|X_1,X_2}) \right)},
\end{align*}
and outer bounded by
 \begin{align*}
\mathcal{C}_{\text{O}} (P_{Y_1,Y_2|X_1,X_2}) \triangleq \overline{\text{co}} \scalemath{1}{\left(\bigcup_{P_{X_1,X_2}} \mathcal{R}(P_{X_1,X_2}, P_{Y_1,Y_2|X_1,X_2}) \right)},
 \end{align*}
 where $\overline{\text{co}}(\cdot)$ denotes taking the closure of the convex hull. 
In general, $\mathcal{C}_{\text{I}}$ and $\mathcal{C}_\text{O}$ are not matched to each other, but if they coincide, then the exact capacity region is obtained.
Our objective is to develop general conditions under which the two bounds coincide. 


In the following, the Shannon \cite{Shannon:1961} and CVA \cite{Chaaban:2017} conditions that imply the equality of $\mathcal{C}_\text{I}$ and $\mathcal{C}_\text{O}$ are summarized. 
In short, the Shannon condition focuses on the permutation invariance structure of the channel transition matrix $[P_{Y_1, Y_2|X_1, X_2}(\cdot, \cdot|\cdot, \cdot)]$, while the CVA condition involves the existence of independent inputs which can achieve the outer bound.
Throughout the paper, we use $I^{(l)}(X_k; Y_j|X_j)$ and $H^{(l)}(Y_j|X_1, X_2)$ to denote the conditional mutual information and the conditional entropy evaluated under input distribution $P^{(l)}_{X_1, X_2}$ for $j, k=1, 2$ with $j\neq k$. 
For $P^{(l)}_{X_1, X_2}=P^{(l)}_{X_j}\cdot P^{(l)}_{X_k|X_j}$ with $j\ne k$, the conditional entropy $H^{(l)}(Y_j|X_j)$ is evaluated using the marginal distribution $P^{(l)}_{Y_j|X_j}(y_j|x_j)=\sum_{x_k}P_{X_k|X_j}^{(l)}(x_k|x_j)\cdot P_{Y_j|X_j, X_k}(y_j|x_j, x_k)$.
Also, for a finite set $\mathcal{A}$, let $\pi^\mathcal{A}: \mathcal{A} \to \mathcal{A}$ denote a permutation (bijection), and for any two symbols $a'$ and $a''$ in $\mathcal{A}$, let $\tau^{\mathcal{A}}_{a',a''}: \mathcal{A} \to \mathcal{A}$ denote the transposition which swaps $a'$ and $a''$ in $\mathcal{A}$, but leaves the other symbols unaffected. 
Finally, let $\mathcal{P}(\mathcal{X}_j)$ denote the set of all probability distributions on $\mathcal{X}_j$, and define $P^{\text{U}}_{\mathcal{X}_j}$ as the uniform probability distribution on $\mathcal{X}_j$ for $j=1, 2$.  

\begin{proposition}[Shannon's One-Sided Symmetry Condition \cite{Shannon:1961}]\label{thm:SC}
For a memoryless TWC with transition probability $P_{Y_1,Y_2|X_1,X_2}$, we have that $\mathcal{C}=\mathcal{C}_\text{I}=\mathcal{C}_\text{O}$ if for any pair of distinct input symbols $x'_1, \allowbreak x''_1\in\mathcal{X}_1$, there exists a pair of permutations 
$(\pi^{\mathcal{Y}_1}[x'_1,x''_1], \pi^{\mathcal{Y}_2}[x'_1,x''_1])$ on $\mathcal{Y}_1$ and $\mathcal{Y}_2$, respectively, (which depend on $x'_1$ and $x''_1$) such that for all $x_1$, $x_2$, $y_1$, $y_2$, 
\begin{IEEEeqnarray}{l}
P_{Y_1,Y_2|X_1,X_2}(y_1,y_2|x_1,x_2)=\scalemath{1}{P_{Y_1,Y_2|X_1,X_2}(\pi^{\mathcal{Y}_1}[x'_1,x''_1](y_1),\pi^{\mathcal{Y}_2}[x'_1,x''_1](y_2)|\tau^{\mathcal{X}_1}_{x'_1,x''_1} (x_1),x_2)}.\IEEEeqnarraynumspace\label{SCcond}
\end{IEEEeqnarray}
Under this condition, the capacity region is given by 
\begin{IEEEeqnarray}{l}
\mathcal{C}=\overline{\text{co}} \left( \bigcup_{P_{X_2}}\mathcal{R}\Big(P^{\text{U}}_{\mathcal{X}_1}{\cdot}P_{X_2},P_{Y_1,Y_2|X_1,X_2}\Big) \right).\label{eq:shannoncap}
\end{IEEEeqnarray}
\end{proposition}

In \cite{Shannon:1961}, the proof of Proposition~\ref{thm:SC} is only sketched. 
To make the paper self-contained and facilitate the understanding of a technique used to derive one of our results (Theorem~\ref{thm:exSC}), we provide a full proof in Appendix \ref{appendix:SCproof}. 
Note that Proposition~\ref{thm:SC} describes a channel symmetry property with respect to the channel input of user~1, but an analogous condition can be obtained by exchanging the roles of users~1 and~2.
The proposition below immediately follows from Proposition~1. 

\begin{proposition}[Shannon's Two-Sided Symmetry Condition \cite{Shannon:1961}]\label{thm:SC2}
For a memoryless TWC with transition probability $P_{Y_1,Y_2|X_1,X_2}$, we have that $\mathcal{C}=\mathcal{C}_\text{I}=\mathcal{C}_\text{O}$ if the TWC satisfies the one-sided symmetry condition with respect to both channel inputs. 
In this case, the capacity region is rectangular and given by
$\mathcal{C}=\mathcal{R}(P^{\text{U}}_{\mathcal{X}_1}{\cdot}P^{\text{U}}_{\mathcal{X}_2},P_{Y_1,Y_2|X_1,X_2}).$
\end{proposition}

\begin{proposition}[CVA Condition \cite{Chaaban:2017}]\label{thm:CVA}
\sloppy For a memoryless TWC with transition probability $P_{Y_1,Y_2|X_1,X_2}$, we have that $\mathcal{C}=\mathcal{C}_\text{I}=\mathcal{C}_\text{O}$ if $H(Y_j|X_1, X_2)$, $j=1, 2$, does not depend on $P_{X_1|X_2}$ for any fixed $P_{X_2}$ and $P_{Y_j|X_1, X_2}$, and for any $P^{(1)}_{X_1,X_2} = P^{(1)}_{X_2}\cdot P^{(1)}_{X_1|X_2}$ there exists $\tilde{P}_{X_1}\in\mathcal{P}(\mathcal{X}_1)$ such that $H^{(1)}(Y_j|X_j)\le H^{(2)}(Y_j|X_j)$ for $j=1, 2$, where $P^{(2)}_{X_1, X_2}=\tilde{P}_{X_1}\cdot P^{(1)}_{X_2}$.
\end{proposition}

Thus, if a TWC satisfies any one of the above conditions, the capacity region can be determined by considering independent inputs:  $P_{X_1, X_2}=P_{X_1}\cdot P_{X_2}$. 
This result implies that adaptive coding, where channel inputs are generated by interactively adapting to the previously received signals, cannot improve the users' achievable rates and that Shannon's random coding scheme is optimal.  
The class of memoryless ISD-TWCs \cite{Chaaban:2017} satisfies the CVA condition (but do not necessarily satisfy the Shannon condition) and hence adaptive coding is useless for such channels. 
A TWC with independent $q$-ary additive noise \cite{Song:2016} is an  example of a channel that satisfies both the Shannon and CVA conditions. 
Although the CVA condition does not require any permutation invariance on the channel marginal distribution $P_{Y_j|X_1, X_2}$, the invariance requirement of $H(Y_j|X_1, X_2)$'s in Proposition~\ref{thm:CVA} does in fact impose a certain symmetry constraint on $P_{Y_j|X_1, X_2}$. 
More details about these conditions will be provided in the proof of Theorem~\ref{thm:CVAto24} and Section~\ref{subsec:summary}.

\subsection{Conditions for the Tightness of Shannon's Inner and Outer Bounds}\label{subsec:newcondset}
In this section, we present conditions that guarantee the tightness of Shannon's inner bound by considering a TWC as two interacting state-dependent one-way channels. 
For example, the state-dependent one-way channel from user~$1$ to user~$2$ is governed by the marginal distribution $P_{Y_2|X_1, X_2}$ (derived from the channel probability $P_{Y_1, Y_2|X_1, X_2}$), where $X_1$ and $Y_2$ are respectively the input and the output of the channel with state $X_2$. 

Let $P_X$ and $P_{Y|X}$ be probability distributions on $\mathcal{X}$ and $\mathcal{Y}$, respectively. 
To simplify the presentation, we use 
\begin{IEEEeqnarray}{l}
\mathcal{I}(P_{X}, P_{Y|X}) =\sum\limits_{x, y}P_{X}(x)P_{Y|X}(y|x)\log
\frac{P_{Y|X}(y|x)}{\sum_{x'}P_{X}(x')P_{Y|X}(y|x')},\nonumber\label{SMI}
\end{IEEEeqnarray}
as an alternative way of writing the mutual information $I(X; Y)$ between input $X$ (governed by $P_X$) and corresponding output $Y$ of a channel with transition probability $P_{Y|X}$.
A useful fact is that $\mathcal{I}(\cdot, \cdot)$ is concave in the first argument when the second argument is fixed.  
Moreover, the conditional mutual information $I(X_1; Y_2|X_2=x_2)$ can be expressed as $\mathcal{I}(P_{X_1|X_2=x_2}, P_{Y_2|X_1, X_2=x_2})$. 

Since the TWC is viewed as two state-dependent one-way channels, each of the following theorems consists of two conditions, one for each direction of the two-way transmission. 
By symmetry, these theorems are valid if the roles of users~1 and~2 are swapped. 


\begin{theorem}
\label{thm:23}
For a memoryless TWC, if conditions (i) and (ii) below are satisfied, then $\mathcal{C}_\text{I}=\mathcal{C}_\text{O}$. 
\begin{itemize}
\item[(i)] There exists $P^*_{X_1}\in\mathcal{P}(\mathcal{X}_1)$ such that \[\argmax_{P_{X_1|X_2=x_2}} I(X_1; Y_2|X_2=x_2)=P^*_{X_1}\] for all $x_2\in\mathcal{X}_2$; 
\item[(ii)] $\mathcal{I}(P_{X_2}, P_{Y_1|X_1=x_1, X_2})$ does not depend on $x_1\in\mathcal{X}_1$ for any fixed $P_{X_2}\in\mathcal{P}(\mathcal{X}_2)$. 
\end{itemize}
\end{theorem}
\begin{IEEEproof}
For any $P^{(1)}_{X_1, X_2}=P^{(1)}_{X_2}\cdot P^{(1)}_{X_1|X_2}$, let $P^{(2)}_{X_1, X_2}=P^*_{X_1}\cdot P^{(1)}_{X_2}$, where $P^*_{X_1}$ is given by (i). 
In light of (i), we have
\begin{IEEEeqnarray}{rCl}
I^{(1)}(X_1; Y_2|X_2) &=&\sum_{x_2}P^{(1)}_{X_2}(x_2)\cdot I^{(1)}(X_1; Y_2|X_2=x_2)\label{s3}\\
  &\le & \sum_{x_2}P^{(1)}_{X_2}(x_2)\cdot\scalemath{1}{\left[\max_{P_{X_1|X_2=x_2}}I(X_1; Y_2|X_2=x_2)\right]}\\
  &=&  \sum_{x_2}P^{(1)}_{X_2}(x_2)\cdot \mathcal{I}(P^*_{X_1}, P_{Y_2|X_1, X_2=x_2})\label{s5}\\
  &=& \sum_{x_2}P^{(1)}_{X_2}(x_2)\cdot I^{(2)}(X_1; Y_2|X_2=x_2)\\
  &=& I^{(2)}(X_1; Y_2|X_2)\label{s4}.
\end{IEEEeqnarray}
Moreover, 
\begin{IEEEeqnarray}{rCl}
I^{(1)}(X_2; Y_1|X_1)&=&\sum_{x_1}P^{(1)}_{X_1}(x_1)\cdot I^{(1)}(X_2; Y_1|X_1=x_1)\nonumber\label{cond3s1}\\
 &=&\sum_{x_1}P^{(1)}_{X_1}(x_1)\cdot\mathcal{I}(P^{(1)}_{X_2|X_1=x_1}, P_{Y_1|X_1=x_1, X_2})\nonumber\\
 &=&\sum_{x_1}P^{(1)}_{X_1}(x_1)\cdot\mathcal{I}(P^{(1)}_{X_2|X_1=x_1}, P_{Y_1|X_1=x'_1, X_2})\label{cond3s2}\\
 &\le &\scalemath{1}{\mathcal{I}\left(\sum_{x_1}P^{(1)}_{X_1}(x_1)P^{(1)}_{X_2|X_1}(x_2|x_1), P_{Y_1|X_1=x'_1, X_2}\right)}\label{cond3s3}\\
 &=& \mathcal{I}(P^{(1)}_{X_2}, P_{Y_1|X_1=x'_1, X_2})\nonumber\\
 &=& \sum_{x'_1}{P}^*_{X_1}(x'_1)\cdot\mathcal{I}(P^{(1)}_{X_2}, P_{Y_1|X_1=x'_1, X_2})\label{cond3s5}\\
 &=& I^{(2)}(X_2; Y_1|X_1)\label{cond3s6}, 
\end{IEEEeqnarray}
where (\ref{cond3s2}) holds by the invariance assumption in (ii) and $x'_1\in\mathcal{X}_1$ is arbitrary, (\ref{cond3s3}) holds since the functional $\mathcal{I}(\cdot, \cdot)$ is concave in the first argument, and (\ref{cond3s5}) is obtained from the invariance assumption in (ii). 
Combining the above yields $\mathcal{R}(P^{(1)}_{X_1,X_2}, P_{Y_1,Y_2|X_1,X_2})\subseteq \mathcal{R}(P^*_{X_1}\cdot P^{(1)}_{X_2}, P_{Y_1,Y_2|X_1,X_2})$, which implies that $C_{\text{O}}\subseteq C_{\text{I}}$ and hence $C_{\text{I}}=C_{\text{O}}$.  
\end{IEEEproof}

Instead of relying on the permutation invariance (row, column, or both) of the channel transition matrix, the symmetry property in the theorem is characterized by a combination of two symmetry properties for state-dependent one-way channels in terms of mutual information: (1) common capacity-achieving input distribution; (2) invariance of input-output mutual information. A special case where condition (i) of Theorem~\ref{thm:23} trivially holds is when each one-way channel $P_{Y_2|X_1, X_2=x_2}$, $x_2\in\mathcal{X}_2$, is $T$-symmetric\footnote{A point-to-point one way channel is called $T$-symmetric if the optimal input distribution (that maximizes the channel's mutual information) is uniform.} \cite{Grant:2004}; in this case we have $P^*_{X_1}=P^{\text{U}}_{\mathcal{X}_1}$.  

We next apply condition (ii) of Theorem~\ref{thm:23} for both directions of the two-way transmission.  
\begin{theorem}
\label{thm:33}
For a memoryless TWC, if conditions (i) and (ii) below are satisfied, then $\mathcal{C}_\text{I}=\mathcal{C}_\text{O}$. 
\begin{itemize}
\item[(i)] $\mathcal{I}(P_{X_1}, P_{Y_2|X_1, X_2=x_2})$ does not depend on $x_2\in\mathcal{X}_2$ for any fixed $P_{X_1}\in\mathcal{P}(\mathcal{X}_1)$;
\item[(ii)] $\mathcal{I}(P_{X_2}, P_{Y_1|X_1=x_1, X_2})$ does not depend on $x_1\in\mathcal{X}_1$ for any fixed $P_{X_2}\in\mathcal{P}(\mathcal{X}_2)$.
\end{itemize}
\end{theorem}
\begin{IEEEproof}
From conditions (i) and (ii), we know that $\max_{P_{X_1|X_2=x_2}} I(X_1; Y_2|X_2=x_2)$ has a common maximizer $P^*_{X_1}$ for all $x_2\in\mathcal{X}_2$ and that $\max_{P_{X_2|X_1=x_1}} I(X_2; Y_1|X_1=x_1)$ has a common maximizer $P^*_{X_2}$ for all $x_1\in\mathcal{X}_1$.  
For any $P^{(1)}_{X_1, X_2}=P^{(1)}_{X_1}\cdot P^{(1)}_{X_2|X_1}$, let $P^{(2)}_{X_1, X_2}=P^*_{X_1}\cdot P^*_{X_2}$. 
Using the same argument as in (\ref{s3})-(\ref{s4}) and applying condition (ii) to \eqref{s5}, we conclude that $I^{(1)}(X_1; Y_2|X_2)\le I^{(2)}(X_1; Y_2|X_2)$ and $I^{(1)}(X_2; Y_1|X_1)\le \allowbreak I^{(2)}(X_2; Y_1|X_1)$. 
Thus, $\mathcal{R}(P^{(1)}_{X_1,X_2}, P_{Y_1,Y_2|X_1,X_2})\subseteq \mathcal{R}(P^*_{X_1}\cdot P^*_{X_2}, P_{Y_1,Y_2|X_1,X_2})$, which yields $\mathcal{C}_\text{I}=\mathcal{C}_\text{O}$. 
\end{IEEEproof}

To verify condition (i) in Theorem \ref{thm:23}, one should find optimal input distributions for the one-way channel from users~1 to~2 for each state $x_2\in\mathcal{X}_2$, say, via the Blahut-Arimoto algorithm \cite{Blahut:1972}.
This process can sometimes be simplified by testing whether the uniform input distribution is optimal via the Karush-Kuhn-Tucker (KKT) conditions for one-way channel capacity \cite{Gallager:1968}. 
However, verifying condition (ii) in Theorem~\ref{thm:23} may necessitate the evaluation of $\mathcal{I}(P_{X_2},P_{Y_1|X_1,X_2}(\cdot|x_1,\cdot))$ for all $P_{X_2}\in\mathcal{P}(\mathcal{X}_2)$ and $x_1 \in \mathcal{X}_1$. 
In practice, such a verification is often complex, especially when the size of the input alphabet is large. 
Similar difficulties arise when ascertaining the conditions of Theorem~\ref{thm:33}.  
In the following results, conditions that are easier to check are presented.

\begin{theorem}\label{thm:nc2}
For a memoryless TWC, if conditions (i) and (ii) below are satisfied, then $\mathcal{C}_\text{I}=\mathcal{C}_\text{O}$. 
\begin{itemize}
\item[(i)] There exists $P^*_{X_1}\in\mathcal{P}(\mathcal{X}_1)$ such that \[\argmax_{P_{X_1|X_2=x_2}} I(X_1; Y_2|X_2=x_2)=P^*_{X_1}\] for all $x_2\in\mathcal{X}_2$ and $\mathcal{I}(P^*_{X_1}, P_{Y_2|X_1, X_2=x_2})$ does not depend on $x_2\in\mathcal{X}_2$; 
\item[(ii)] There exists $P^*_{X_2}\in\mathcal{P}(\mathcal{X}_2)$ such that \[\argmax_{P_{X_2|X_1=x_1}} I(X_2; Y_1|X_1=x_1)=P^*_{X_2}\] for all $x_1\in\mathcal{X}_1$ and $\mathcal{I}(P^*_{X_2}, P_{Y_1|X_1=x_1, X_2})$ does not depend on $x_1\in\mathcal{X}_1$. 
\end{itemize}
\end{theorem}
\begin{IEEEproof}
\sloppy For any $P^{(1)}_{X_1, X_2}=P^{(1)}_{X_2}\cdot P^{(1)}_{X_1|X_2}$, consider $P^{(2)}_{X_1, X_2}=P^*_{X_1}\cdot P^*_{X_2}$, where $P^*_{X_1}$ and $P^*_{X_2}$ are given by (i) and (ii), respectively. 
Following the same steps as in \eqref{s3}-\eqref{s4} and using the second part of condition (i), we obtain that $I^{(1)}(X_1; Y_2|X_2) \le I^{(2)}(X_1; Y_2|X_2)$. 
By a similar argument, we obtain the inequality $I^{(1)}(X_2; Y_1|X_1)\allowbreak \le I^{(2)}(X_2; Y_1|X_1)$. 
Hence, $\mathcal{R}(P^{(1)}_{X_1,X_2}, P_{Y_1,Y_2|X_1,X_2})$ $\subseteq \mathcal{R}(P^*_{X_1}{\cdot}P^*_{X_2}, P_{Y_1,Y_2|X_1,X_2})$ which implies $\mathcal{C}_\text{I}=\mathcal{C}_\text{O}$. 
\end{IEEEproof}

Unlike condition (ii) of Theorem~\ref{thm:23} and the conditions in Theorem~\ref{thm:33}, Theorem~\ref{thm:nc2} only requires checking the existence of a common maximizer and testing whether $\mathcal{I}(P^*_{X_1}, P_{Y_2|X_1, X_2=x_2})$ is invariant with respect to $x_2\in\mathcal{X}_2$ and $\mathcal{I}(P^*_{X_2}, P_{Y_1|X_1=x_1, X_2})$ is invariant with respect to $x_1\in\mathcal{X}_1$, thus significantly reducing the validation computational complexity vis-a-vis Theorems~\ref{thm:23} and~\ref{thm:33}. 

The next two corollaries provide even simpler conditions. 
Let $[P_{Y_2|X_1,X_2}(\cdot|\cdot,x_2)]$ denote the transition matrix of the channel from users~1 to~2 when the input of user~2 is fixed to be $x_2$. The matrix $[P_{Y_2|X_1,X_2}(\cdot|\cdot,x_2)]$ has size $|\mathcal{X}_1| \times |\mathcal{Y}_2|$ and its entry at the $x_1$th row and $y_2$th column is $P_{Y_2|X_1,X_2}(y_2|x_1,x_2)$. Similarly, let $[P_{Y_2|X_1,X_2}(\cdot|x_1,\cdot)]$ denote the transition matrix of the channel from users~2 to~1 when the input of user~1 is fixed to be $x_1$.
\begin{corollary}\label{cor:nc1}
For a memoryless TWC, if conditions (i) and (ii) below are satisfied, then $\mathcal{C}_\text{I}=\mathcal{C}_\text{O}$. 
\begin{itemize}
\item[(i)] The channel with transition matrix $[P_{Y_2|X_1,X_2}(\cdot|\cdot,x_2)]$ is quasi-symmetric\footnote{A discrete memoryless channel with transition matrix $[P_{Y|X}(\cdot|\cdot)]$ is said to be  weakly-symmetric if the rows are permutations of each other 
and all the column sums  are identical \cite{Cover:2006}. 
A discrete memoryless channel is said to be quasi-symmetric if its transition matrix $[P_{Y|X}(\cdot|\cdot)]$  can be partitioned along its columns into  weakly-symmetric
sub-matrices \cite{Alajaji}.}
for all $x_2 \in \mathcal{X}_2$; 
\item[(ii)] The matrices $[P_{Y_1|X_1, X_2}(\cdot|x_1, \cdot)]$, $x_1\in\mathcal{X}_1$, are column permutations of each other.
\end{itemize}
\end{corollary}
\begin{IEEEproof}
It suffices to show that conditions (i) and (ii) imply the conditions of Theorem~\ref{thm:23}. 
Under condition (i), we obtain a common maximizer given by $P^*_{X_1}=P^{\text{U}}_{\mathcal{X}_1}$ since the optimal input distribution for a quasi-symmetric channel is the uniform distribution \cite{Alajaji}; this  implies condition~(i) of Theorem~\ref{thm:23}. 
Furthermore, we observe that $\mathcal{I}(P_{X_2}, P_{Y_1|X_1,X_2}(\cdot|x_1,\cdot))$ is invariant with respect to column permutations of the transition matrix $P_{Y_1|X_1,X_2}(\cdot|x_1,\cdot)$ for given $P_{X_2}$. 
Since the matrices $[P_{Y_1|X_1,X_2}(\cdot|x_1,\cdot)]$, $x_1\in\mathcal{X}_1$, are column permutations of each other, we conclude that $\mathcal{I}(P_{X_2}, P_{Y_1|X_1=x_1, X_2})$ does not depend on $x_1\in\mathcal{X}_1$ for any fixed $P_{X_2}\in\mathcal{P}(\mathcal{X}_2)$, which is the second condition of Theorem~\ref{thm:23}. 
\end{IEEEproof} 

\begin{corollary}
\label{cor:77}
For a memoryless TWC, if conditions (i) and (ii) below are satisfied, then $\mathcal{C}_\text{I}=\mathcal{C}_\text{O}$. 
\begin{itemize}
\item[(i)] The matrices $[P_{Y_2|X_1, X_2}(\cdot|\cdot, x_2)]$, $x_2\in\mathcal{X}_2$, are column permutations of each other; 
\item[(ii)] The matrices $[P_{Y_1|X_1, X_2}(\cdot|x_1, \cdot)]$, $x_1\in\mathcal{X}_1$, are column permutations of each other.
\end{itemize}
\end{corollary}
\begin{IEEEproof}
It suffices to show that conditions (i) and (ii) imply the conditions of Theorem~\ref{thm:33}.
This can be done using a similar argument as in the second part of the proof of Corollary~\ref{cor:nc1}, and hence the details are omitted. 
\end{IEEEproof}

If the transition probability $P_{Y_1, Y_2|X_1, X_2}$ satisfies conditions (i) and (ii) of Theorem~\ref{thm:23}, the capacity region is given by
\begin{equation}
\label{cap:thm1}
\mathcal{C} = \overline{\text{co}} \left( \bigcup_{P_{X_2}} \mathcal{R}(P^*_{X_1}{\cdot}P_{X_2}, P_{Y_1,Y_2|X_1,X_2}) \right),
\end{equation}
where $P^*_{X_1}$ is given by condition (i). 
For example, condition (i) trivially holds when each one-way channel with fixed state $x_2\in\mathcal{X}_2$ from users~1 to~2 is $T$-symmetric. In this case, we have $P^*_{X_1}=P^\text{U}_{\mathcal{X}_1}$ and the capacity region becomes 
\begin{equation}
\label{cap:cor1}
\mathcal{C} = \overline{\text{co}} \left( \bigcup_{P_{X_2}} \mathcal{R}(P^{\text{U}}_{\mathcal{X}_1}{\cdot}P_{X_2}, P_{Y_1,Y_2|X_1,X_2}) \right). 
\end{equation}
In fact, this is also the capacity region for memoryless TWCs which satisfy Corollary \ref{cor:nc1} because condition (ii) of Corollary~\ref{cor:nc1} implies condition (ii) of Theorem~\ref{thm:23} (this follows from the proof of Corollary~\ref{cor:nc1}). 
Moreover, the proof of Theorem~\ref{thm:33} demonstrates that a common maximizer exists for each direction of the two-way transmission under the conditions of Theorem~\ref{thm:33}.
Let $\argmax_{P_{X_1|X_2=x_2}} I(X_1; Y_2|X_2=x_2)=P^*_{X_1}$ for all $x_2\in\mathcal{X}_2$ and $\argmax_{P_{X_2|X_1=x_1}} I(X_2; Y_1|X_1=x_1)=P^*_{X_2}$ for all $x_1\in\mathcal{X}_1$.   
A TWC which satisfies the conditions of Theorem~\ref{thm:33} has the capacity region
\begin{equation}
\label{cap:thm2}
\mathcal{C}= \mathcal{R}(P^*_{X_1}{\cdot}P^*_{X_2}, P_{Y_1,Y_2|X_1,X_2}).
\end{equation}
The region is rectangular which suggests that such a two-way transmission inherently comprises two independent one-way transmissions. 
A memoryless TWC that satisfies the conditions in either Theorem~\ref{thm:nc2} or Corollary~\ref{cor:77} also has a capacity region given by (\ref{cap:thm2}).
 
To end this section, we remark that it is possible to combine different conditions to determine the capacity region of a broader class of memoryless TWCs as shown below. 

\begin{theorem}
\label{thm:24}
For a memoryless TWC, if both of the following conditions are satisfied, then $\mathcal{C}=\mathcal{C}_\text{I}=\mathcal{C}_\text{O}$ with $\mathcal{C}$ given by \eqref{cap:thm1}:
\begin{itemize}
\item[(i)] There exists $P^*_{X_1}\in\mathcal{P}(\mathcal{X}_1)$ such that  \[\argmax_{P_{X_1|X_2=x_2}} I(X_1; Y_2|X_2=x_2)=P^*_{X_1}\] for all $x_2\in\mathcal{X}_2$;  
\item[(ii)] $H(Y_1|X_1, X_2)$ does not depend on $P_{X_1|X_2}$ given $P_{X_2}$ and $P_{Y_1|X_1, X_2}$, and $P^*_{X_1}$ given in (i) satisfies  
$H^{(1)}(Y_1|X_1)\le H^{(2)}(Y_1|X_1)$ for any $P^{(1)}_{X_1, X_2}=P^{(1)}_{X_2}\cdot P^{(1)}_{X_1|X_2}$, where $P^{(2)}_{X_1, X_2}=P^*_{X_1}\cdot P^{(1)}_{X_2}$.
\end{itemize}
\end{theorem}
Here, condition (i) is directly from Theorem~\ref{thm:23}; condition (ii) is obtained by extracting the CVA condition related to the channel from user~2 to user~1.  
In order that the two conditions jointly determine the capacity region, the $\tilde{P}_{X_1}$ required by the CVA condition is forced to be $P^*_{X_1}$. 
\begin{IEEEproof}[Proof of Theorem~\ref{thm:24}]
Given any $P^{(1)}_{X_1, X_2}=P^{(1)}_{X_2}\cdot P^{(1)}_{X_1|X_2}$, let $P^{(2)}_{X_1, X_2}=P^*_{X_1}\cdot P^{(1)}_{X_2}$. 
Invoking the same argument as in (\ref{s3})-(\ref{s4}), we obtain that $I^{(1)}(X_1; Y_2|X_2)\le I^{(2)}(X_1; Y_2|X_2)$ using condition (i). 
Moreover, condition (ii) implies that $I^{(1)}(X_2; Y_1|X_1)=H^{(1)}(Y_1|X_1)-H^{(1)}(Y_1|X_1, X_2)\le H^{(2)}(Y_1|X_1)-H^{(2)}(Y_1|X_1, X_2)=I^{(2)}(X_2; Y_1|X_1)$.
Combining the above then completes the proof.
\end{IEEEproof}

\subsection{Examples}\label{subsec:example}
We next illustrate the proposed conditions via examples.   
\begin{example}[Memoryless Binary Additive-Noise TWCs with Erasures]\label{ex:BANE}
Let $\mathcal{X}_1=\mathcal{X}_2=\{0, 1\}$ and $\mathcal{Y}_1=\mathcal{Y}_2=\mathcal{Z}=\{0, 1, \mathbf{E}\}$, where $\mathbf{E}$ denotes channel erasure. 
A binary additive noise TWC with erasures is defined by the channel equations
\begin{align*}
\scalemath{1}{Y_{1,i} = (X_{1,i} \oplus_2 X_{2,i} \oplus_2 Z_{1,i}){\cdot}\mathbf{1}\{Z_{1, i}\neq \mathbf{E}\} + \mathbf{E}{\cdot}\mathbf{1}\{Z_{1, i}= \mathbf{E}\}},\\
\scalemath{1}{Y_{2,i} = (X_{1,i} \oplus_2 X_{2,i} \oplus_2 Z_{2,i}){\cdot}\mathbf{1}\{Z_{2, i}\neq \mathbf{E}\} + \mathbf{E}{\cdot}\mathbf{1}\{Z_{2, i}= \mathbf{E}\}},
\end{align*}  
where $\oplus_2$ denotes modulo-$2$ addition, $\{(Z_{1,i},Z_{2,i})\}_{i=1}^\infty$ is a memoryless joint noise-erasure process that is independent of the users' messages and has components $Z_{1,i}, Z_{2,i}\in\mathcal{Z}$ such that $\Pr(Z_{j,i}=\mathbf{E})=\varepsilon_j$ and $\Pr(Z_{j,i}=1)=\alpha_j$, where $0\le \varepsilon_j+\alpha_j\le 1 $ for $j=1,2$, and $\mathbf{1}\{\cdot\}$ denotes the indicator function. Here, we adopt the convention $\mathbf{E}\cdot 0=0$ and $\mathbf{E}\cdot 1=\text{E}$ to simplify the representation of the channel equations.\footnote{Strictly speaking, $X_{1,i} \oplus_2 X_{2,i} \oplus_2 Z_{j,i}$ is undefined when $Z_{j,i}=\mathbf{E}$, but we set $(X_{1,i} \oplus_2 X_{2,i} \oplus_2 \mathbf{E})\cdot 0=0$.} 
The channel equations yield the following transition matrices for the one-way channels:
\begin{align*}
& [P_{Y_2|X_1,X_2}(\cdot|\cdot,0)] = 
\begin{pmatrix}
1-\varepsilon_2-\alpha_2 & \alpha_2 & \varepsilon_2\\
\alpha_2 & 1-\varepsilon_2-\alpha_2 & \varepsilon_2
\end{pmatrix},\\
& [P_{Y_2|X_1,X_2}(\cdot|\cdot,1)] = 
\begin{pmatrix}
\alpha_2 & 1-\varepsilon_2-\alpha_2 &\varepsilon_2\\
1-\varepsilon_2-\alpha_2 & \alpha_2 &  \varepsilon_2
\end{pmatrix},\\
& [P_{Y_1|X_1,X_2}(\cdot|0,\cdot)] = 
\begin{pmatrix}
1-\varepsilon_1-\alpha_1 & \alpha_1 & \varepsilon_1\\
\alpha_1 & 1-\varepsilon_1-\alpha_1 & \varepsilon_1
\end{pmatrix},\\
& [P_{Y_1|X_1,X_2}(\cdot|1,\cdot)] = 
\begin{pmatrix}
\alpha_1 & 1-\varepsilon_1-\alpha_1 &\varepsilon_1\\
1-\varepsilon_1-\alpha_1 & \alpha_1 &  \varepsilon_1
\end{pmatrix}, 
\end{align*}
where the rows are indexed by $0$ and $1$ (from top to bottom) and the columns are indexed by $0$, $1$, and $\mathbf{E}$ (from left to right).
As all our proposed conditions are only based on the marginal transition probabilities, the relationship between $Z_{1, i}$ and $Z_{2, i}$ can be arbitrary. 
By Corollary \ref{cor:77}, we obtain that the optimal channel input distribution is $P^*_{X_1}\cdot P^*_{X_2}=P^{\text{U}}_{\mathcal{X}_1}\cdot P^{\text{U}}_{\mathcal{X}_2}$ since the marginal channel transition matrices not only exhibit column permutation properties but also are quasi-symmetric. 
The capacity region is given by
\begin{IEEEeqnarray}{rCl}
\mathcal{C}=\Bigg\{ (R_1,R_2): R_1 & \le & (1-\varepsilon_2){\cdot}\Big(1-H_{\text{b}}\Big(\frac{\alpha_2}{1-\varepsilon_2}\Big)\Big),\nonumber\\
\qquad\qquad\qquad\qquad R_2  &\le & (1-\varepsilon_1){\cdot}\Big(1- H_{\text{b}}\Big(\frac{\alpha_1}{1-\varepsilon_1}\Big) \Big)\Bigg\},\nonumber
\end{IEEEeqnarray}
where $H_\text{b}(\cdot)$ denotes the binary entropy function.
One can verify that this TWC also satisfies the conditions of Theorems~\ref{thm:23}-\ref{thm:nc2} and Corollary~\ref{cor:nc1}.
\end{example}

\begin{remark}
Various TWCs are special cases of this TWC model:  
\begin{enumerate}
\item
If $\alpha_1=\alpha_2=0$, then the memoryless binary additive TWC with erasures is recovered:   
\begin{align*}
Y_{1,i} = (X_{1,i} \oplus_2 X_{2,i}){\cdot}\mathbf{1}\{Z_{1, i}\neq \mathbf{E}\} + \mathbf{E}{\cdot}\mathbf{1}\{Z_{1, i}= \mathbf{E}\},\\
Y_{2,i} = (X_{1,i} \oplus_2 X_{2,i}){\cdot}\mathbf{1}\{Z_{2, i}\neq \mathbf{E}\} + \mathbf{E}{\cdot}\mathbf{1}\{Z_{2, i}= \mathbf{E}\}.
\end{align*} 
The capacity region is given by $$\mathcal{C}=\{ (R_1,R_2): R_1 \le 1-\varepsilon_2, R_2 \le 1-\varepsilon_1 \}.$$
\item
If $\varepsilon_1 = \varepsilon_2 = 0$, then the memoryless binary additive-noise TWC is obtained:  
\begin{align*}
Y_{1,i} = X_{1,i} \oplus_2 X_{2,i} \oplus_2 Z_{1,i},\\
Y_{2,i} = X_{1,i} \oplus_2 X_{2,i} \oplus_2 Z_{2,i}.
\end{align*} 
The capacity region of this channel is given by $$\mathcal{C}=\{ (R_1,R_2): R_1 \le 1-H_{\text{b}}(\alpha_2), R_2 \le 1-H_{\text{b}}(\alpha_1) \}. $$
\item
If $\varepsilon_1 = \varepsilon_2 = 0$ and $\alpha_1=\alpha_2=0$, then we obtain the memoryless binary additive TWC given by $Y_{1,i} = X_{1,i} \oplus_2 X_{2,i}$ and $Y_{2,i} = X_{1,i} \oplus_2 X_{2,i}$. 
The capacity region is given by $\mathcal{C}=\{ (R_1,R_2): R_1 \le 1, R_2 \le 1\}$ \cite{Shannon:1961}, \cite{Devroye:2014}. 
\end{enumerate}
\end{remark}

\begin{remark}
Example~1 can be generalized to a non-binary setting: for some integer $q> 2$, $\mathcal{X}_1=\mathcal{X}_2=\{0, 1, \dots, q-1\}$ and $\mathcal{Y}_1=\mathcal{Y}_2=\mathcal{Z}=\{0, 1, \dots, q-1, \mathbf{E}\}$,  the $q$-ary channel model obeys the same equations as in Example~1 with modulo-$2$ addition replaced with the modulo-$q$ operation $\oplus_q$.
Furthermore, the channel noise-erasure variables have marginal distributions given by $\Pr(Z_{j,i}=\mathbf{E})=\varepsilon_j$ and $\Pr(Z_{j,i}=z)=\alpha_j/(q-1)$ for $z=1, 2, \dots, q-1$, where $0\le \alpha_j+\varepsilon_j\le 1$ for $j=1, 2$. 
By Corollary~\ref{cor:77}, we directly have that $\mathcal{C}_{\text{I}}=\mathcal{C}_{\text{O}}$, and 
\begin{IEEEeqnarray}{l}
\mathcal{C}=\Bigg\{ (R_1,R_2):
R_1 \le (1-\varepsilon_2){\cdot}\left(\log_2 q-H_q\left(\frac{\alpha_2}{(q-1)(1-\varepsilon_2)}\right)\right), \qquad\nonumber\\
\quad\quad\quad\qquad\qquad R_2  \le (1-\varepsilon_1){\cdot}\left(
\log_2 q- H_q\left( \frac{\alpha_1}{(q-1)(1-\varepsilon_1)}\right)\right)\Bigg\},\nonumber
\end{IEEEeqnarray}
where $H_q(x)\triangleq x{\cdot}\log_2(q-1)-x{\cdot}\log_2 x - (1-x){\cdot}\log_2 (1-x)$. 
\end{remark}

\begin{example}[Data Access TWCs]
Let $q=2^m$ for some integer $m\ge 1$ and consider the alphabets $\mathcal{X}_1=\mathcal{X}_2=\mathcal{X}=\{0, 1, \dots, q-1\}$, $\mathcal{Y}_1=\mathcal{Y}_2=\{0, 1, \dots, q-1, \mathbf{E}\}$, and $\mathcal{Z}=\{0, 1, 2\}$. 
A data access TWC linking two storage devices is described by
\begin{IEEEeqnarray}{l}
\scalemath{1}{Y_{1,i} = (X_{1, i}\boxplus_q X_{2,i})\cdot\mathbf{1}\{Z_{1, i}=0\}}
\scalemath{1}{+ ((q-1)\boxplus_q X_{1, i} \boxplus_q X_{2,i}){\cdot}\mathbf{1}\{Z_{1, i}=1\}+\mathbf{E}{\cdot}\mathbf{1}\{Z_{1, i}= 2\}},\nonumber\\[+0.1cm]
\scalemath{1}{Y_{2,i} = (X_{1,i}\boxplus_q X_{2,i})\cdot\mathbf{1}\{Z_{2, i}=0\}}
\scalemath{1}{+ ((q-1)\boxplus_q X_{1,i}\boxplus_q X_{2, i}){\cdot}\mathbf{1}\{Z_{2, i}=1\}+\mathbf{E}{\cdot}\mathbf{1}\{Z_{2, i}= 2\}},\nonumber
\end{IEEEeqnarray}  
where $a \boxplus_q b$ denotes bit-wise addition for the length-$q$ standard binary representation of $a, b\in\mathcal{X}$, and $\{(Z_{1,i},Z_{2,i})\}_{i=1}^\infty$ is a memoryless joint noise-erasure process that is independent of the stored messages and has components $Z_{1,i}, Z_{2,i}\in\mathcal{Z}$ such that $\Pr(Z_{j,i}=1)=\alpha_j$ and $\Pr(Z_{j,i}=\mathbf{E})=\varepsilon_j$, where $0\le \alpha_j+\varepsilon_j\le 1$ for $j=1, 2$. 
This channel model can capture the effect of user signal superpositions (when $Z_{j, i}=0$), bit-level burst errors which flip all bits of $X_{1, i} \boxplus_q X_{2,i}$ (when $Z_{j, i}=1$), and data package losses (when $Z_{j, i}=2$). 

For this channel, an application of Corollary \ref{cor:77} immediately gives the capacity region:
\begin{IEEEeqnarray}{rCl}
\mathcal{C}=\Bigg\{ (R_1,R_2): R_1 &\le & (1-\varepsilon_2){\cdot}\left(m-H_{\text{b}}\left(\frac{\alpha_2}{1-\varepsilon_2}\right)\right), \nonumber\\
 R_2 &\le & (1-\varepsilon_1){\cdot}\left(m- H_{\text{b}}\left(\frac{\alpha_1}{1-\varepsilon_1}\right)\right)\Bigg\}.\nonumber
\end{IEEEeqnarray}
\end{example}

The next example redervies a known result in \cite{Chaaban:2017} based on our approach. 
\begin{example}[Memoryless Injective Semi-Deterministic TWCs \cite{Chaaban:2017}]\label{ex:ex2}
Let $\mathcal{T}_j$ and $\mathcal{Z}_j$ denote finite sets. 
A memoryless ISD-TWC is defined in \cite{Chaaban:2017} by the channel equations 
\begin{align}
Y_{j,i}=h_j(X_{j,i},T_{j,i}) \text{ and } T_{j,i}=\tilde{h}_j(X_{k,i},Z_{j,i}) \label{eq:ISD}
\end{align}
for $j, k =1, 2$ with $j{\neq}k$, where $h_j: \mathcal{X}_j \times \mathcal{T}_j \to \mathcal{Y}_j$ is invertible in $\mathcal{T}_j$ and $\tilde{h}_j: \mathcal{X}_{k} \times \mathcal{Z}_j \to \mathcal{T}_j$ is invertible in $\mathcal{Z}_j$, i.e., for every $x_j \in \mathcal{X}_j$, $h_j(x_j,t_j)$ is one-to-one in $t_j \in \mathcal{T}_j$ and for every $x_{k} \in \mathcal{X}_{k}$,  $\tilde{h}_j(x_{k},z_j)$ is one-to-one in $z_j \in \mathcal{Z}_j$. 
Here, $\{(Z_{1,i},Z_{2,i})\}_{i=1}^\infty$ is a memoryless joint noise process that is independent of users' messages. 
For this channel, we have \cite{Chaaban:2017} 
\begin{align}
I(X_1;Y_2|X_2=x_2) & \le \max_{P_{X_1}} H(\tilde{h}_2(X_1,Z_2))-H(Z_2).\nonumber
\end{align}
This upper bound does not depend on $X_2$, and hence a common maximizer exists, i.e., $P^*_{X_1}=\arg \max_{P_{X_1}} H(\tilde{h}_2(X_1,Z_2))$. 
Moreover, the value of $\max_{P_{X_1}}I(X_1;Y_2|X_2=x_2)$ is identical for all $x_2\in\mathcal{X}_2$. 
We immediately observe that condition (i) in Theorem~\ref{thm:nc2} holds. 
By a similar argument, condition (ii) in Theorem~\ref{thm:nc2} also holds, implying that Shannon's inner and outer bounds coincide.  
The capacity region is given by 
\begin{IEEEeqnarray}{rCl}
\mathcal{C}=\Bigg\{ (R_1,R_2): R_1 &\le & \max_{P_{X_1}} H(\tilde{h}_2(X_1,Z_2))-H(Z_2),\nonumber\\
 R_2 &\le & \max_{P_{X_2}} H(\tilde{h}_1(X_2,Z_1))-H(Z_1)\Bigg\}. \nonumber
\end{IEEEeqnarray}
\end{example}

\begin{figure}[tb]
\begin{centering}
\includegraphics[scale=0.4, draft=false]{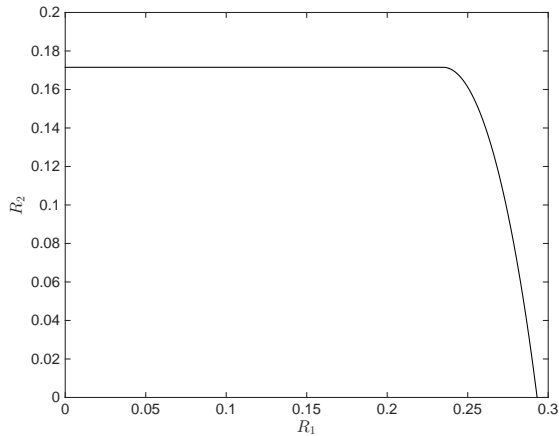}
\caption{The capacity region of the point-to-point memoryless TWC in Example 4.\label{fig:capacityExm}}
\end{centering}
\end{figure}

\begin{example}\label{ex:ex4}
Consider the TWC with $\mathcal{X}_1=\mathcal{X}_2=\mathcal{Y}_1=\mathcal{Y}_2=\{0, 1\}$ and transition probability
\begin{IEEEeqnarray}{l}
[P_{Y_1,Y_2|X_1,X_2}] = 
\scalemath{1}{ 
 \bordermatrix{
& 00& 01 & 10 & 11\cr
00& 0.783 & 0.087 & 0.117 & 0.013\cr
01& 0.36279 & 0.05421 & 0.50721 & 0.07579\cr
10& 0.261 & 0.609 & 0.039 & 0.091 \cr
11& 0.173889 & 0.243111 & 0.243111 & 0.339889}.}\nonumber
\end{IEEEeqnarray}
The one-way channel marginal distributions are
\[
\scalemath{1}{[P_{Y_2|X_1,X_2}(\cdot|\cdot,0)]= 
 \bordermatrix{
& 0 & 1\cr
0 & 0.9 & 0.1\cr
1 & 0.3 & 0.7
}}\]
and 
\[
\scalemath{1}{[P_{Y_2|X_1,X_2}(\cdot|\cdot,1)]= 
 \bordermatrix{
 & 0 & 1 \cr
0 & 0.87 & 0.13\cr
1 & 0.417 &  0.583
}}\\
\]
with $[P_{Y_1|X_1,X_2}(\cdot|0,\cdot)]{=}[P_{Y_1|X_1,X_2}(\cdot|1,\cdot)]{=}[P_{Y_2|X_1,X_2}(\cdot|\cdot,1)]$.

Shannon's symmetry condition in Proposition~\ref{thm:SC} does not hold for this channel since there are no permutations of $\mathcal{Y}_1$ and $\mathcal{Y}_2$ which can result in (\ref{SCcond}). Furthermore, since $H(Y_2|X_1=0, X_2=0)=H_{\text{b}}(0.1)$ and $H(Y_2|X_1=1, X_2=0)=H_{\text{b}}(0.3)$, $H(Y_2|X_1, X_2)$ depends on $P_{X_1|X_2}$ for fixed $P_{X_2}$. 
Thus, the CVA condition in Proposition~\ref{thm:CVA} does not hold either. 
However, the conditions of Theorem~\ref{thm:23} are satisfied since a common maximizer exists for the one-way channel from users~1 to~2 given by $P^*_{X_1}(0)=0.471$, and condition~(ii) trivially holds. 
By considering all input distributions of the form $P_{X_1, X_2}=P^*_{X_1}\cdot P_{X_2}$, where $P_{X_2}\in\mathcal{P}(\mathcal{X}_2)$, one can compute the capacity region as shown in Fig.~\ref{fig:capacityExm}. 
We note that, with some extra effort, one can show that the conditions of Theorem~\ref{thm:24} also hold \cite{JJW:2018}.  
\end{example}

Finally, we point out (without proof) that the channels in the examples in \cite[Fig. 2 \& Table II]{Shannon:1961} and \cite[Section~IV-B]{Chaaban:2017} satisfy the conditions of Theorem~\ref{thm:23}.

\subsection{Comparison with Prior Results}\label{subsec:comp}
In this section, we show that Theorems~\ref{thm:23} and~\ref{thm:33} generalize the Shannon results in Propositions~\ref{thm:SC} and~\ref{thm:SC2}, respectively, and that Theorem~\ref{thm:24} subsumes the CVA result in Proposition~\ref{thm:CVA} as a special case. 

\begin{theorem}
A TWC that satisfies the Shannon's one-sided symmetry condition of Proposition~\ref{thm:SC} must satisfy the conditions of Theorem~\ref{thm:23}. 
\label{thm:23toSC}
\end{theorem}
\begin{IEEEproof}
If a TWC satisfies the Shannon condition in Proposition~\ref{thm:SC}, the capacity-achieving input distribution is of the form $P_{X_1, X_2}=P^{\text{U}}_{\mathcal{X}_1}\cdot P_{X_2}$ for some $P_{X_2}\in\mathcal{P}(\mathcal{X}_2)$ \cite{Shannon:1961}. 
This implies that condition (i) of Theorem~\ref{thm:23} is satisfied because a common maximizer exists for all $x_2\in\mathcal{X}$ and is given by $P^*_{X_1}=P^{\text{U}}_{\mathcal{X}_1}$. 
To prove that condition (ii) is also satisfied, we consider the transition matrices $[P_{Y_1|X_1, X_2}(\cdot|x'_1, \cdot)]$ and $[P_{Y_1|X_1, X_2}(\cdot|x''_1, \cdot)]$ for arbitrary $x'_1, x''_1\in\mathcal{X}_1$ and show that these are column permutations of each other and hence $\mathcal{I}(P_{X_2}, P_{Y_1|X_1=x'_1, X_2})=\mathcal{I}(P_{X_2}, P_{Y_1|X_1=x''_1, X_2})$. 
The first claim is true because
\begin{IEEEeqnarray}{rCl}
P_{Y_1|X_1,X_2}(y_1|x'_1,x_2)&=& P_{Y_1|X_1,X_2}(\pi^{\mathcal{Y}_1}[x_1',x_1''](y_1)|\tau^{\mathcal{X}_1}_{x_1',x_1''}(x'_1),x_2)\label{eq:com11}\\
&= &P_{Y_1|X_1,X_2}(\pi^{\mathcal{Y}_1}[x_1',x_1''](y_1)|x''_1,x_2),\nonumber
\end{IEEEeqnarray}
where \eqref{eq:com11} is obtained by marginalizing over $Y_2$ on both sides of (\ref{SCcond}). 
For the second claim, we have 
\begin{IEEEeqnarray}{l}
\mathcal{I}(P_{X_2}, P_{Y_1|X_1=x'_1, X_2}) \nonumber\\
\ \ = \scalemath{1}{\sum_{x_2, y_1}P_{X_2}(x_2)\cdot P_{Y_1|X_1, X_2}(y_1|x'_1, x_2)}\scalemath{1}{\cdot\log\frac{P_{Y_1|X_1, X_2}(y_1|x'_1, x_2)}{\sum_{\tilde{x}_2}P_{X_2}(\tilde{x}_2){\cdot}P_{Y_1|X_1, X_2}(y_1|x'_1, \tilde{x}_2)}}\nonumber\\
\ \ = \scalemath{1}{\sum_{x_2, y_1}P_{X_2}(x_2)\cdot P_{Y_1|X_1, X_2}(\pi^{\mathcal{Y}_1}[x'_1, x''_1](y_1)|x''_1, x_2)}\nonumber\\
\qquad\qquad\qquad\qquad\quad\quad\  \scalemath{1}{\cdot\log\frac{P_{Y_1|X_1, X_2}(\pi^{\mathcal{Y}_1}[x'_1, x''_1](y_1)|x''_1, x_2)}{\sum_{\tilde{x}_2}P_{X_2}(\tilde{x}_2){\cdot}P_{Y_1|X_1, X_2}(\pi^{\mathcal{Y}_1}[x'_1, x''_1](y_1)|x''_1, \tilde{x}_2)}}\label{C7toC3s1}\\
\ \ = \scalemath{1}{\sum_{x_2, \tilde{y}_1}P_{X_2}(x_2)\cdot P_{Y_1|X_1, X_2}(\tilde{y}_1|x''_1, x_2)}\scalemath{1}{\cdot\log\frac{P_{Y_1|X_1, X_2}(\tilde{y}_1|x''_1, x_2)}{\sum_{\tilde{x}_2}P_{X_2}(\tilde{x}_2){\cdot}P_{Y_1|X_1, X_2}(\tilde{y}_1|x''_1, \tilde{x}_2)}}\label{C7toC3s2}\nonumber\\
\ \  =\mathcal{I}(P_{X_2}, P_{Y_1|X_1=x''_1, X_2})\nonumber\label{C7toC3s3},
\end{IEEEeqnarray}
where (\ref{C7toC3s1}) holds by the first claim. 
\end{IEEEproof}

\begin{remark}
Since the optimal input distribution of user~1 in Theorem~\ref{thm:23} is not necessarily uniform as illustrated in Example~\ref{ex:ex4}, Theorem~\ref{thm:23} is more general than Proposition~\ref{thm:SC}. 
\end{remark}

\begin{theorem}
A TWC that satisfies the Shannon two-sided symmetry condition of Proposition~2 must satisfy the conditions of Theorem~\ref{thm:33}. 
\label{thm:23toSC2}
\end{theorem}
This theorem is immediate, and hence the proof is omitted. 
Together with Example~5 given in the next section, Theorem~\ref{thm:33} is shown to be more general than Proposition~\ref{thm:SC2}. 
We next show that the symmetry properties identified by the conditions of Theorem~\ref{thm:24} are more general than those in the CVA condition.  

\begin{theorem}
A TWC that satisfies the CVA condition in Proposition~\ref{thm:CVA} must satisfy the conditions in Theorem~\ref{thm:24}. 
\label{thm:CVAto24}
\end{theorem}
\begin{IEEEproof}
Suppose that the condition of Proposition~\ref{thm:CVA} is satisfied.  
To prove the theorem, we show that for $j=1, 2$, $H(Y_j|X_1=x'_1, X_2=x_2)=H(Y_j|X_1=x''_1, X_2=x_2)$ for all $x'_1, x''_1\in\mathcal{X}_1$ and $x_2\in\mathcal{X}_2$. 
Given arbitrary pairs $(x'_1, x_2)$ and $(x''_1, x_2)$, consider the probability distributions 
\begin{equation}
P^{(1)}_{X_1, X_2}(a, b)=\left\{
\begin{array}{ll}
1,&\ \text{if}\ a=x'_1\ \text{and}\ b=x_2,\\
0,&\ \text{otherwise,}
\end{array}
\right.\nonumber
\end{equation}
and
\begin{equation}
P^{(2)}_{X_1, X_2}(a, b)=\left\{
\begin{array}{ll}
1,&\ \text{if}\ a=x''_1\ \text{and}\ b=x_2,\\
0,&\ \text{otherwise.}
\end{array}
\right.\nonumber
\end{equation}
Noting that $P^{(1)}_{X_2}=P^{(2)}_{X_2}$, we have $H(Y_j|X_1=x'_1, X_2=x_2)=H^{(1)}(Y_j|X_1, X_2)=H^{(2)}(\scalemath{1}{Y_j|X_1, X_2})\allowbreak=H(Y_j|X_1=x''_1, X_2=x_2)$, 
where the first and last equality are due to the definitions of $P^{(1)}_{X_1, X_2}$ and $P^{(2)}_{X_1, X_2}$, respectively, and the second equality follows from the CVA condition since $P^{(1)}_{X_2}=P^{(2)}_{X_2}$. 
Thus $H(Y_j|X_1=x_1, X_2=x_2)$ does not depend on $x_1$ for fixed $x_2$ as claimed.
Also, since $H(Y_j|X_1, X_2=x_2)=\sum_{x_1}P_{X_1|X_2}(x_1|x_2)\cdot\allowbreak H(Y_j|X_1=x_1, X_2=x_2)$, $H(Y_j|X_1, X_2=x_2)$ does not depend on $P_{X_1|X_2=x_2}$. 

Next, we show that condition (i) of Theorem~\ref{thm:24} holds by constructing a common maximizer from the CVA condition. 
For fixed $x_2\in\mathcal{X}_2$, let 
\begin{IEEEeqnarray}{rCl}
P^*_{X_1|X_2=x_2}&=&\argmax_{P_{X_1|X_2=x_2}}I(X_1; Y_2|X_2=x_2)\nonumber\\
&=& \argmax_{P_{X_1|X_2=x_2}} \Big(H(Y_2|X_2=x_2)-H(Y_2|X_1, X_2=x_2)\Big),\nonumber
\end{IEEEeqnarray}
and define $P^{(1)}_{X_1, X_2}=P^{(1)}_{X_2}\cdot P^*_{X_1|X_2}$ for some $P^{(1)}_{X_2}\in\mathcal{P}(\mathcal{X}_2)$. 
Since $H(Y_j|X_1, X_2=x_2)$ does not depend on $P_{X_1|X_2=x_2}$, $P^*_{X_1|X_2=x_2}$ is in fact a maximizer of $H(Y_2|X_2=x_2)$. 
Note that the maximizer $P^*_{X_1|X_2=x_2}$ is not necessarily unique, but any choice works for our purposes.
Now for $P^{(1)}_{X_1, X_2}$, by the CVA condition, there exists $\tilde{P}_{X_1}\in\mathcal{P}(\mathcal{X}_1)$ such that $H^{(1)}(Y_2|X_2)\le H^{(2)}(Y_2|X_2)$, where $P^{(2)}_{X_1, X_2}=\tilde{P}_{X_1}\cdot P^{(1)}_{X_2}$.
Since $P^*_{X_1|X_2=x_2}$ is the maximizer for $H(Y_2|X_2=x_2)$, we have
\begin{IEEEeqnarray}{rCl}
H^{(1)}(Y_2|X_2)&=&\sum_{x_2}P^{(1)}_{X_2}(x_2)\cdot H^{(1)}(Y_2|X_2=x_2)\nonumber\\
&=&\sum_{x_2}P^{(1)}_{X_2}(x_2)\cdot\scalemath{1}{\left[\max_{P_{X_1|X_2=x_2}}H(Y_2|X_2=x_2)\right]}\nonumber\\
&\ge & \sum_{x_2}P^{(1)}_{X_2}(x_2)\cdot H^{(2)}(Y_2|X_2=x_2)\nonumber\label{thm6:s4}\\
&=& H^{(2)}(Y_2|X_2)\nonumber
\end{IEEEeqnarray}
Thus, $H^{(1)}(Y_2|X_2)=H^{(2)}(Y_2|X_2)$, i.e., 
\begin{IEEEeqnarray}{l}
\scalemath{1}{\sum\limits_{x_2}P^{(1)}_{X_2}(x_2){\cdot}H^{(1)}(Y_2|X_2=x_2){=}\sum\limits_{x_2}P^{(1)}_{X_2}(x_2){\cdot}H^{(2)}(Y_2|X_2=x_2)}.\nonumber
\end{IEEEeqnarray}
Since $H^{(2)}(Y_2|X_2=x_2)\le H^{(1)}(Y_2|X_2=x_2)$ for each $x_2\in\mathcal{X}_2$, we obtain $H^{(1)}(Y_2|X_2=x_2)=H^{(2)}(Y_2|X_2=x_2)$, i.e., $\tilde{P}_{X_1}$ achieves the same value for $H(Y_2|X_2=x_2)$ as $P^*_{X_1|X_2=x_2}$ for all $x_2\in\mathcal{X}_2$. 
Consequently, $\tilde{P}_{X_1}$ is a common maximizer and thus condition (i) of Theorem~\ref{thm:24} is satisfied.
Moreover, since the common maximizer $\tilde{P}_{X_1}$ is from the CVA condition, we have that $H^{(1)}(Y_1|X_1)\le H^{(2)}(Y_1|X_1)$, which together with the fact that $H(Y_1|X_1, X_2)$ does not depend on $P_{X_1|X_2}$ given $P_{X_2}$ and $P_{Y_1|X_1, X_2}$ (guaranteed by the CVA condition) implies that condition (ii) of Theorem~\ref{thm:24} holds.
\end{IEEEproof}

\begin{remark}
As illustrated by Example~4, a TWC that satisfies the conditions of Theorem~\ref{thm:24} does not necessarily satisfy the CVA condition in Proposition~\ref{thm:CVA}.  
Therefore, Theorem~\ref{thm:24} is a more general result than Proposition~\ref{thm:CVA}. 
We note that the main difference between Theorem~\ref{thm:24} and Proposition~\ref{thm:CVA} lies in the fact that we allow $H(Y_2|X_1, X_2)$ to depend on $P_{X_1|X_2}$, given $P_{X_2}$. 
\end{remark}

\subsection{Connection Between the Shannon and CVA Conditions}\label{subsec:summary}
In this section, we connect Shannon's result to the CVA condition. 
First, the proof in Appendix~\ref{appendix:SCproof} shows that Shannon's symmetry conditions are more than sufficient for $\mathcal{C}_\text{I}$ and $\mathcal{C}_\text{O}$ to coincide. In fact, assume that the marginal channels $P_{Y_j|X_1, X_2}$'s (derived from $P_{Y_1, Y_2|X_1, X_2}$) satisfy the following extended Shannon's symmetry condition: for any pair of distinct input symbols $x'_1, \allowbreak x''_1\in\mathcal{X}_1$, there exists a pair of permutations 
$(\pi^{\mathcal{Y}_1}[x'_1,x''_1], \pi^{\mathcal{Y}_2}[x'_1,x''_1])$ on $\mathcal{Y}_1$ and $\mathcal{Y}_2$, respectively, (which depend on $x'_1$ and $x''_1$) such that for all $x_1$, $x_2$, $y_1$, $y_2$, 
\begin{equation}
\left\{
\begin{array}{l}
P_{Y_1|X_1,X_2}(y_1|x_1,x_2)= P_{Y_1|X_1,X_2}(\pi^{\mathcal{Y}_1}[x'_1,x''_1](y_1)|\tau^{\mathcal{X}_1}_{x'_1,x''_1} (x_1),x_2),\\
P_{Y_2|X_1,X_2}(y_2|x_1,x_2) = P_{Y_2|X_1,X_2}(\pi^{\mathcal{Y}_2}[x'_1,x''_1](y_2)|\tau^{\mathcal{X}_1}_{x'_1,x''_1} (x_1),x_2),
\end{array}\right.\\ \label{exSCcond}
\end{equation}
then $\mathcal{C}_\text{I}=\mathcal{C}_\text{O}=\mathcal{C}$ with $\mathcal{C}$ given by \eqref{eq:shannoncap}.

The extended Shannon's symmetry conditions are more general than their original versions since \eqref{SCcond} implies \eqref{exSCcond} but the reverse implication is not true as shown below.

\begin{example}
Consider the TWC with $\mathcal{X}_1=\mathcal{X}_2=\mathcal{Y}_1=\mathcal{Y}_2=\{0, 1\}$ and transition probability
\begin{align*}
[P_{Y_1,Y_2|X_1,X_2}] = \scalemath{1}{
 \bordermatrix{
& 00& 01 & 10 & 11\cr
00& 0.25 & 0.5 & 0.25 & 0\cr
01& 0.375 & 0375 & 0.125 & 0.125\cr
10& 0.125 & 0.125 & 0.375 & 0.375 \cr
11& 0.125 & 0.125 & 0.375 & 0.375}.}
\end{align*}
The marginal distributions are
\[
\scalemath{1}{[P_{Y_1|X_1,X_2}] = 
 \bordermatrix{
& 0& 1 \cr
00& 0.75 & 0.25\cr
01& 0.75 & 0.25\cr
10& 0.25 & 0.75\cr
11& 0.25 & 0.75}}\]
and 
\[
\hspace{-0.35cm}\scalemath{1}{[P_{Y_2|X_1,X_2}] = 
 \bordermatrix{
& 0& 1 \cr
00& 0.5 & 0.5\cr
01& 0.5 & 0.5\cr
10& 0.5 & 0.5\cr
11& 0.5 & 0.5}.}
\]
Clearly, neither of the Shannon conditions in Proposition~\ref{thm:SC} or~\ref{thm:SC2} holds, but the extended condition in \eqref{exSCcond} holds.   
\end{example}

We now show that the above extended symmetry condition implies the CVA condition. 

\begin{theorem}\label{thm:SCtoCVA}
A TWC that satisfies the condition in \eqref{exSCcond} must satisfy the CVA condition of Proposition~\ref{thm:CVA}. 
\end{theorem}
\begin{IEEEproof}
If the marginal channels $P_{Y_1|X_1, X_2}$ and $P_{Y_2|X_1, X_2}$ satisfy the extended one-sided symmetry condition, then $H(Y_j|X_1=x_1, X_2=x_2)$ does not depend on $x_1\in\mathcal{X}_1$ for any fixed $x_2\in\mathcal{X}_2$ since the rows of $[P_{Y_j|X_1, X_2}(\cdot| \cdot, x_2)]$ are permutations of each other. 
Hence, $H(Y_j|X_1, X_2)$ does not depend on $P_{X_1|X_2}$ given $P_{X_2}\in\mathcal{P}(\mathcal{X}_2)$ as required by the CVA condition. 

Next, for any given joint distribution $P^{(1)}_{X_1, X_2}=P^{(1)}_{X_2}\cdot P^{(1)}_{X_1|X_2}$, we show that $P^{(2)}_{X_1, X_2}=\tilde{P}_{X_1}\cdot P^{(1)}_{X_2}$ with the choice $\tilde{P}_{X_1}=P^{\text{U}}_{\mathcal{X}_1}$ meets the remaining requirements of the CVA condition in Proposition~\ref{thm:CVA}.  
Since the TWC satisfies the extended Shannon condition, Lemma~\ref{lem:Sl3} in Appendix~\ref{appendix:SCproof} gives the two inequalities: $I^{(1)}(X_1; Y_2|X_2)\le I^{(2)}(X_1; Y_2|X_2)$ and $I^{(1)}(X_2; Y_1|X_1)\le I^{(2)}(X_2; Y_1|X_1)$. 
Observing that $I^{(1)}(X_1; Y_2|X_2)= H^{(1)}(Y_2|X_2)- H^{(1)}(Y_2|X_1, X_2)=H^{(1)}(Y_2|X_2)- H^{(2)}(Y_2|X_1, X_2)$, we immediately obtain that $H^{(1)}(Y_2|X_2)\le H^{(2)}(Y_2|X_2)$ since $I^{(1)}(X_1; Y_2|X_2)\le I^{(2)}(X_1; Y_2|X_2)$. 
Moreover, since $H^{(1)}(Y_1|X_1, X_2)=H^{(2)}(Y_1|X_1, X_2)$ and $I^{(1)}(X_2; Y_1|X_1)\le I^{(2)}(X_2; Y_1|X_1)$, we have that $H^{(1)}(Y_1|X_1)\le H^{(2)}(Y_1|X_1)$.
Thus, the CVA condition is fulfilled. 
\end{IEEEproof}

\begin{remark}
In \cite{Chaaban:2017}, the existence of examples showing that the Shannon and CVA results are not equivalent was posed as an open question. 
The example below shows that the CVA condition is more general than the extended (one-sided) Shannon's symmetry condition \eqref{exSCcond}. 
Together with Example~5, we conclude that the CVA result is more general than the Shannon result.
\end{remark}

\begin{example}
Consider the TWC with $\mathcal{X}_1=\mathcal{Y}_1=\mathcal{Y}_2=\{0, 1, 2\}$ and $\mathcal{X}_2=\{0, 1\}$ and marginal distributions given by
\begin{IEEEeqnarray}{l}
[P_{Y_1|X_1,X_2}(\cdot|\cdot,0)]= 
\bordermatrix{
& 0 & 1 & 2\cr
0 & 0.3 & 0.2 & 0.5\cr
1 & 0.5 & 0.3 & 0.2\cr
2 & 0.2 & 0.5 & 0.3}.\nonumber
\end{IEEEeqnarray}
with $[P_{Y_1|X_1,X_2}(\cdot|\cdot,1)]{=}[P_{Y_2|X_1,X_2}(\cdot|\cdot,0)]{=}[P_{Y_2|X_1,X_2}(\cdot|\cdot,1)]\allowbreak {=}[P_{Y_1|X_1,X_2}(\cdot|\cdot,0)]$.
Clearly, there are no relabeling functions for $\mathcal{Y}_1$ and $\mathcal{Y}_2$  which recover $[P_{Y_1|X_1,X_2}(\cdot|\cdot,0)]$ after exchanging the labels of $X_1=0$ and $X_1=1$, so that the extended one-sided symmetry condition does not hold. 
To check the CVA condition, we first observe that $H(Y_j|X_1=x_1, X_2=x_2)$ does not depend on $x_1\in\mathcal{X}_1$ and $x_2\in\mathcal{X}_2$; thus $H(Y_j|X_1, X_2)$ does not depend on $P_{X_1, X_2}$ for $j=1, 2$.  
Furthermore, for any given $P^{(1)}_{X_1, X_2}=P^{(1)}_{X_2}\cdot P^{(1)}_{X_1|X_2}$, consider $P^{(2)}_{X_1, X_2}=\tilde{P}_{X_1}\cdot P^{(1)}_{X_2}$ with $\tilde{P}_{X_1}=P^{\text{U}}_{\mathcal{X}_1}$. Then, we have $
I^{(1)}(X_1; Y_2|X_2)=\sum_{x_2}P^{(1)}_{X_2}(x_2)\cdot I^{(1)}(X_1; Y_2|X_2=x_2)\le \sum_{x_2}P^{(1)}_{X_2}(x_2)\cdot I^{(2)}(X_1; Y_2|X_2=x_2) = I^{(2)}(X_1; Y_2|X_2)$, where the inequality follows from the fact that $P^{\text{U}}_{\mathcal{X}_1}$ is the capacity-achieving input distribution for all one-way channels from users~1 to~2. 
On the other hand, since the matrices $[P_{Y_1|X_1, X_2}(\cdot|x_1, \cdot)]$, $x_1\in\mathcal{X}_1$, are column permutations of each other, $\mathcal{I}(P_{X_2}, P_{Y_1|X_1=x_1, X_2})$ does not depend on $x_1\in\mathcal{X}_1$ for any fixed $P_{X_2}\in\mathcal{P}(\mathcal{X}_2)$.  
One can then follow the proof of Theorem~\ref{thm:23} to obtain that $I^{(1)}(X_2; Y_1|X_1)\le I^{(2)}(X_2; Y_1|X_1)$. 
Now, since $H(Y_j|X_1, X_2)$ does not depend on the input distribution, we conclude that $H^{(1)}(Y_j|X_j)\le H^{(2)}(Y_j|X_j)$ for $j=1, 2$, and thus the CVA condition is satisfied.   
\end{example}

\begin{remark}
The channel in the above example in fact also satisfies the conditions of Theorem~\ref{thm:23}. Nevertheless, the connection between the conditions of Theorem~\ref{thm:23} and the CVA condition is still unclear. 
\end{remark}

We close this section by noting that the symmetry properties induced by our proposed conditions are not necessarily specific to two-user memoryless TWCs as we will see  in Section~IV.
It is also worth mentioning that the proposed conditions can be used to investigate whether or not Shannon-type random coding schemes (under independent and non-adaptive inputs) provide tight bounds for other classical communication scenarios such as MACs with feedback and one-way compound channels. 
In particular, our conditions can be used to identify compound channels where the availability of channel state information at the transmitter (in addition to the receiver) cannot improve capacity.   

\section{Two-Way Symmetric Channels with Memory}\label{sec:memory}
We here consider point-to-point TWCs with memory whose inputs and outputs are related via functions $F_1$ and $F_2$ as follows:
\begin{IEEEeqnarray}{rCl}
Y_{1,i}&=&F_1(X_{1,i}, X_{2, i}, Z_{1, i}),\label{eq:symF1}\\
Y_{2,i}&=&F_2(X_{1,i}, X_{2, i}, Z_{2, i}),\label{eq:symF2}
\end{IEEEeqnarray}
where $\{(Z_{1, i}, Z_{2, i})\}_{i=1}^{\infty}$ is a stationary and ergodic noise process which is independent of the users' messages $M_1$ and $M_2$. 
Note that this model is a special case of the general model introduced in Section~\ref{subsec:p2pmodel}; it is also a generalization of the discrete additive-noise TWC considered in \cite{Song:2016}.

We first state (without proof) an inner bound for arbitrary (time-invariant) functions $F_1$ and $F_2$.
The bound can be proved via Shannon's standard random coding scheme (under non-adaptive independent inputs) for information stable one-way channels with memory, applied in each direction of the two-way transmission. 

\begin{lemma}[Inner Bound]{\label{thm:memin1}}
For the channel described in \eqref{eq:symF1} and \eqref{eq:symF2}, a rate pair $(R_1, R_2)$ is achievable if there exist two sequences of codes $(f_1^n, g_1)$ and $(f_2^n, g_2)$ with message sets $\mathcal{M}_1=\{1, 2, \dots, 2^{nR_1}\}$ and $\mathcal{M}_2=\{1, 2, \dots, 2^{nR_2}\}$, respectively, such that
\begin{IEEEeqnarray}{rCl}
R_1&\le & \lim_{n\rightarrow\infty} \frac{1}{n}I(X_1^n; Y_2^n|X_2^n),\nonumber\\
R_2&\le & \lim_{n\rightarrow\infty} \frac{1}{n}I(X_2^n; Y_1^n|X_1^n),\nonumber
\end{IEEEeqnarray} 
where the mutual information terms are evaluated under a sequence of product input probability distributions $\{P_{X_1^n}{\cdot}P_{X_2^n}\}_{n=1}^{\infty}$ and the inputs $X_j^n$ are independent of $\{(Z_{1, i}, Z_{2, i})\}_{i=1}^n$, $j=1, 2$. 
\end{lemma}

We say that $F_j(X_1, X_2, Z_j)$ is invertible in $Z_j$ if $F_{j}(x_1, x_2, \cdot)$ is one-to-one for any fixed $x_1\in\mathcal{X}_1$ and $x_2\in\mathcal{X}_2$. 
Under this invertibility condition, we obtain the following corollary.
\begin{corollary}\label{cor:memin1}
If $F_j$ is invertible in $Z_j$ for $j=1, 2$,  
a rate pair $(R_1, R_2)$ is achievable if 
\begin{IEEEeqnarray}{rCl}
R_1&\le & \lim_{n\rightarrow\infty} \frac{1}{n}H(Y_2^n|X_2^n) - \bar{H}(Z_2),\label{eq:memin1}\\
R_2&\le & \lim_{n\rightarrow\infty} \frac{1}{n}H(Y_1^n|X_1^n) - \bar{H}(Z_1),\label{eq:memin2}
\end{IEEEeqnarray}
for product distributions $\{P_{X_1^n}{\cdot}P_{X_2^n}\}_{n=1}^{\infty}$, where $\bar{H}(Z_j)$ denotes the entropy rate of the noise process $\{Z_{j, i}\}_{i=1}^{\infty}$ and the inputs $X_j^n$ are independent of $\{(Z_{1, i}, Z_{2, i})\}_{i=1}^n$, $j=1, 2$. 
\end{corollary}
\begin{IEEEproof}
The proof follows from the fact that 
\begin{IEEEeqnarray}{rCl}
I(X_1^n; Y_2^n|X_2^n)&=&H(Y_2^n|X_2^n)-H(Y_2^n|X_1^n, X_2^n)\nonumber\\
&=& H(Y_2^n|X_2^n)-H(Z_2^n|X_1^n, X_2^n)\nonumber\\
&=& H(Y_2^n|X_2^n)-H(Z_2^n)\nonumber, 
\end{IEEEeqnarray}
where the second equality holds since $F_2$ is invertible in $Z_2$ and the last equality holds since the channel inputs are generated independently of the noise process $\{(Z_{2, 1}, Z_{2, i})\}_{i=1}^{\infty}$. 
Applying a similar argument to $I(X_1^n; Y_2^n|X_2^n)$ completes the proof.
\end{IEEEproof}

{Let $F^{-1}_j$ denote the inverse of $F_j$ for fixed $(x_1, x_2)$ so that $z_j=F^{-1}_j(x_1, x_2, y_j)$, $j=1, 2$. If we further assume that $z_j=F^{-1}_j(x_1, x_2, y_j)$ is one-to-one in $x_{j'}$ for any fixed $x_j\in\mathcal{X}_j$ and $y_j\in\mathcal{Y}_j$, where $j, j'=1, 2$ with $j\neq j'$,} and impose cardinality constraints on the alphabets, we can simplify the expressions in \eqref{eq:memin1} and \eqref{eq:memin2} as follows. 
\begin{corollary}\label{cor:memin2}
Suppose that $F_j$ is invertible in $Z_j$ and $F^{-1}_j$ is one-to-one for $j, j'=1, 2$ with $j\neq j'$. Also, $|\mathcal{X}_2|=|\mathcal{Y}_1|=|\mathcal{Z}_1|=q_1$ and $|\mathcal{X}_1|=|\mathcal{Y}_2|=|\mathcal{Z}_2|=q_2$ for some integers $q_1, q_2 \ge 2$. 
Then, a rate pair $(R_1, R_2)$ is achievable if 
\begin{IEEEeqnarray}{rCl}
R_1&\le & \log q_2 - \bar{H}(Z_2),\nonumber\\
R_2&\le & \log q_1 - \bar{H}(Z_1).\nonumber
\end{IEEEeqnarray}
\end{corollary}
\begin{IEEEproof}
The proof hinges on noting that $H(Y_j^n|X_j^n)\le n\cdot\log q_j$ and that the uniform input distribution $P_{X_1^n, X_2^n}=(P^{\text{U}}_{\mathcal{X}_1}\cdot P^{\text{U}}_{\mathcal{X}_2})^n$ achieves the upper bound.
More specifically, we have to show that if $P_{X_1^n, X_2^n}$ is the uniform distribution, then $P_{Y^n_{j}|X_j^n}(y_j^n|x_j^n)$ is uniform on $\mathcal{Y}_j^n$ for any given $X_j^n=x_j^n$, and hence $H(Y_j^n|X_j^n=x_j^n)=n\cdot\log q_j$.  
By symmetry, we only provide the details for $H(Y_2^n|X_2^n)$.
Suppose that $P_{X_1^n, X_2^n}$ is the uniform distribution on $\mathcal{X}_1^n\times \mathcal{X}_2^n$. 
Then, for any $x_2^n$ we have
\begin{IEEEeqnarray}{rCl}
P_{Y_2^n|X_2^n}(y_2^n|x_2^n) &=& \sum_{x_1^n}P_{Y_2^n|X_1^n, X_2^n}(y_2^n|x_1^n, x_2^n)P_{X_1^n|X_2^n}(x_1^n|x_2^n)\nonumber\\
&=& \left(\frac{1}{q_2}\right)^n\cdot\sum_{x_1^n}P_{Y_2^n|X_1^n, X_2^n}(F_2(x_1^n, x_2^n, z_2^n)|x_1^n, x_2^n)\nonumber\\
&=& \left(\frac{1}{q_2}\right)^n\cdot\sum_{x_1^n}P_{Z_2^n|X_1^n, X_2^n}(F^{-1}_2(x_1^n, x_2^n, y_2^n)|x_1^n, x_2^n)\nonumber\\
&=& \left(\frac{1}{q_2}\right)^n\cdot\sum_{z_2^n}P_{Z_2^n}(z_2^n)\label{eq:x2}\IEEEeqnarraynumspace\\
&=& \left(\frac{1}{q_2}\right)^n,\nonumber
\end{IEEEeqnarray}
where \eqref{eq:x2} holds since $(X_1^n, X_2^n)$ is independent of $Z_2^n$ and {$F_2^{-1}(X_1, X_2, Y_2)$ is onto in $X_1$} due to the cardinality constraint. 
Clearly, $P_{Y_2^n|X_2^n=x_2^n}$ is the uniform distribution for any $x_2^n$, implying that $H(Y_2^n|X_2^n)=n\cdot\log q_2$. 
\end{IEEEproof}

Next we consider ISD-TWCs as in Example~\ref{ex:ex2} and \cite{Chaaban:2017}, but with the assumption that the noise process $\{(Z_{1,i}, Z_{2, i})\}_{i=1}^{\infty}$ can have memory. 
Note that any ISD-TWC with memory is a special case of the system model in \eqref{eq:symF1} and \eqref{eq:symF2} satisfying the invertibility condition in $Z_1$ and $Z_2$. 
Thus, Corollary~\ref{cor:memin1} applies to ISD-TWCs with memory to obtain the following result. 
\begin{corollary}\label{cor:inner2}
For the ISD-TWC with memory, a rate pair $(R_1,R_2)$ is achievable if 
\begin{IEEEeqnarray}{rCl}
R_1 &\le & \lim_{n\rightarrow\infty}\frac{1}{n}\max_{P_{X_1^n}}H(\tilde{h}_2(X_1^n, Z_2^n)) - \bar{H}(Z_2),\nonumber\\
R_2 &\le & \lim_{n\rightarrow\infty}\frac{1}{n}\max_{P_{X_2^n}}H(\tilde{h}_1(X_2^n, Z_1^n)) - \bar{H}(Z_1),\nonumber
\end{IEEEeqnarray}
where $\bar{H}(Z_j)$ denotes the entropy rate of the process $\{Z_{j, i}\}_{i=1}^{\infty}$ for $j=1, 2$.  
\end{corollary}

We note that Corollary~\ref{cor:memin2} also applies to ISD-TWCs with memory under identical alphabet size constraints so that any rate pair in $\{(R_1, R_2): R_1 \le \log q_2 - \bar{H}(Z_2), R_2 \le \log q_1 - \bar{H}(Z_1)\}$ is achievable for ISD-TWCs with memory.
We next derive converses to Corollaries~\ref{cor:memin2} and~\ref{cor:inner2}. 

\begin{lemma}[Outer Bound for Noise-Invertible TWCs with Memory]\label{lem:outer}
Suppose that $|\mathcal{Y}_j|=q_j$ for some integer $q_j\ge 2$. 
If $F_j$ is invertible in $Z_j$ for $j=1, 2$, any achievable rate pair $(R_1, R_2)$ must satisfy
\begin{align*}
R_1 \le \log q_2 - \lim_{n \to \infty} \frac{1}{n} \sum_{i=1}^n H(Z_{2,i}|Z_1^{i-1}, Z_2^{i-1}),\\
R_2 \le \log q_1 - \lim_{n \to \infty} \frac{1}{n} \sum_{i=1}^n H(Z_{1,i}|Z_1^{i-1}, Z_2^{i-1}),
\end{align*}
where the limits exist because $\{ (Z_{1,i},Z_{2,i}) \}_{i=1}^\infty$ is stationary.
\end{lemma}
\begin{IEEEproof}
For an achievable rate pair $(R_1,R_2)$, we have
\begin{IEEEeqnarray}{rCl}
n{\cdot}R_1  &=& H(M_1|M_2) \nonumber\\
&=& I(M_1;Y_2^n|M_2) + H(M_1|Y_2^n,M_2)\nonumber\\
& \le &  I(M_1;Y_2^n|M_2) + n{\cdot}\epsilon_n \label{eq:sstwc1}\\
&= &\scalemath{0.94}{\sum_{i=1}^n \Big[ H(Y_{2,i}|M_2,Y_2^{i-1}) - H(Y_{2,i}|M_1,M_2,Y_2^{i-1})\Big]{+}n{\cdot}\epsilon_n} \nonumber \\* \label{eq:sstwc6}\\
& \le & \sum_{i=1}^n \Big[\log q_2 - H(Y_{2,i}|M_1,M_2,Y_2^{i-1})\Big] + n{\cdot}\epsilon_n \label{eq:sstwc2}\\
& \le  &\scalemath{0.94}{\sum_{i=1}^n \Big[ \log q_2 {-} H(Y_{2,i}|M_1,M_2,Y_1^{i-1},Y_2^{i-1}, X_{1,i}, X_{2, i})\Big]{+}n{\cdot}\epsilon_n}\nonumber\\
& = &\scalemath{0.94}{\sum_{i=1}^n \Big[\log q_2 {-} H(Z_{2, i}|M_1,M_2,Y_1^{i-1},Y_2^{i-1},X_1^{i},X_2^{i})\Big]{+}n{\cdot}\epsilon_n}\nonumber\\*\IEEEeqnarraynumspace\label{eq:sstwc7}\\
& = &\sum_{i=1}^n \Big[\log q_2 \scalemath{0.94}{- H(Z_{2,i}|M_1,M_2,Y_1^{i-1},Y_2^{i-1},X_1^{i},X_2^{i},Z_1^{i-1},Z_2^{i-1})\Big]}+ n{\cdot}\epsilon_n \label{eq:sstwc3}\\
& = &\sum_{i=1}^n \Big[\log q_2 - H(Z_{2,i}|Z_1^{i-1},Z_2^{i-1})\Big] + n{\cdot}\epsilon_n\label{eq:sstwc4}\\
& = & n \cdot\log q_2 - \sum_{i=1}^n H(Z_{2,i}|Z_1^{i-1},Z_2^{i-1}) + n{\cdot}\epsilon_n,\label{eq:sstwc5}
\end{IEEEeqnarray} 
where \eqref{eq:sstwc1} is due to Fano's inequality with $\epsilon_n \to 0$ as $n \to \infty$, \eqref{eq:sstwc2} follows from $|\mathcal{Y}_2|=q_2$, \eqref{eq:sstwc7} and \eqref{eq:sstwc3} hold since $F_j$ is invertible in $Z_j$ given $(X_{1,i}, X_{2, i})$, and \eqref{eq:sstwc4} holds since
\begin{IEEEeqnarray}{rCl}
H(Z_{2,i}|Z_1^{i-1},Z_2^{i-1}) &=& H(Z_{2,i}|M_1,M_2,Z_1^{i-1},Z_2^{i-1}) \label{eq:memory21}\\
&=& H(Z_{2,i}|M_1,M_2,Z_1^{i-1},Z_2^{i-1},X_{1,1},X_{2,1}) \label{eq:memory22} \\
&=& H(Z_{2,i}|M_1,M_2,Z_1^{i-1},Z_2^{i-1},X_{1,1},X_{2,1}, Y_{1,1},Y_{2,1}) \label{eq:memory23}\IEEEeqnarraynumspace\\
&=& H(Z_{2,i}|M_1,M_2,Z_1^{i-1},Z_2^{i-1},X_{1}^2,X_{2}^2, Y_{1,1},Y_{2,1}) \label{eq:memory24}\\
&=& H(Z_{2,i}|M_1,M_2,Z_1^{i-1},Z_2^{i-1}, X_1^{i},X_2^{i},Y_1^{i-1},Y_2^{i-1}) \label{eq:memory25}
\end{IEEEeqnarray}
where \eqref{eq:memory21} is due to the fact that  $\{ (Z_{1,i},Z_{2,i}) \}_{i=1}^\infty$ is independent of $(M_1,M_2)$, \eqref{eq:memory22} and \eqref{eq:memory24} hold since $X_{j,i} = f_{j,i}(M_j, Y_j^{i-1})$ for $j=1,2$, \eqref{eq:memory23} follows from the identity $Y_{j,i}=F_j(X_{1, i}, X_{2, i}, Z_{j, i})$, and \eqref{eq:memory25} is obtained by recursively using the same argument as in \eqref{eq:memory22}-\eqref{eq:memory24}. 
Similarly, we have
\begin{align}
n{\cdot}R_2 \le n{\cdot}\log q_1 - \sum_{i=1}^n H(Z_{1,i}|Z_1^{i-1},Z_2^{i-1}) + n{\cdot}\hat{\epsilon}_n.\label{eq:sstwc8}
\end{align}
The proof is completed by dividing both sides of \eqref{eq:sstwc5} and \eqref{eq:sstwc8} by $n$ and letting $n \to \infty$.
\end{IEEEproof}

\begin{lemma}[Outer Bound for ISD-TWCs with Memory]\label{lem:outer2}
For the ISD-TWC with memory, any achievable rate pair $(R_1,R_2)$ must satisfy
\begin{IEEEeqnarray}{rCl}
R_1 &\le& \scalemath{0.93}{\lim_{n \to \infty} \frac{1}{n}\left[\max_{P_{X_1^n}}H(\tilde{h}_2(X_1^n, Z_2^n))  -  \sum_{i=1}^n H(Z_{2,i}|Z_1^{i-1}, Z_2^{i-1})\right]},\nonumber\\
R_2 &\le& \scalemath{0.93}{\lim_{n \to \infty} \frac{1}{n}\left[\max_{P_{X_2^n}}H(\tilde{h}_1(X_2^n, Z_1^n))  - \sum_{i=1}^n H(Z_{1,i}|Z_1^{i-1}, Z_2^{i-1})\right]}.\nonumber
\end{IEEEeqnarray}
\end{lemma}
\begin{IEEEproof}
The proof is similar to the proof of the previous lemma and hence the details are omitted. 
The main difference is that the first term in \eqref{eq:sstwc6} is now upper bounded as follows
\begin{IEEEeqnarray}{rCl}
\sum_{i=1}^n H(Y_{2, i}|M_2, Y_2^{i-1})&=& \sum_{i=1}^n H(h_2(X_{2, i}, T_{2, i})|M_2, Y_2^{i-1}, X_2^i, T_2^{i-1})\nonumber\\
&\le &\sum_{i=1}^n H(T_{2, i}|T_2^{i-1})\nonumber\\
&= &H(T_2^n)\nonumber\\
&\le &\max_{P_{X_1^n}}H(\tilde{h}_2(X_1^n, Z_2^n)),\nonumber
\end{IEEEeqnarray}
where the first equality holds since $X_2^i$ is a function of $M_2$ and $Y_2^{i-1}$ and $Y_2=h_2(X_2, T_2)$ is invertible in $T_2$ given $X_2$. 
\end{IEEEproof}

Based on the preceding inner and outer bounds, the capacity region for two classes of TWCs with memory (whose component noise processes are independent of each other) can be exactly determined as follows.
\begin{theorem}{\label{thm:mem1}}
For a TWC with memory such that $\{Z_{1,i}\}_{i=1}^\infty$ and $\{Z_{2,i}\}_{i=1}^\infty$ are stationary ergodic and mutually independent, $F_j$ is invertible in $Z_j$ {and $F^{-1}_j$ is one-to-one in $X_{j'}$ for $j, j'=1, 2$ with $j\neq j'$}, and $|\mathcal{X}_2|=|\mathcal{Y}_1|=|\mathcal{Z}_1|=q_1$ and $|\mathcal{X}_1|=|\mathcal{Y}_2|=|\mathcal{Z}_2|=q_2$ for some integers $q_1, q_2 \ge 2$, the capacity region is given by 
\begin{IEEEeqnarray}{l}
\mathcal{C}=\big\{(R_1, R_2): R_1\le\log q_2 - \bar{H}(Z_2), R_2\le\log q_1 - \bar{H}(Z_1)\big\}. \IEEEeqnarraynumspace\label{eq:MTWCCap}
\end{IEEEeqnarray}
\end{theorem}

\begin{theorem}{\label{thm:ISDCap}}
For a ISD-TWC with memory such that $\{Z_{1,i}\}_{i=1}^\infty$ and $\{Z_{2,i}\}_{i=1}^\infty$ are stationary ergodic and mutually independent, the capacity region is given by 
\begin{IEEEeqnarray}{l}
\mathcal{C}=\Big\{(R_1, R_2):R_1\le \lim_{n\rightarrow\infty}\frac{1}{n}\max_{P_{X_1^n}}H(\tilde{h}_2(X_1^n, Z_2^n)) - \bar{H}(Z_2),\qquad\nonumber\\
 \qquad\qquad\quad\quad\quad  R_2\le\lim_{n\rightarrow\infty}\frac{1}{n}\max_{P_{X_2^n}}H(\tilde{h}_1(X_2^n, Z_1^n))- \bar{H}(Z_1)\Big\}.\IEEEeqnarraynumspace\label{eq:ISCCap}
\end{IEEEeqnarray}
\end{theorem}

\begin{remark}
Theorem~\ref{thm:ISDCap} generalizes \cite[Corollary 1]{Chaaban:2017} for memoryless ISD-TWCs. 
If one further has $|\mathcal{X}_2|=|\mathcal{T}_1|=|\mathcal{Z}_1|=q_1$ and $|\mathcal{X}_1|=|\mathcal{T}_2|=|\mathcal{Z}_2|=q_2$ for some integers $q_1, q_2 \ge 2$, then $\lim_{n\rightarrow\infty}\frac{1}{n}\max_{P_{X_1^n}}H(\tilde{h}_2(X_1^n, Z_2^n))=\log q_1$ and that $\lim_{n\rightarrow\infty}\frac{1}{n}\max_{P_{X_2^n}}H(\tilde{h}_1(X_2^n, Z_1^n))=\log q_2$.
\end{remark}

The next example shows that if the noise processes are {\it{dependent}}, then Shannon's random coding scheme is not optimal.   
\begin{example}[Adaptation is Useful]
Let $q_1=q_2=2$ and suppose that the channel is given by 
\begin{align*}
Y_{1,i} & = F_1(X_{1,i}, X_{2,i}, Z_{1,i})=X_{1,i} \oplus_2 X_{2,i} \oplus_2 Z_{1,i}, \\
Y_{2,i} & = F_2(X_{1,i}, X_{2,i}, Z_{2,i})=X_{1,i} \oplus_2 X_{2,i} \oplus_2 Z_{2,i}, 
\end{align*}
where $\{Z_{1,i}\}_{i=1}^n$ is assumed to be memoryless with $Z_{1,i}$ uniformly distributed on $\mathcal{Z}_1=\{0, 1\}$ for $i=1, 2, \dots, n$, and $\{Z_{2, i}\}_{i=1}^n$ is given by $Z_{2,1}=0$ and $Z_{2,i}=Z_{1,i-1}$ for $i=2, 3, \dots, n$. 
Since the functions $F_1$ and $F_2$ are invertible in $Z_1$ and $Z_2$, the outer bound in Lemma~\ref{lem:outer} indicates that
\begin{align*}
R_1 & \le \log 2 - \lim_{n \to \infty} \frac{1}{n} \sum_{i=1}^n H(Z_{2,i}|Z_1^{i-1}, Z_2^{i-1}) \\
& = 1 - 0 =1, \\
R_2 & \le \log 2 - \lim_{n \to \infty} \frac{1}{n} \sum_{i=1}^n H(Z_{1,i}|Z_1^{i-1}, Z_2^{i-1}) \\
& =  1 - H(Z_{1,i}) = 0.
\end{align*}
We claim that the rate pair $(R_1, R_2)=(1,0)$ can be achieved by an adaptive coding scheme.
Let $\{M_{1, i}\}_{i=1}^n$ denote the binary messages to be sent from users~1 to~2. 
For $i=1, 2, \dots, n$, set the encoding function of user~1 as $X_{1,i}=f_{1, i}(\{M_{1, i}\}_{i=1}^n, Y_1^{i-1})\triangleq M_{1, i} \oplus_2 X_{1,i-1} \oplus_2 Y_{1,i-1}$ with initial conditions $X_{1,0}=X_{2,0}=Y_{1,0}=0$, and set the encoder output of user~2 to be zero, i.e., $X_{2,i}=0$ for all $i$. 
With this coding scheme, the received signal at user~2 is given by
\begin{align*}
 Y_{2,i} & = X_{1,i} \oplus_2 X_{2,i} \oplus_2 Z_{2,i} \\
        & = M_{1, i} \oplus_2 X_{1,i-1} \oplus_2 Y_{1,i-1} \oplus_2 Z_{2,i}\\
        & = M_{1, i} \oplus_2 X_{1,i-1} \oplus_2 X_{1,i-1} \oplus_2 Z_{1,i-1} \oplus_2 Z_{2,i} = M_{1, i},
\end{align*}
and thus the rate pair $(1,0)$ is achievable. 
This achievability result together with the outer bound imply that the channel capacity is given by $\mathcal{C} = \{ (R_1,R_2): R_1 \le 1, R_2=0 \}$.
However, the Shannon-type random coding scheme only provides $R_1\le 1-\bar{H}(Z_2)=0$ and $R_2\le 1-\bar{H}(Z_2)=0$ by Corollary~\ref{cor:memin2}. 
\end{example}  

\section{Multiple access/degraded broadcast TWCs}\label{sec:MADBC}
This section considers a three-user two-way communication scenario combining multiaccess and broadcasting. 
We first introduce the channel model and derive inner and outer bounds for the capacity region. 
Then, sufficient conditions for the two bounds to coincide are provided, along with illustrative examples. 
\subsection{Channel Model}\label{subsec:MADBC}

\begin{figure}[!t]
\begin{centering}
\includegraphics[draft=false, scale=0.55]{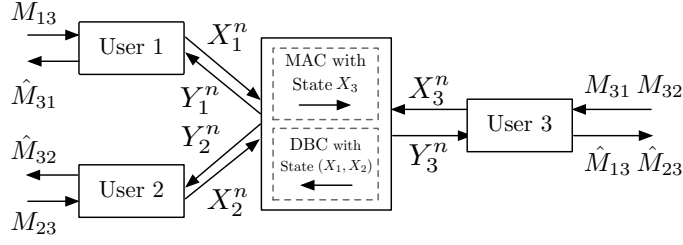}
\caption{The information flow of MA/DB TWCs.\label{fig:MADBC}}
\end{centering}
\end{figure}

Two-way communication over a discrete additive-noise MA/DB TWC comprises three users as depicted in Fig. \ref{fig:MADBC}.
Users~1 and~2 want to transmit messages $M_{13}$ and $M_{23}$, respectively, to user~3 through the TWC that acts as a MAC in the forward direction.
User~3 wishes to broadcast messages $M_{31}$ and $M_{32}$ to users~1 and~2, respectively, through the TWC that acts as a DBC in the reverse direction. 
The messages are assumed to be independent of each other and uniformly distributed over their alphabets. 
The joint distribution of all the variables for $n$ channel uses is given by
\begin{IEEEeqnarray}{l}
P_{M_{\{13, 23, 31, 32\}}, X_{\{1,2,3\}}^n, Y_{\{1,2,3\}}^n}=P_{M_{13}}\cdot P_{M_{23}}\cdot P_{M_{31}}\cdot P_{M_{32}}  \scalemath{1}{\cdot\left(\prod\limits_{i=1}^n P_{X_{1,i}|M_{13}, Y_{1}^{i-1}}\right)} \nonumber\\
\qquad\qquad\qquad\scalemath{1}{\cdot\left(\prod\limits_{i=1}^n P_{X_{2,i}|M_{23}, Y_{2}^{i-1}}\right)\cdot\left(\prod\limits_{i=1}^n P_{X_{3, i}|M_{\{31, 32\}}, Y_3^{i-1}}\right){\cdot}\left(\prod\limits_{i=1}^n P_{Y_{1,i}, Y_{2, i}, Y_{3, i}|X_{\{1, 2, 3\}}^i, Y_{\{1, 2, 3\}}^{i-1}}\right),}\nonumber
\end{IEEEeqnarray}
\sloppy where $M_{\{13, 23, 31, 32\}}\triangleq \{M_{13}, M_{23}, M_{31}, M_{32}\}$, $X_{\{1, 2, 3\}}^n\triangleq \{X_1^n, X_2^n, X_3^n\}$, and $Y_{\{1, 2, 3\}}^n\triangleq \{Y_1^n, Y_2^n, Y_3^n\}$. 
Thus, the $n$ transmissions can be described by the sequence of input-output conditional probabilities $\{ P_{Y_{1,i},Y_{2,i},Y_{3, i}|X_{\{1, 2, 3\}}^i,Y_{\{1, 2, 3\}}^{i-1}}\}_{i=1}^n$.

To simplify our analysis, we assume that the channel is memoryless in the sense that given current channel inputs, the current channel outputs are independent of past signals, i.e., $P_{Y_{1,i},Y_{2,i},Y_{3, i}|X_{\{1, 2, 3\}}^i,Y_{\{1, 2, 3\}}^{i-1}}=P_{Y_{1,i},Y_{2,i},Y_{3, i}|X_{1, i}, X_{2, i}, X_{3, i}}$ for all $i$. 
Furthermore, the two directions of transmission are assumed to interact in a way such that $P_{Y_{1,i},Y_{2,i},Y_{3, i}|X_{1, i}, X_{2, i}, X_{3, i}}=P_{Y_{1,i},Y_{2,i}|X_{1, i}, X_{2, i}, X_{3, i}}\cdot P_{Y_{3,i}|X_{1, i}, X_{2, i}, X_{3, i}}$. 
Let all channel input and output alphabets other than $\mathcal{Y}_{3}$ equal $\mathcal{Q} \triangleq \{0,1,...,q-1\}$ for some $q \ge 2$. 
The MA/DB TWC is defined by the transition probability $P_{Y_3|X_1,X_2,X_3}$ in the MA direction and the transmission equations in the DB direction are given by 
\begin{align}
Y_{1,i} &= X_{1,i} \oplus_q X_{3,i} \oplus_q Z_{1,i},\label{eq:MADB1}\\
Y_{2,i} &= X_{2,i} \oplus_q X_{3,i} \oplus_q Z_{1,i} \oplus_q Z_{2,i},\label{eq:MADB2}
\end{align}
for $i=1,2,...,n$, where $Z_{1,i}, Z_{2,i}\in \mathcal{Q}$ denote additive noise variables, the components of the memoryless and independent noise processes $\{Z_{1,i}\}_{i=1}^n$ and $\{Z_{2,i}\}_{i=1}^n$, respectively. We also assume that the channel noise processes are independent of all users' messages. 
Thus, the channel transition probability of this MA/DB TWC at time $i$ can be written as  
\begin{IEEEeqnarray}{l}
P_{Y_{1,i},Y_{2,i},Y_{3,i}|X_{1}^i,X_{2}^i,X_{3}^i,Y_{1}^{i-1},Y_{2}^{i-1},Y_{3}^{i-1}} (y_{1,i},y_{2,i},y_{3,i}|x_{1}^i,x_{2}^i, x_{3}^i, y_{1}^{i-1},y_{2}^{i-1},y_{3}^{i-1})\nonumber\\
\ \ =  P_{Y_{1,i},Y_{2,i},Y_{3,i}|X_{1,i},X_{2,i},X_{3,i}} (y_{1,i},y_{2,i},y_{3,i}|x_{1,i},x_{2,i},x_{3,i})\nonumber\\
\ \ = P_{Y_{3,i}|X_{1,i},X_{2,i},X_{3,i} } (y_{3,i}|x_{1,i},x_{2,i},x_{3,i})\cdot P_{Y_{1,i}|X_{1,i},X_{2,i},X_{3,i},Y_{3,i} } (y_{1,i}|x_{1,i},x_{2,i},x_{3,i},y_{3,i})\nonumber\\
\quad\ \ \ \cdot P_{Y_{2,i}|X_{1,i},X_{2,i},X_{3,i},Y_{1,i},Y_{3,i}} (y_{2,i}|x_{1,i},x_{2,i},x_{3,i},y_{1,i},y_{3,i})\nonumber\\
\ \ = P_{Y_{3}|X_{1},X_{2},X_{3} } (y_{3,i}|x_{1,i},x_{2,i},x_{3,i} \cdot P_{Z_1}(y_{1,i} \ominus_q x_{1,i} \ominus_q x_{3,i}) \cdot P_{Z_2}( y_{2,i} \ominus_q x_{2,i} \ominus_q y_{1,i} \oplus_q x_{1,i}),\nonumber
\end{IEEEeqnarray}
where $\ominus_q$ denotes modulo-$q$ subtraction.  

We next define channel codes, achievable rates, and channel capacity for the MA/DB TWC.  
\begin{definition}
An $(n, R_{13}, R_{23},R_{31}, R_{32})$ channel code for the memoryless MA/DB TWC consists of four message sets $\mathcal{M}_{13}=\{1,2,...,2^{nR_{13}}\}$, $\mathcal{M}_{23}=\{1,2,...,2^{nR_{23}}\}$, $\mathcal{M}_{31}=\{1,2,...,2^{nR_{31}}\}$, $\mathcal{M}_{32}=\{1,2,...,2^{nR_{32}}\}$, three sequences of encoding functions: $f_{1}^n=(f_{1,1},f_{1,2},...,f_{1,n})$, $f_2^n=(f_{2,1},f_{2,2},...,f_{2,n})$, $f_3^n=(f_{3,1},f_{3,2},...,f_{3,n})$ such that
\begin{IEEEeqnarray}{rClrCl}
X_{1, 1}&=& f_{1,1}(M_{13}), &\ \ X_{1,i} &=&f_{1,i}(M_{13},Y_1^{i-1}), \label{eq:encoder31}\\
X_{2, 1}&=& f_{2,1}(M_{23}), &\ \ X_{2,i} &=&f_{2,i}(M_{23},Y_2^{i-1}), \label{eq:encoder32}\\
X_{3, 1}&=& f_{3,1}(M_{31}, M_{32}), &\ \ X_{3,i} &=& f_{3,i}(M_{31},M_{32},Y_3^{i-1}),\label{eq:encoder33}\IEEEeqnarraynumspace
\end{IEEEeqnarray}
for $i=2, 3, \dots ,n$, and three decoding functions $g_1$, $g_2$, and $g_3$, such that $\hat{M}_{31} =g_{1}(M_{13},Y_1^n)$, $\hat{M}_{32} = g_{2}(M_{23},Y_2^n)$, and $(\hat{M}_{13},\hat{M}_{23} ) = g_{3}(M_{31},M_{32},Y_3^n)$.
\end{definition}

When messages are encoded via the channel code, the probability of decoding error is defined as $P^{(n)}_{\text{e}}(f_1^n, f_2^n, f_3^n, g_1, g_2, g_3)=\Pr\{\hat{M}_{13} \neq M_{13}\ \text{or}\ \hat{M}_{23} \neq M_{23}\ \text{or}\ \hat{M}_{31} \neq M_{31}\ \text{or}\ \hat{M}_{32} \neq M_{32} \}.$
\begin{definition}
A rate quadruple $(R_{13},R_{23},R_{31},R_{32})$ is said to be achievable for the memoryless MA/DB TWC if there exists a sequence of $(n, R_{13}, R_{23},R_{31}, R_{32})$ codes with  $\lim_{n \to \infty} P^{(n)}_{\text{e}}=0.$
\end{definition}

\begin{definition}
The capacity region $\mathcal{C}^{\text{MA-DBC}}$ of the memoryless MA/DB TWC is the closure of the convex hull of all achievable rate quadruples $(R_{13},R_{23},R_{31},R_{32})$. 
\end{definition}
  
\subsection{Capacity Inner and Outer Bounds for the Memoryless MA/DB TWCs}\label{sec2B}
Let $\mathcal{R}^{\text{MA-DBC}}(P_{X_1,X_2,X_3,V},P_{Y_3|X_1,X_2,X_3}, P_{Z_1}, P_{Z_2})$ denote the set of rate quadruples $(R_{13},R_{23},R_{31},R_{32})$ which satisfy the constraints
\begin{IEEEeqnarray}{rCl}
R_{13} & \le & I(X_1;Y_3|X_2,X_3),\nonumber\\
R_{23} & \le & I(X_2;Y_3|X_1,X_3),\nonumber\\
R_{13} + R_{23} & \le & I(X_1,X_2;Y_3|X_3),\nonumber\\
R_{31} & \le & I(X_3;X_3 \oplus_q Z_1|V),\nonumber\\
R_{32} & \le & I(V;X_3 \oplus_q Z_1 \oplus_q Z_2),\nonumber
\end{IEEEeqnarray}
where $V$ is an auxiliary random variable with alphabet $\mathcal{V}$ such that $|\mathcal{V}|\le q+1$ and the mutual information terms are evaluated according to the joint probability distribution $P_{X_1,X_2,X_3,V,Y_3,Z_1,Z_2}=P_{X_1,X_2,X_3,V}{\cdot}P_{Y_3|X_1,X_2,X_3}{\cdot}P_{Z_1}{\cdot}P_{Z_2}$.
We next establish a Shannon-type inner bound and an outer bound for the capacity of MA/DB TWCs in Theorems~\ref{the:3inner} and~\ref{the:3outer}, respectively. Note that the achievable scheme in Theorem~\ref{the:3inner} is given by combining Shannon's standard (non-adaptive) coding schemes for the MAC \cite[Theorem~4.2]{Gamal:2012} and the DBC \cite[Theorem~5.2]{Gamal:2012}, and hence the proof is omitted here. The derivation for the outer bound is given in Appendix~\ref{Appendix3}.
\begin{theorem}[Inner Bound] \label{the:3inner}
For a memoryless MA/DB TWC with MA transition probability $P_{Y_3|X_1,X_2,X_3}$ and DB noise distributions $P_{Z_1}$ and $P_{Z_2}$, any rate quadruple $(R_{13},R_{23},R_{31},R_{32})\in\mathcal{C}^{\text{MA-DBC}}_{\text{I}}(P_{Y_3|X_1,X_2,X_3},P_{Z_1},P_{Z_2})$ is achievable, where 
\begin{IEEEeqnarray}{l}
\mathcal{C}_{\text{I}}^{\text{MA-DBC}}(P_{Y_3|X_1,X_2,X_3},P_{Z_1},P_{Z_2}) \triangleq\scalemath{1}{\overline{\text{co}} \Bigg( \scalemath{0.88}{\bigcup_{P_{X_1}, P_{X_2}, P_{V,X_3}}\mathcal{R}^{\text{MA-DBC}}(P_{X_1}{\cdot}P_{X_2}{\cdot}P_{V,X_3}, P_{Y_3|X_1,X_2,X_3}, P_{Z_1}, P_{Z_2})} \Bigg)}.\nonumber
\end{IEEEeqnarray}
\end{theorem}

\begin{theorem}[Outer Bound] \label{the:3outer}
For a memoryless MA/DB TWC with MA transition probability $P_{Y_3|X_1,X_2}$ and DB noise distributions $P_{Z_1}$ and $P_{Z_2}$, all achievable rate quadruples $(R_{13},R_{23},R_{31},R_{32})$ belong to $\mathcal{C}^{\text{MA-DBC}}_{\text{O}}(P_{Y_3|X_1,X_2,X_3},P_{Z_1},P_{Z_2})$, where 
\begin{IEEEeqnarray}{l}
\mathcal{C}_{\text{O}}^{\text{MA-DBC}}(P_{Y_3|X_1,X_2,X_3},P_{Z_1},P_{Z_2}) \triangleq\scalemath{1}{\overline{\text{co}} \Bigg( \scalemath{0.9}{\bigcup_{P_{X_1,X_2,X_3,V}}}\mathcal{R}^{\text{MA-DBC}}(P_{X_1,X_2,X_3,V}, P_{Y_3|X_1,X_2,X_3},P_{Z_1},P_{Z_2}{)} \Bigg)}.\nonumber
\end{IEEEeqnarray}
\end{theorem}

\subsection{Conditions for the Tightness of the Inner and Outer Bounds}
The inner and outer bounds derived in the previous section are of the same form but have different restrictions on the joint distribution $P_{X_1, X_2, X_3, V}$, and hence they do not match. 
Here, we establish conditions under which the two bounds have matching input distributions, implying that they coincide and yield the capacity region. 
The proofs of Theorems~\ref{thm:exMain}-\ref{thm:exSC} are given in Appendices~\ref{Appendix4}-\ref{Appendix2}, respectively.

\begin{theorem}\label{thm:exMain}
The inner and outer capacity bounds in Theorems~\ref{the:3inner} and~\ref{the:3outer} coincide if for every conditional input distribution $P^{(1)}_{X_1, X_2|X_3}$, there exists a product input distribution $P^{(2)}_{X_1, X_2|X_3}=\tilde{P}_{X_1}{\cdot}\tilde{P}_{X_2}$ (which depends on $P^{(1)}_{X_1, X_2|X_3}$) such that 
\begin{IEEEeqnarray}{rCl}
I^{(1)}(X_1;Y_3|X_2,X_3=x_3)&\le & I^{(2)}(X_1;Y_3|X_2,X_3=x_3)\label{eq:3include1}\IEEEeqnarraynumspace \\
I^{(1)}(X_2;Y_3|X_1,X_3=x_3)&\le & I^{(2)}(X_2;Y_3|X_1,X_3=x_3)\label{eq:3include2}\IEEEeqnarraynumspace \\
I^{(1)}(X_1, X_2;Y_3|X_3=x_3)&\le & I^{(2)}(X_1, X_2;Y_3|X_3=x_3)\label{eq:3include3}\IEEEeqnarraynumspace
\end{IEEEeqnarray}
hold for all $x_3\in\mathcal{X}_3$. 
Under this condition, the capacity region is given by
\begin{IEEEeqnarray}{l}
\mathcal{C}^{\text{MA-DBC}}= \scalemath{1}{\overline{\text{co}} \Bigg( \scalemath{0.87}{\bigcup_{P_{X_1}, P_{X_2}, P_{V,X_3}}\mathcal{R}^{\text{MA-DBC}}\Big(P_{X_1}{\cdot}P_{X_2}{\cdot}P_{V,X_3},P_{Y_3|X_1,X_2,X_3},P_{Z_1},P_{Z_2}\Big)} \Bigg)}. \nonumber
\end{IEEEeqnarray}
\end{theorem}

A special case of the above theorem is when $\tilde{P}_{X_1}{\cdot}\cdot\tilde{P}_{X_2}$ does not depend on $P_{X_1, X_2|X_3}$. 
This case may happen when $P_{Y_3|X_1, X_2, X_3}$ has a strong symmetry property.    
\begin{corollary}\label{the:exMain1}
The inner and outer capacity bounds in Theorems~\ref{the:3inner} and~\ref{the:3outer} coincide if there exists an input distributions $P^{(2)}_{X_1, X_2}=P^*_{X_1}\cdot P^*_{X_2}$ such that for all $P^{(1)}_{X_1,X_2|X_3}$ and $x_3 \in \mathcal{X}_3$ the inequalities given in \eqref{eq:3include1}-\eqref{eq:3include3} hold. 
In this case, the capacity region is given by
\begin{IEEEeqnarray}{l}
\mathcal{C}^{\text{MA-DBC}} = \scalemath{1}{\overline{\text{co}} \Bigg( \scalemath{0.9}{\bigcup_{P_{V,X_3}}\mathcal{R}^{\text{MA-DBC}}\Big(P^*_{X_1}{\cdot}P^*_{X_2}{\cdot}P_{V,X_3},P_{Y_3|X_1,X_2,X_3},P_{Z_1},P_{Z_2}\Big)} \Bigg)}. \nonumber
\end{IEEEeqnarray}
\end{corollary}

The next result is derived by treating the channel as a composition of state-dependent one-way channels. 
\begin{theorem}\label{thm:exMain2}
The inner and outer capacity bounds in Theorems~\ref{the:3inner} and~\ref{the:3outer} coincide if the following conditions hold: 
\begin{itemize}
\item[(i)] There exists $P^*_{X_1}\in\mathcal{P}(\mathcal{X}_1)$ such that \[\argmax_{P_{X_1|X_2=x_2, X_3=x_3}} I(X_1; Y_3|X_2=x_2, X_3=x_3)=P^*_{X_1}\] for all $x_2 \in \mathcal{X}_2$ and $x_3 \in \mathcal{X}_3$, and \[\mathcal{I}(P^*_{X_1}, P_{Y_3|X_1, X_2=x_2, X_3=x_3})\] does not depend on $x_2$ for every fixed $x_3$; 
\item[(ii)] For any $P_{X_2}\in\mathcal{P}(\mathcal{X}_2)$, $\mathcal{I}(P_{X_2}, P_{Y_3|X_1=x_1, X_2, X_3=x_3})$ does not depend on $x_1\in\mathcal{X}_1$ and $x_3\in\mathcal{X}_3$; 
\item[(iii)] For any fixed $P_{X_1, X_2}$, we have that the mutual information $\mathcal{I}(P_{X_1, X_2}, P_{Y_3|X_1, X_2, X_3=x_3})$ does not depend on $x_3\in\mathcal{X}_3$, and for each $x_3\in\mathcal{X}_3$ we have that 
\begin{IEEEeqnarray}{l}
\mathcal{I}(P_{X_1, X_2}, P_{Y_3|X_1, X_2, X_3=x_3}) \le \mathcal{I}(P^*_{X_1}\cdot P_{X_2}, P_{Y_3|X_1, X_2, X_3=x_3}),\nonumber
\end{IEEEeqnarray} 
where $P^*_{X_1}$ is given by condition (i) and $P_{X_2}(x_2)=\sum_{x_1}P_{X_1, X_2}(x_1, x_2)$ for $x_2\in\mathcal{X}_2$. 
\end{itemize}
Under this condition, the capacity region is given by
\begin{IEEEeqnarray}{l}
\mathcal{C}^{\text{MA-DBC}} =\scalemath{1}{\overline{\text{co}} \Bigg( \scalemath{0.9}{\bigcup_{P_{X_2}, P_{V,X_3}}\mathcal{R}^{\text{MA-DBC}}\Big(P^*_{X_1}{\cdot}P_{X_2}{\cdot}P_{V,X_3},P_{Y_3|X_1,X_2,X_3},P_{Z_1},P_{Z_2}\Big)} \Bigg)}. \nonumber
\end{IEEEeqnarray}
\end{theorem}

Next, we derive our last sufficient condition by generalizing Shannon's condition (in Proposition~\ref{thm:SC}) to the three-user setting.  
This new condition is easier to verify than the previous ones.  
\begin{theorem}\label{thm:exSC}
The inner and outer capacity bounds in Theorems~\ref{the:3inner} and~\ref{the:3outer} coincide if the following conditions hold:
\begin{itemize}
\item[(i)] For any relabeling $\tau^{\mathcal{X}_1}_{x'_1,x''_1}$ on $\mathcal{X}_1$, there exists a permutation $\pi^{\mathcal{Y}_3}[x'_1,x''_1]$ on $\mathcal{Y}_3$ such that for all $x_1$, $x_2$, $x_3$, and $y_3$, we have
\begin{IEEEeqnarray}{l}
P_{Y_3|X_1,X_2,X_3}(y_3|x_1,x_2,x_3)\scalemath{1}{=P_{Y_3|X_1,X_2,X_3}\big(\pi^{\mathcal{Y}_3}[x'_1,x''_1](y_3)\big|\tau^{\mathcal{X}_1}_{x'_1,x''_1} (x_1), x_2, x_3 \big)};\IEEEeqnarraynumspace \label{eq:md1}
\end{IEEEeqnarray}
\item[(ii)] For any relabeling $\tau^{\mathcal{X}_2}_{x'_2,x''_2}$  on $\mathcal{X}_2$, there exists a permutation on $\pi^{\mathcal{Y}_3}[x'_2,x''_2]$ on $\mathcal{Y}_3$ such that for all $x_1$, $x_2$, $x_3$, and $y_3$, we have
\begin{IEEEeqnarray}{l}
P_{Y_3|X_1,X_2,X_3}(y_3|x_1,x_2,x_3)\scalemath{1}{=P_{Y_3|X_1,X_2,X_3}\big(\pi^{\mathcal{Y}_3}[x'_1,x''_1](y_3)\big|x_1, \tau^{\mathcal{X}_2}_{x'_2,x''_2} (x_2), x_3 \big)}. \IEEEeqnarraynumspace\label{eq:md2}
\end{IEEEeqnarray}
\end{itemize}
Under these conditions, the capacity region is given by 
\begin{IEEEeqnarray}{l}
\mathcal{C}^{\text{MA-DBC}} \scalemath{1}{= \overline{\text{co}} \Bigg( \scalemath{0.9}{\bigcup_{P_{V,X_3}}\mathcal{R}^{\text{MA-DBC}}\Big(P^{\text{U}}_{\mathcal{X}_1}{\cdot}P^{\text{U}}_{\mathcal{X}_2}{\cdot}P_{V,X_3},P_{Y_3|X_1,X_2,X_3},P_{Z_1},P_{Z_2}\Big)}\Bigg)},\IEEEeqnarraynumspace\label{cap:macbc}
\end{IEEEeqnarray}
\end{theorem}
where $P^{\text{U}}_{\mathcal{X}_i}$ denotes uniform probability distribution on $\mathcal{X}_i$ for $i=1, 2$. 

\subsection{Examples}
We next illustrate Theorems~\ref{thm:exMain}-\ref{thm:exSC} via three examples.

\begin{example}[Additive-Noise MA/DB TWC]
Consider a discrete memoryless additive-noise MA/DB TWC in which the inputs and outputs of the DBC are described by \eqref{eq:MADB1} and \eqref{eq:MADB2} and the inputs and outputs of MAC are related via
\begin{align}
Y_{3,i}= X_{1,i} \oplus_q X_{2,i} \oplus_q X_{3,i} \oplus_q  Z_{3,i}, \label{eq:MADB3} 
\end{align}
where $\{Z_{3,i}\}_{i=1}^\infty$ with $Z_{3,i} \in\mathcal{Q}$ is a discrete memoryless noise process which is independent of all user messages and the noise processes $\{Z_{1,i}\}_{i=1}^\infty$ and $\{Z_{2,i}\}_{i=1}^\infty$. 
For any $x_3\in\mathcal{X}_3$, we have the following bounds: 
\begin{IEEEeqnarray}{l}
I(X_1;Y_3|X_2,X_3=x_3) =H(Y_3|X_2,X_3=x_3)- H(Y_3|X_1,X_2,X_3=x_3) \le \log_2 q - H_{\text{b}}(Z_3),\nonumber\\
I(X_2;Y_3|X_1,X_3=x_3)  = H(Y_3|X_1,X_3=x_3) - H(Y_3|X_1,X_2,X_3=x_3) \le \log_2 q - H_{\text{b}}(Z_3),\nonumber\\
I(X_1,X_2;Y_3|X_3=x_3)   = H(Y_3|X_3=x_3) - H(Y_3|X_1,X_2,X_3=x_3) \le \log_2 q - H_{\text{b}}(Z_3),\nonumber
\end{IEEEeqnarray}
where equalities hold when $P_{X_1,X_2}=P^{\text{U}}_{\mathcal{X}_1}\cdot P^{\text{U}}_{\mathcal{X}_2}$. 
Choosing $\tilde{P}_{X_1}=P^{\text{U}}_{\mathcal{X}_1}$ and $\tilde{P}_{X_2}=P^{\text{U}}_{\mathcal{X}_2}$, it is clear that \eqref{eq:3include1}-\eqref{eq:3include3} in Theorem~\ref{thm:exMain} hold, and hence the capacity region given by  
\begin{IEEEeqnarray}{rCl}
\mathcal{C}^{\text{MA-DBC}} &=& \overline{\text{co}} \scalemath{0.9}{\left( \bigcup_{P_{V,X_3}}  
\mathcal{R}^{\text{MA-DBC}}\Big(P^{\text{U}}_{\mathcal{X}_1}{\cdot}P^{\text{U}}_{\mathcal{X}_2}{\cdot}P_{U,X_3},P_{Y_3|X_1,X_2,X_3},P_{Z_1},P_{Z_2}\Big) \right)} \nonumber\\
&= & \overline{\text{co}} \Bigg(  \bigcup_{P_{V,X_3}}  
 \big\{ (R_{13},R_{23}, R_{31},R_{32} ): R_{13} + R_{23} \le  \log_2 q - H_{\text{b}}(Z_3), \nonumber\\
& & \qquad \qquad \qquad \qquad \qquad \quad \quad \quad \quad \qquad \hspace{+0.08cm} R_{31} \le  I(X_1;X_3\oplus_2 Z_1|V), \nonumber\\
& & \qquad \qquad \qquad \qquad \qquad \quad \quad \quad \quad \qquad \hspace{+0.08cm} R_{32} \le I(X_2\oplus Z_1\oplus Z_2;V)\big\} \Bigg). \nonumber
\end{IEEEeqnarray}
\end{example}

\begin{example}
Suppose that $\mathcal{X}_1=\mathcal{X}_2=\mathcal{X}_3=\{0, 1\}$, $\mathcal{Y}_1=\mathcal{Y}_2=\{0, 1\}$, and $\mathcal{Y}_3=\{0, 1, 2\}$. 
We consider a discrete memoryless MA/DB TWC in which the DB direction is described by \eqref{eq:MADB1} and \eqref{eq:MADB2} and the channel transition matrix $[P_{Y_3|X_1, X_2, X_3}(\cdot|\cdot, \cdot, \cdot)]$ for the MA direction is given by
\[
\scalemath{1}{
\bordermatrix{
& 0 & 1 & 2\cr
000& 1-\varepsilon & 0 & \varepsilon\cr
100& 1-\varepsilon & 0 & \varepsilon\cr
010& 0 & 1-\varepsilon & \varepsilon\cr
110& 0 & 1-\varepsilon & \varepsilon \cr
001& 0 & \varepsilon & 1-\varepsilon \cr
101& 0 & \varepsilon & 1-\varepsilon \cr
011& 1-\varepsilon & \varepsilon & 0 \cr
111& 1-\varepsilon & \varepsilon & 0}}
\]
where $0\le \varepsilon\le 1$. 
Since each marginal channel governed by the transition matrix $[P_{Y_3|X_1, X_2, X_3}(\cdot|\cdot, x_2, x_3)]$ is quasi-symmetric, we immediately have that $P^*_{X_1}=P^{\text{U}}_{\mathcal{X}_1}$. 
Also, since $[P_{Y_3|X_1, X_2, X_3}(\cdot|\cdot, x_2, x_3)]$, $x_2\in\mathcal{X}_2$ and $x_3\in\mathcal{X}_3$, are column permutations of each other, for any fixed $x_3\in\mathcal{X}_3$, $\mathcal{I}(P^*_{X_1}, P_{Y_3|X_1, X_2=x_2, X_3=x_3})$ does not depend on $x_2\in\mathcal{X}_2$. 
Thus, condition~(i) of Theorem~\ref{thm:exMain2} holds.  
Moreover, condition (ii) holds since the matrices $[P_{Y_3|X_1, X_2, X_3}(\cdot|x_1, \cdot, x_3)]$, $x_1\in\mathcal{X}_1$ and $x_3\in\mathcal{X}_3$, are column permutations of each other. 

Verifying condition (iii) involves several steps.
We first observe that $\mathcal{I}(P_{X_1, X_2}, P_{Y_3|X_1, X_2, X_3=x_3})$ does not depend on $x_3\in\mathcal{X}_3$ for any fixed $P_{X_1, X_2}$ since the matrices $[P_{Y_3|X_1, X_2, X_3}(\cdot|\cdot, \cdot, x_3)]$, $x_3\in\mathcal{X}_3$, are column permutations of each other.  
From \eqref{eq:macnew6} and \eqref{eq:macnew7} in Appendix~\ref{Appendix5}, it suffices to consider input distributions of this form: $P_{X_1, X_2, X_3, V}=P_{X_1,X_2}\cdot P_{X_3, V}$. 
Thus, given any $P^{(1)}_{X_1, X_2, X_3, V}=P^{(1)}_{X_1,X_2}\cdot P^{(1)}_{X_3, V}$, we define $P^{(2)}_{X_1, X_2, X_3, V}(x_1, x_2, x_3, v)=P^{(1)}_{X_1, X_2, X_3, V}(x_1\oplus_2 1, x_2, x_3, v)$ for all $x_1, x_2, x_3, v$. 
Also, let $P^{(3)}_{X_1, X_2, X_3, V}=\frac{1}{2}(P^{(1)}_{X_1, X_2, X_3, V}+P^{(2)}_{X_1, X_2, X_3, V})$ so that we have $P^{(3)}_{X_1, X_2, X_3, V}=P^{(3)}_{X_1}\cdot P^{(1)}_{X_2}\cdot P^{(1)}_{X_3, V}$ with $P^{(3)}_{X_1}=P^{\text{U}}_{\mathcal{X}_1}=P^{*}_{X_1}$. 
Now, since \eqref{eq:md1} holds in this example, one can directly obtain   that $I^{(1)}(X_1, X_2; Y_3|X_3=x_3)\le I^{(3)}(X_1, X_2; Y_3|X_3=x_3)$ from the proof of Lemma~\ref{lma:mcbase}. 
As a result, this TWC satisfies all conditions of Theorem~\ref{thm:exMain2} and has capacity region given by
\begin{IEEEeqnarray}{l}
\mathcal{C}^{\text{MA-DBC}} \scalemath{1}{= \overline{\text{co}} \Bigg( \scalemath{1}{\bigcup_{P_{X_2}, P_{V,X_3}}\mathcal{R}^{\text{MA-DBC}}\Big(P^{\text{U}}_{\mathcal{X}_1}{\cdot}P_{X_2}{\cdot}P_{V,X_3},P_{Y_3|X_1,X_2,X_3},P_{Z_1},P_{Z_2}\Big)} \Bigg)}. \nonumber
\end{IEEEeqnarray}
\end{example}

\begin{example}[Binary MA/DB TWC with Erasures]
Suppose that $\mathcal{X}_1=\mathcal{X}_2=\mathcal{X}_3=\{0, 1\}$, $\mathcal{Y}_1=\mathcal{Y}_2=\{0, 1\}$, and $\mathcal{Y}_3=\{0, 1, \mathbf{E}\}$, where $\mathbf{E}$ denotes erasure symbol. 
We consider a discrete memoryless MA/DB TWC in which the DBC direction is described by \eqref{eq:MADB1} and \eqref{eq:MADB2} and the MAC direction is described by
\begin{align}
Y_{3,i}= (X_{1,i} \oplus_2 X_{2,i} \oplus_2 X_{3,i}){\cdot}\mathbf{1}\{Z_{3, i}\neq \mathbf{E}\}{+}\mathbf{E}{\cdot}\mathbf{1}\{Z_{3,i}=\mathbf{E}\}, \label{eq:MADB3} 
\end{align}
where $\{Z_{3,i}\}_{i=1}^\infty$ with $Z_{3,i}\in \{0, \mathbf{E}\}$ is a discrete memoryless noise process which is independent of all users' messages and the noise processes $\{Z_{1,i}\}_{i=1}^\infty$ and $\{Z_{2,i}\}_{i=1}^\infty$. 
Also, we assume that $\Pr(Z_{3,i}=\mathbf{E})=\varepsilon$ for all $i$, thereby obtaining the channel transition matrix $[P_{Y_3|X_1, X_2, X_3}(\cdot|\cdot, \cdot, \cdot)]$:
\[
\scalemath{1}{
\bordermatrix{
& 0 & 1 & \mathbf{E}\cr
000& 1-\varepsilon & 0 & \varepsilon\cr
100& 0 & 1-\varepsilon & \varepsilon\cr
010& 0 & 1-\varepsilon & \varepsilon\cr
110& 1-\varepsilon & 0 & \varepsilon \cr
001& 0 & 1-\varepsilon & \varepsilon \cr
101& 1-\varepsilon & 0 & \varepsilon \cr
011& 1-\varepsilon & 0 & \varepsilon \cr
111& 0 & 1-\varepsilon & \varepsilon}}.
\]
It can be directly verified that \eqref{eq:md1} and \eqref{eq:md2} in Theorem~\ref{thm:exSC} hold. 
Hence, the inner and outer bounds coincide and the capacity region is given by
\begin{IEEEeqnarray}{rCl}
\mathcal{C}^{\text{MA-DBC}}  &=& \overline{\text{co}} \scalemath{0.9}{\Bigg( \bigcup_{P_{V,X_3}}\mathcal{R}^{\text{MA-DBC}}\Big(P^{\text{U}}_{\mathcal{X}_1}{\cdot}P^{\text{U}}_{\mathcal{X}_2}{\cdot}P_{V,X_3},P_{Y_3|X_1,X_2,X_3},P_{Z_1},P_{Z_2}\Big) \Bigg)}\nonumber\\
&=&  \overline{\text{co}} \Bigg(  \bigcup_{P_{V,X_3}}  
 \big\{ (R_{13},R_{23}, R_{31},R_{32} ): R_{13} + R_{23} \le  1 - H_{\text{b}}(\epsilon), \nonumber\\
& & \qquad \qquad \qquad \quad \quad \quad \quad \quad \quad \quad \quad \quad \quad \hspace{+0.08cm} R_{31} \le  I(X_1;X_3\oplus_2 Z_1|V), \nonumber\\
& & \qquad \qquad \qquad \quad \quad \quad \quad \quad \quad \quad \quad \quad \quad \hspace{+0.08cm} R_{32} \le I(X_2\oplus_2 Z_1 \oplus_2 Z_2;V)\big\} \Bigg). \nonumber
\end{IEEEeqnarray}
\end{example}   

\begin{remark}
Examples~9 and~10 also satisfy Theorem~\ref{thm:exMain} since the product distribution $\tilde{P}_{X_1}\cdot\tilde{P}_{X_2}$ required by Theorem~\ref{thm:exMain} are explicitly given in these examples. 
Moreover, it is straightforward to show that Examples~9 and~10 do not satisfy the conditions of Theorems~\ref{thm:exSC} and~\ref{thm:exMain2}, respectively. 
In other words, Theorems~\ref{thm:exMain2} and~\ref{thm:exSC} are neither equivalent nor special cases of each other.
\end{remark}

\section{Conclusion}\label{sec:conclusion}
We have identified salient symmetry conditions for three types of two-way noisy networks: two-user TWCs with and without memory, and three-user MA/DB TWCs, under which Shannon-type random coding inner bounds exactly yield channel capacity. 
These tightness results, which subsume previously established symmetry properties as special cases, delineate large families of TWCs for which user interactive adaptive coding is not beneficial in terms of improving capacity. 
Future research directions include identifying necessary conditions for the tightness of Shannon-type inner bounds and deriving conditions under which Han's adaptive coding inner bound \cite{Han:1984} is tight. 
An additional interesting avenue of investigation is to examine whether adaptive coding is useful for the (almost) lossless and lossy transmission of correlated sources over TWCs whose capacity are achievable by the Shannon-type random coding scheme.

\appendix
\subsection{Proof of Proposition \ref{thm:SC} (Shannon's One-sided Symmetry Condition)}\label{appendix:SCproof}
The proof of Proposition \ref{thm:SC} is based on the following lemmas. 
\begin{lemma}\label{lem:Sl1}
If a memoryless TWC satisfies the conditions in Proposition \ref{thm:SC}, then for any input distribution $P^{(1)}_{X_1,X_2}$, any $x_1'$, $x_1''\in \mathcal{X}_1$, and $P^{(2)}_{X_1,X_2}(\cdot, \cdot)\triangleq P^{(1)}_{X_1,X_2}(\tau^{\mathcal{X}_1}_{x_1',x_1''}(\cdot), \cdot)$, 
the following hold: 
\begin{align}
I^{(1)}(X_1;Y_2|X_2) & =I^{(2)}(X_1;Y_2|X_2),\label{eq:r1}\\
I^{(1)}(X_2;Y_1|X_1) & = I^{(2)}(X_2;Y_1|X_1),\label{eq:r2}\\
\mathcal{R}(P^{(1)}_{X_1,X_2},P_{Y_1,Y_2|X_1,X_2})&=\mathcal{R}(P^{(2)}_{X_1,X_2},P_{Y_1,Y_2|X_1,X_2}). \label{eq:r3}
\end{align}
\end{lemma}
\begin{IEEEproof}
For any $P^{(1)}_{X_1,X_2}$ and $P^{(2)}_{X_1,X_2}(\cdot, \cdot)\triangleq P^{(1)}_{X_1,X_2}(\tau^{\mathcal{X}_1}_{x_1',x_1''}(\cdot), \cdot)$, we have 
\begin{IEEEeqnarray}{l}
I^{(2)}(X_1;Y_2|X_2)\nonumber\\
\ = \sum_{x_2} P^{(2)}_{X_2}(x_2)\cdot I^{(2)}(X_1;Y_2|X_2=x_2)\nonumber\\
\ =  \sum_{x_2} P^{(2)}_{X_2}(x_2)\sum_{x_1,y_2} P^{(2)}_{X_1|X_2}(x_1|x_2)\cdot P_{Y_2|X_1,X_2}(y_2|x_1,x_2)\cdot\log\frac{P_{Y_2|X_1,X_2}(y_2|x_1,x_2)}{P^{(2)}_{Y_2|X_2}(y_2|x_2)}\nonumber\\
=  \sum_{x_1,x_2,y_2} P^{(2)}_{X_1,X_2}(x_1,x_2)\cdot P_{Y_2|X_1,X_2}(y_2|x_1,x_2)\cdot \log\frac{P_{Y_2|X_1,X_2}(y_2|x_1,x_2)}{\sum_{\tilde{x}_1}P^{(2)}_{X_1|X_2}(\tilde{x}_1|x_2){\cdot}P_{Y_2|X_1,X_2}(y_2|\tilde{x}_1,x_2)}\nonumber\\[+1em]
=  \sum_{x_1,x_2,y_2} P^{(1)}_{X_1,X_2}(\tau^{\mathcal{X}_1}_{x_1',x_1''}(x_1),x_2)\cdot P_{Y_2|X_1,X_2}(\pi^{\mathcal{Y}_2}[x_1',x_1''](y_2)|\tau^{\mathcal{X}_1}_{x_1',x_1''}(x_1),x_2)\nonumber\\
\qquad\qquad\qquad  \cdot \scalemath{1}{\log\frac{P_{Y_2|X_1,X_2}(\pi^{\mathcal{Y}_2}[x_1',x_1''](y_2)|\tau^{\mathcal{X}_1}_{x_1',x_1''}(x_1),x_2)}{\sum_{\tilde{x}_1}P^{(1)}_{X_1|X_2}(\tau^{\mathcal{X}_1}_{x_1',x_1''}(\tilde{x}_1)|x_2){\cdot}P_{Y_2|X_1,X_2}(\pi^{\mathcal{Y}_2}[x_1',x_1''](y_2)|\tau^{\mathcal{X}_1}_{x_1',x_1''}(\tilde{x}_1),x_2)}}\label{eq:s11}\\ 
 =  \sum_{x_1,x_2,y_2} P^{(1)}_{X_1,X_2}(\tau^{\mathcal{X}_1}_{x_1',x_1''}(x_1),x_2)\cdot P_{Y_2|X_1,X_2}(\pi_2^{\mathcal{Y}_2}[x_1',x_1''](y_2)|\tau^{\mathcal{X}_1}_{x_1',x_1''}(x_1),x_2)\nonumber\\
\qquad\qquad\qquad \cdot \scalemath{0.95}{\log\frac{P_{Y_2|X_1,X_2}(\pi^{\mathcal{Y}_2}[x_1',x_1''](y_2)|\tau^{\mathcal{X}_1}_{x_1',x_1''}(x_1),x_2)}{\sum_{\tilde{x}_1} P^{(1)}_{X_1|X_2}(\tilde{x}_1|x_2) P_{Y_2|X_1,X_2}(\pi^{\mathcal{Y}_2}[x_1',x_1''](y_2)|\tilde{x}_1,x_2) }}\label{eq:s12}\\
 = \sum_{x_1,x_2,y_2}{\cdot}P^{(1)}_{X_1,X_2}(\tau^{\mathcal{X}_1}_{x_1',x_1''}(x_1),x_2)\cdot P_{Y_2|X_1,X_2}(\pi_2^{\mathcal{Y}_2}[x_1',x_1''](y_2)|\tau^{\mathcal{X}_1}_{x_1',x_1''}(x_1),x_2) \nonumber\\
 \qquad\qquad\qquad \cdot  \log\frac{P_{Y_2|X_1,X_2}(\pi^{\mathcal{Y}_2}[x_1',x_1''](y_2)|\tau^{\mathcal{X}_1}_{x_1',x_1''}(x_1),x_2)}{P^{(1)}_{Y_2|X_2}(\pi^{\mathcal{Y}_2}[x_1',x_1''](y_2)|x_2) }\nonumber\\
 =  \sum_{x_1,x_2,\tilde{y}_2} \scalemath{1}{P^{(1)}_{X_1,X_2}(\tau^{\mathcal{X}_1}_{x_1',x_1''}(x_1),x_2)\cdot P_{Y_2|X_1,X_2}(\tilde{y}_2|\tau^{\mathcal{X}_1}_{x_1',x_1''}(x_1),x_2)} \cdot\log\frac{P_{Y_2|X_1,X_2}(\tilde{y}_2|\tau^{\mathcal{X}_1}_{x_1',x_1''}(x_1),x_2)}{P^{(1)}_{Y_2|X_2}(\tilde{y}_2|x_2) }\label{eq:s13}\IEEEeqnarraynumspace\\
 =  \sum_{\tilde{x}_1, x_2, \tilde{y}_2} P^{(1)}_{X_1,X_2}(\tilde{x}_1,x_2) \cdot P_{Y_2|X_1,X_2}(\tilde{y}_2|\tilde{x}_1,x_2)\cdot\log\frac{P_{Y_2|X_1,X_2}(\tilde{y}_2|\tilde{x}_1,x_2)}{P^{(1)}_{Y_2|X_2}(\tilde{y}_2|x_2) }\label{eq:s14}\\\
 = I^{(1)}(X_1;Y_2|X_2), \label{eq:s15}
\end{IEEEeqnarray}
where \eqref{eq:s11} holds by the definition of $P^{(2)}_{X_1,X_2}(x_1,x_2)$ and the fact that $P_{Y_2|X_1,X_2}(y_2|x_1,x_2)=\allowbreak P_{Y_2|X_1,X_2}(\pi^{\mathcal{Y}_2}[x_1',x_1''](y_2)|\tau^{\mathcal{X}_1}_{x_1',x_1''}(x_1),x_2)$ due to the Shannon condition in \eqref{SCcond}, \eqref{eq:s12} and \eqref{eq:s14} hold since $\tau^{\mathcal{X}_1}_{x_1',x_1''}$ is a bijection, and \eqref{eq:s13} holds since $\pi^{\mathcal{Y}_2}[x'_1,x''_1]$ is a bijection.

By a similar argument, we can verify that $I^{(1)}(X_2;Y_1|X_1)=\allowbreak I^{(2)}(X_2;Y_1|X_1)$. 
The proof is then completed by noting that the identity $\mathcal{R}(P^{(1)}_{X_1,X_2},P_{Y_1,Y_2|X_1,X_2})=\allowbreak\mathcal{R}(P^{(2)}_{X_1,X_2},P_{Y_1,Y_2|X_1,X_2})$ follows from the definition of $\mathcal{R}$ in \eqref{eq:rateR}. 
\end{IEEEproof}

\begin{lemma}\label{lem:Sl2}
If a memoryless TWC satisfies the condition in Proposition~\ref{thm:SC}, then for any input distribution $P^{(1)}_{X_1,X_2}$, any $x_1'$, $x_1''\in \mathcal{X}_1$, and $P^{(2)}_{X_1,X_2}(\cdot, \cdot)\triangleq P^{(1)}_{X_1,X_2}(\tau^{\mathcal{X}_1}_{x_1',x_1''}(\cdot), \cdot)$, we have
\begin{align}
\mathcal{R}(P^{(1)}_{X_1,X_2},P_{Y_1,Y_2|X_1,X_2})&\subseteq \mathcal{R}(P^{(3)}_{X_1,X_2},P_{Y_1,Y_2|X_1,X_2})\label{eq:r4}
\end{align}
where $P^{(3)}_{X_1,X_2}(x_1,x_2){\triangleq}\frac{1}{2}(P^{(1)}_{X_1,X_2}(x_1,x_2){+}P^{(2)}_{X_1,X_2}(x_1,x_2))$.
\end{lemma}

\begin{IEEEproof}
The proof relies on the concavity of $I(X_1;Y_2|X_2)$ and $I(X_2;Y_1|X_1)$ in $P_{X_1,X_2}$ \cite{Shannon:1961}. 
For any given $P^{(1)}_{X_1, X_2}$ and $P^{(2)}_{X_1,X_2}(\cdot, \cdot)= P^{(1)}_{X_1,X_2}(\tau^{\mathcal{X}_1}_{x_1',x_1''}(\cdot), \cdot)$, let $P^{(3)}_{X_1,X_2}=\frac{1}{2}(P^{(1)}_{X_1,X_2}+P^{(2)}_{X_1,X_2})$. The concavity property then implies that
\begin{IEEEeqnarray}{rCl}
I^{(3)}(X_1;Y_2|X_2) &\ge & \frac{1}{2}\big( I^{(1)}(X_1;Y_2|X_2) + I^{(2)}(X_1;Y_2|X_2)\big)\label{eq:sr11}\IEEEeqnarraynumspace\\
&= & I^{(1)}(X_1;Y_2|X_2),\label{eq:sr12}
\end{IEEEeqnarray}
and that
\begin{IEEEeqnarray}{rCl}
I^{(3)}(X_2;Y_1|X_1)&\ge & \frac{1}{2}\big( I^{(1)}(X_2;Y_1|X_1) + I^{(2)}(X_2;Y_1|X_1)\big)\label{eq:sr21}\IEEEeqnarraynumspace\\
&= & I^{(1)}(X_2;Y_1|X_1), \label{eq:sr22}
\end{IEEEeqnarray}
where \eqref{eq:sr12} and \eqref{eq:sr22} follow from Lemma~\ref{lem:Sl1}. 
The proof is completed by invoking the definition of $\mathcal{R}$ in \eqref{eq:rateR}. 
\end{IEEEproof}

\begin{lemma}\label{lem:Sl3}
If a memoryless TWC satisfies the condition in Proposition~\ref{thm:SC}, then for any given input distribution $P_{X_1,X_2}=P_{X_1|X_2}P_{X_2}$, we have
\begin{IEEEeqnarray}{l}
\mathcal{R}(P_{X_1,X_2},P_{Y_1,Y_2|X_1,X_2}) \subseteq \mathcal{R}\Big(P^{\text{U}}_{\mathcal{X}_1}{\cdot}P_{X_2}, P_{Y_1,Y_2|X_1,X_2}\Big),\IEEEeqnarraynumspace\label{eq:rave2}
\end{IEEEeqnarray}
where $P^{\text{U}}_{\mathcal{X}_1}$ denotes the uniform probability distribution on $\mathcal{X}_1$.
\end{lemma}
\begin{IEEEproof}
Without loss of generality, we assume that $\mathcal{X}_1 \triangleq \{ 1,2,..., \kappa \}$.  
Define $\mathcal{P}_{m}=\{P_{X_1, X_2}\in\mathcal{P}(\mathcal{X}_1\times\mathcal{X}_2): P_{X_1, X_2}(1, x_2)=P_{X_1, X_2}(2, x_2)=\cdots=P_{X_1, X_2}(m, x_2)\ \text{for all}\ x_2\in\mathcal{X}_2\}$, where $1\le m\le \kappa$. 
Lemma \ref{lem:Sl2} shows that for any $P^{(1)}_{X_1, X_2}\in\mathcal{P}_1$, one can construct $P^{(3)}_{X_1, X_2}\in\mathcal{P}_2$ in such a way that \eqref{eq:r4} holds. 
We now extend this result by induction on $m$ showing that for any $P^{(1)}_{X_1,X_2}\in\mathcal{P}_{m}$ with $2\le m< \kappa$, there exists a $P^{(m+2)}_{X_1, X_2}\in\mathcal{P}_{m+1}$ such that $\mathcal{R}(P^{(1)}_{X_1,X_2},P_{Y_1,Y_2|X_1,X_2})\subseteq \mathcal{R}(P^{(m+2)}_{X_1,X_2},P_{Y_1,Y_2|X_1,X_2})$. 

Suppose that the above claim is true up to $m$ for some $1\le m<\kappa$, where the base case $m=1$ was proved in Lemma \ref{lem:Sl2}. 
We next prove the claim for $m+1$. 
For any $P^{(1)}_{X_1,X_2}\in\mathcal{P}_{m}$, define 
\[
P^{(m+2)}_{X_1, X_2}(x_1, x_2)\triangleq \frac{1}{m+1}\sum_{i=1}^{m+1} P^{(i)}_{X_1, X_2}(x_1, x_2),
\label{eq:mixinduc}
\]
where $P^{(i)}_{X_1, X_2}(\cdot, \cdot)\triangleq P^{(1)}_{X_1, X_2}(\tau^{\mathcal{X}_1}_{i-1, m+1}(\cdot), \cdot)$ for $2\le i\le m+1$. 
Due to the Shannon's one-sided symmetry condition and Lemma~\ref{lem:Sl1}, we have that $I^{(i)}(X_1;Y_2|X_2)=I^{(1)}(X_1;Y_2|X_2)$ and that $I^{(i)}(X_2;Y_1|X_1)=I^{(1)}(X_2;Y_1|X_1)$ for $2\le i\le m+1$. Concavity then implies that
\begin{align}
I^{(m+2)}(X_1;Y_2|X_2)
& \ge \frac{1}{m+1}\sum_{i=1}^{m+1} I^{(i)}(X_1;Y_2|X_2)\nonumber\\
& = I^{(1)}(X_1;Y_2|X_2).\nonumber
\end{align} 
Similarly, we obtain that $I^{(m+2)}(X_2;Y_1|X_1){\ge}I^{(1)}(X_2;Y_1|X_1)$. 
Moreover, since $P^{(1)}_{X_1, X_2}\in \mathcal{P}_{m}$, we have that $P^{(m+2)}_{X_1, X_2}(x_1, x_2)=(m\cdot P_{X_1, X_2}^{(1)}(1, x_2)+P_{X_1, X_2}^{(1)}(m+1, x_2))/(m+1)$ for $1\le x_1\le m+1$ and all $x_2\in\mathcal{X}_2$, i.e., $P^{(m+2)}_{X_1, X_2}\in\mathcal{P}_{m+1}$, thereby proving the claim. 

Since any $P_{X_1, X_2}=P_{X_1|X_2}{\cdot}P_{X_2}\in\mathcal{P}_{\kappa}$ can be expressed as $P^{\text{U}}_{\mathcal{X}_1}{\cdot}P_{X_2}$, in view of the definition of $\mathcal{R}$ the proof is completed.
\end{IEEEproof}

We are now ready to prove Proposition~\ref{thm:SC}. 
\begin{IEEEproof}[Proof of Proposition~1]
Note that
\begin{IEEEeqnarray}{rCl}
\mathcal{C}_{\text{O}}(P_{Y_1,Y_2|X_1,X_2})& =& \overline{\text{co}} \left( \bigcup_{ P_{X_1,X_2}}\mathcal{R}(P_{X_1,X_2},P_{Y_1,Y_2|X_1,X_2}) \right)\nonumber\\
& \subseteq & \overline{\text{co}} \scalemath{0.9}{\left( \bigcup_{ P_{X_2}}\mathcal{R}\Big(P^{\text{U}}_{\mathcal{X}_1}{\cdot}P_{X_2},P_{Y_1,Y_2|X_1,X_2}\Big) \right)}\IEEEeqnarraynumspace\label{eq:s3}\\
& \subseteq &\mathcal{C}_{\text{I}}(P_{Y_1,Y_2|X_1,X_2}),\label{eq:s4}
\end{IEEEeqnarray}
where \eqref{eq:s3} follows from Lemma \ref{lem:Sl3}. 
Together with $\mathcal{C}_\text{I}(P_{Y_1,Y_2|X_1,X_2}) \subseteq \mathcal{C}_\text{O}(P_{Y_1,Y_2|X_1,X_2})$, this gives: 
\begin{align}
\mathcal{C}& =\mathcal{C}_\text{I}(P_{Y_1,Y_2|X_1,X_2})\nonumber\\
& = \mathcal{C}_\text{O}(P_{Y_1,Y_2|X_1,X_2})\nonumber\\
& =\overline{\text{co}} \left( \bigcup_{ P_{X_2}}\mathcal{R}\Big(P^{\text{U}}_{\mathcal{X}_1}{\cdot}P_{X_2},P_{Y_1,Y_2|X_1,X_2}\Big) \right). \label{eq:s5}
\end{align}
\end{IEEEproof}

We remark that, based on the proof of Proposition~\ref{thm:SC}, it is straightforward to prove Shannon's two-sided symmetry condition in Proposition \ref{thm:SC2}. 

\subsection{Proof of Theorem~\ref{the:3outer}}\label{Appendix3}
\begin{IEEEproof}
Suppose that $(R_{13},R_{23},R_{31},R_{32})$ is an achievable quadruple. 
We derive the necessary conditions for those rates by the standard converse method.  
For $R_{13}$, we have
\begin{IEEEeqnarray}{l}
 n{\cdot}R_{13} \nonumber\\
\ = H(M_{13}|M_{23},M_{31},M_{32}) \nonumber \\
\ = \scalemath{0.95}{I(M_{13};Y_3^n|M_{23},M_{31},M_{32})- H(M_{13}|Y_3^n,M_{23},M_{31},M_{32})}\nonumber\\
\ \le I(M_{13};Y_3^n|M_{23},M_{31},M_{32}) + n{\cdot}\epsilon_{n} \label{eq:r13s4}\\ 
\ \le I(M_{13};Y_2^n,Y_3^n|M_{23},M_{31},M_{32}) + n{\cdot}\epsilon_{n}\nonumber\\
\ = \sum_{i=1}^n I(M_{13}; Y_{2,i}, Y_{3,i}|Y_2^{i-1},Y_3^{i-1},M_{23},M_{31},M_{32})+ n{\cdot}\epsilon_{n}\nonumber\\
\ = \sum_{i=1}^n \Big( H (Y_{2,i}, Y_{3,i}|X_{2,i},X_{3,i},Y_2^{i-1},Y_3^{i-1},M_{23},M_{31},M_{32}) \nonumber\\
\qquad\qquad - H (Y_{2,i}, Y_{3,i}|X_{2,i},X_{3,i},Y_2^{i-1},Y_3^{i-1},M_{23}, M_{31},M_{32},M_{13}) \Big) + n{\cdot}\epsilon_{n} \label{eq:r13s8}\\
\ \le \sum_{i=1}^n \Big( H (Y_{2,i}, Y_{3,i}|X_{2,i},X_{3,i})- H (Y_{2,i}, Y_{3,i}|X_{1,i},X_{2,i},X_{3,i}) \Big) + n{\cdot}\epsilon_{n}\label{eq:r13s10}\\
\ = \sum_{i=1}^n I (X_{1,i};Y_{2,i}, Y_{3,i}|X_{2,i},X_{3,i})+ n{\cdot}\epsilon_{n} \nonumber\\
\ = \sum_{i=1}^n \scalemath{0.9}{I (X_{1,i};X_{2,i}\oplus_q X_{3,i}\oplus_q Z_{1,i}\oplus_q Z_{2,i}, Y_{3,i}|X_{2,i},X_{3,i})+ n{\cdot}\epsilon_{n}}\nonumber\\
\ = \sum_{i=1}^n I (X_{1,i};Y_{3,i}|X_{2,i},X_{3,i})+ I (X_{1,i}; Z_{1,i}\oplus_q Z_{2,i}| Y_{3,i},X_{2,i},X_{3,i})+ n{\cdot}\epsilon_{n}\nonumber\\
\ = \sum_{i=1}^n I (X_{1,i};Y_{3,i}|X_{2,i},X_{3,i})+ n{\cdot}\epsilon_{n},\label{eq:r13s14}
\end{IEEEeqnarray}
where (\ref{eq:r13s4}) follows from Fano's inequality with $\epsilon_{n}\rightarrow 0$ as $n \to \infty$, (\ref{eq:r13s8}) holds since $X_{2, i}=f_{2, i}(M_{23}, Y_2^{i-1})$ and $X_{3, i}=f_{3, i}(M_{31}, M_{32}, Y_3^{i-1})$, (\ref{eq:r13s10}) follows since the channel is memoryless, and \eqref{eq:r13s14} follows since $(Z_{1,i}, Z_{2,i})$ is independent of $(Y_{3,i},X_{1,i},X_{2,i},X_{3,i})$.
By symmetry, we also have 
\begin{align}
n{\cdot}R_{23} \le \sum_{i=1}^n I (X_{2,i};Y_{3,i}|X_{1,i},X_{3,i}) + n{\cdot}\epsilon_{n}.\label{eq:1c4}
\end{align}
For the sum rate $R_{13}+R_{23}$, we have  
\begin{IEEEeqnarray}{l}
n\cdot (R_{13}+R_{23})\nonumber\\ 
\quad = H(M_{13},M_{23}|M_{31},M_{32}) \nonumber \\
\quad \le I(M_{13},M_{23};Y_3^n|M_{31},M_{32})  + n{\cdot}\epsilon_{n}\nonumber\\ 
\quad = \sum_{i=1}^n \Big( H (Y_{3,i}|X_{3,i},Y_3^{i-1},M_{31},M_{32})- H ( Y_{3,i}|Y_3^{i-1},M_{31},M_{32},M_{13},M_{23}) \Big)+ n{\cdot}\epsilon_{n}\nonumber\\
\quad \le \sum_{i=1}^n \Big( H (Y_{3,i}|X_{3,i}) - H ( Y_{3,i}|Y_3^{i-1},M_{31},M_{32},M_{13},M_{23}) \Big)+ n{\cdot}\epsilon_{n}\nonumber\\
\quad \le \sum_{i=1}^n \Big( H (Y_{3,i}|X_{3,i}) - H ( Y_{3,i}|X_{1,i},X_{2,i},X_{3,i}) \Big) + n{\cdot}\epsilon_{n}\IEEEeqnarraynumspace\nonumber\\ 
\quad = \sum_{i=1}^n I (X_{1,i},X_{2,i}; Y_{3,i}|X_{3,i})+ n{\cdot}\epsilon_{n}, \nonumber
\end{IEEEeqnarray}
where $\epsilon_{n}\rightarrow 0$ as $n \to \infty$ by Fano's inequality. Therefore, for the rates in the MA direction, we have 
\begin{IEEEeqnarray}{rCl}
R_{13} & \le & \frac{1}{n}\sum_{i=1}^n I (X_{1,i};Y_{3,i}|X_{2,i},X_{3,i})+ \epsilon_{n} \nonumber\\
&\le & I (X_{1};Y_{3}|X_{2},X_{3})+\epsilon_{n}\nonumber\\[+0.2cm]
R_{23} & \le &\frac{1}{n} \sum_{i=1}^n I (X_{2,i};Y_{3,i}|X_{1,i},X_{3,i}) + \epsilon_{n} \nonumber\\
&\le & I (X_{2};Y_{3}|X_{1},X_{3})+\epsilon_{n}\nonumber\\[+0.2cm]
R_{13}+R_{23} &  \le &\frac{1}{n} \sum_{i=1}^n I (X_{1,i},X_{2,i}; Y_{3,i}|X_{3,i})+ \epsilon_{n} \nonumber\\
&\le & I (X_{1},X_{2};Y_{3}|X_{3})+\epsilon_{n}\nonumber
\end{IEEEeqnarray}
where the inequalities hold since $I (X_{1};Y_{3}|X_{2},X_{3})$, $I (X_{2};Y_{3}|X_{1},X_{3})$, and $I (X_{1},X_{2};Y_{3}|X_{3})$ are concave\footnote{This follows from the fact that $I(A; C|B)$ is concave in $P_{A, B}$ for fixed $P_{C|A, B}$ \cite{Shannon:1961}.} in the joint input distribution $P_{X_1,X_2,X_3}$, where $P_{X_1,X_2,X_3}= \frac{1}{n} \sum_{i=1}^n P_{X_{1,i},X_{2,i},X_{3,i}}$. 

For the achievable rate $R_{32}$ in the DB direction, we have
\begin{IEEEeqnarray}{l}
n{\cdot}R_{32}\nonumber\\
\ = H(M_{32}|M_{23})\nonumber\\
\ \le I(M_{32};Y_2^n|M_{23}) + n{\cdot}{\epsilon}_{n} \nonumber \\
\ = \sum_{i=1}^n I(M_{32};Y_{2,i}|Y_2^{i-1},M_{23},X_2^{i}) + n{\cdot}{\epsilon}_{n}\nonumber\\
\ = \scalemath{1}{\sum_{i=1}^n I(M_{32};X_{3,i}\oplus_q Z_{1,i}\oplus_q Z_{2,i}|X_3^{i-1}\oplus_q Z_1^{i-1}\oplus_q Z_2^{i-1},} M_{23},X_2^{i}) + n{\cdot}{\epsilon}_{n} \nonumber \\
\ = \scalemath{1}{\sum_{i=1}^n I(M_{32};X_{3,i}\oplus_q Z_{1,i}\oplus_q Z_{2,i}|X_3^{i-1}\oplus_q Z_1^{i-1}\oplus_q Z_2^{i-1},} M_{23}) + n{\cdot}{\epsilon}_{n} \label{eq:2R3} \\
\ \le \sum_{i=1}^n I(M_{32},X_3^{i-1}\oplus_q Z_1^{i-1}\oplus_q Z_2^{i-1},M_{23}; X_{3,i}\oplus_q Z_{1,i}\oplus_q Z_{2,i}) +n{\cdot}{\epsilon}_{n}\label{eq:2R4}\IEEEeqnarraynumspace\\
\ \le \sum_{i=1}^n \scalemath{0.9}{I(M_{32},M_{23},M_{13},X_3^{i-1}\oplus_q Z_1^{i-1}\oplus_q Z_2^{i-1},} X_3^{i-1}\oplus_q Z_1^{i-1}; X_{3,i}\oplus_q Z_{1,i}\oplus_q Z_{2,i} ) + n {\cdot}{\epsilon}_{n} \nonumber\\
\ = \sum_{i=1}^n I(M_{\{32,23,13\}},\tilde{Y}_1^{i-1},\tilde{Y}_2^{i-1}; \tilde{Y}_{2,i}) + n{\cdot}{\epsilon}_{n} \label{eq:2R6}
\end{IEEEeqnarray}
where (\ref{eq:2R3}) holds since $X_2^{i}$ is a function of $(X_3^{i-1}\oplus_q Z_1^{i-1}\oplus_q Z_2^{i-1},M_{23})$, (\ref{eq:2R4}) follows from the chain rule and the non-negativity of mutual information, and (\ref{eq:2R6}) is expressed in terms of $\tilde{Y}_{1, i}\triangleq X_{3, i}\oplus_q Z_{1, i}$, and $\tilde{Y}_{2,i}\triangleq X_{3,i}\oplus_q Z_{1,i}\oplus_q Z_{2,i}=\tilde{Y}_{1, i}\oplus_q Z_{2,i}$.

For $R_{31}$, we have
\begin{IEEEeqnarray}{l}
n{\cdot}R_{31} \nonumber\\ 
\ = H(M_{31}|M_{\{32,23,13\}}) \nonumber \\
\ \le I(M_{31};Y_1^n,Y_2^n|M_{\{32,23,13\}}) + n{\cdot}{\epsilon}_{n} \nonumber\\
\ = \sum_{i=1}^n I(M_{31};Y_{1,i},Y_{2,i}|Y_1^{i-1},Y_2^{i-1}, M_{\{32,23,13\}}) + n{\cdot}{\epsilon}_{n}\nonumber\\
\ \le \sum_{i=1}^n I(M_{31},X_{3,i};Y_{1,i},Y_{2,i}|Y_1^{i-1},Y_2^{i-1}, M_{\{32,23,13\}}) + n{\cdot}{\epsilon}_{n}\nonumber\\
\ = \sum_{i=1}^n \scalemath{1}{I(M_{31},X_{3,i};Y_{1,i},Y_{2,i}|Y_1^{i-1},Y_2^{i-1}, M_{\{32,23,13\}}, X_1^i, X_2^i)+n{\cdot}{\epsilon}_{n}}\nonumber\\*\label{eq:3R1}\\
\ = \sum_{i=1}^n \scalemath{1}{I(M_{31},X_{3,i};\tilde{Y}_{1, i}, \tilde{Y}_{2, i}|Y_1^{i-1},Y_2^{i-1}, M_{\{32,23,13\}}, X_1^i, X_2^i)+ n{\cdot}{\epsilon}_{n}}\nonumber\\
\ = \sum_{i=1}^n I(M_{31},X_{3,i};\tilde{Y}_{1, i}, \tilde{Y}_{2, i}|\tilde{Y}_1^{i-1}, \tilde{Y}_2^{i-1}, M_{\{32,13,23\}}) + n{\cdot}{\epsilon}_{n}\nonumber\\*\label{eq:3R2}\\
\ = \sum_{i=1}^n I(X_{3,i};\tilde{Y}_{1, i}, \tilde{Y}_{2, i}|\tilde{Y}_1^{i-1}, \tilde{Y}_2^{i-1}, M_{\{32,13,23\}})\nonumber\\   
\qquad\qquad\qquad + \scalemath{1}{\sum_{i=1}^n I(M_{31};\tilde{Y}_{1, i}, \tilde{Y}_{2, i}|\tilde{Y}_1^{i-1}, \tilde{Y}_2^{i-1}, M_{\{32,13,23\}},X_{3,i}) + n{\cdot}{\epsilon}_{n}}\nonumber\\
\ = \sum_{i=1}^n I(X_{3,i};\tilde{Y}_{1, i}, \tilde{Y}_{2, i}|\tilde{Y}_1^{i-1}, \tilde{Y}_2^{i-1}, M_{\{32,13,23\}})  +n{\cdot}{\epsilon}_{n} \label{eq:3R3}\\
\ = \sum_{i=1}^n I(X_{3,i};\tilde{Y}_{1,i}|\tilde{Y}_{1}^{i-1}, \tilde{Y}_{2}^{i-1}, M_{\{32,13,23\}}) + n{\cdot}{\epsilon}_{n} \label{eq:3R4}
\end{IEEEeqnarray}
where (\ref{eq:3R1}) holds since $X_{1, i}=f_{1, i}(M_{13}, Y_1^{i-1})$ and $X_{2, i}=f_{2, i}(M_{23}, Y_2^{i-1})$, (\ref{eq:3R2}) holds since $(Y_1^{i-1}, Y_2^{i-1}, X_1^i, X_2^i)$ can be generated knowing $(M_{13}$, $M_{23}, \tilde{Y}_1^{i-1}, \tilde{Y}_2^{i-1})$, (\ref{eq:3R3}) holds because  $M_{31} \markov (\tilde{Y}_1^{i-1}, \tilde{Y}_2^{i-1}, M_{\{32,13,23\}}, X_{3,i})\markov (\tilde{Y}_{1, i}, \tilde{Y}_{2, i})$ form a Markov chain, and 
\eqref{eq:3R4} holds since $\tilde{Y}_{2,i}\markov (\tilde{Y}_{1,i}, \tilde{Y}_1^{i-1}, \tilde{Y}_2^{i-1},M_{\{32,13,23\}})
\markov X_{3,i}$ form a Markov chain.
Note that these Markov chain properties hold since $\{Z_{1,i}\}_{i=1}^n$ and $\{Z_{2,i}\}_{i=1}^n$ are independent memoryless processes and are independent of all user messages. 

Setting $V_i=(\tilde{Y}_1^{i-1}, \tilde{Y}_2^{i-1}, M_{\{32,13,23\}})$, we have that $V_i\markov X_{3,i}\markov (\tilde{Y}_{1, i}, \tilde{Y}_{2, i})$ form a Markov chain. 
From \eqref{eq:2R6} and \eqref{eq:3R4}, we obtain that 
$n{\cdot}R_{32}  \le \sum_{i=1}^n I(V_i;\tilde{Y}_{2,i}) + n{\cdot}{\epsilon}_{n}$ and $n{\cdot}R_{31}  \le \sum_{i=1}^n I(X_{3,i};\tilde{Y}_{1,i}|V_i) + n{\cdot}{\epsilon}_{n}$.
Let $K$ be a time-sharing random variable that is uniform over $\{1,2,...,n\}$ and independent of all messages, inputs, and outputs. 
Setting $V=(K,V_K)$, $X_3=X_{3,K}$, $Z_1=Z_{1,K}$, $Z_2=Z_{2,K}$ $\tilde{Y}_1=X_3\oplus_q Z_1=\tilde{Y}_{1,K}$, $\tilde{Y}_2=X_3\oplus_q Z_1\oplus_q Z_2=\tilde{Y}_{2,K}$, we have
\begin{IEEEeqnarray}{rCl}
n{\cdot}R_{32}&\le &\sum_{i=1}^n I(V_i;\tilde{Y}_{2,i}) + n{\cdot}{\epsilon}_{n}\nonumber\\
&=& n\cdot I(V_K;\tilde{Y}_{2,K}|K)+ n{\cdot}{\epsilon}_{n}\nonumber\\
&\le & n\cdot I(V;\tilde{Y}_2) +n{\cdot}{\epsilon}_{n}\nonumber\\
&=& n\cdot I(V;X_3 \oplus_q Z_1 \oplus_q Z_2) +n{\cdot}{\epsilon}_{n},\nonumber
\end{IEEEeqnarray}
and 
\begin{IEEEeqnarray}{rCl}
\hspace{-0.5cm}n{\cdot}R_{31}&\le &\sum_{i=1}^n I(X_{3,i};\tilde{Y}_{1,i}|V_i) + n{\cdot}{\epsilon}_{n}\nonumber\\
&=& n\cdot I(X_{3};\tilde{Y}_{1}|V)+ n{\cdot}\epsilon_{n} \nonumber\\
&=& n\cdot I(X_{3};X_3 \oplus_q Z_1|V)+ n{\cdot}{\epsilon}_{n}\nonumber
\end{IEEEeqnarray}
for some $P_{Z_1, Z_2, X_3, V}=P_{X_3, V}\cdot P_{Z_1}\cdot P_{Z_2}$. 
Combining the obtained bounds for rates $R_{13}$ and $R_{23}$, the proof is completed by letting $n \to \infty$. 
The bound on the alphabet size of $V$ can be established by the convex cover method \cite{Gamal:2012}.
\end{IEEEproof}

\subsection{Proof of Theorem~\ref{thm:exMain}}\label{Appendix4}
\begin{IEEEproof}
Consider a MA-DB TWC governed by $P_{Y_3|X_1,X_2,X_3}$, $P_{Z_1}$, and $P_{Z_2}$. 
Recall that 
\begin{IEEEeqnarray}{l}
\mathcal{R}^{\text{MA-DBC}}(P_{X_1,X_2,X_3,V},P_{Y_3|X_1,X_2,X_3},P_{Z_1},P_{Z_2})=\Big\{(R_{13},R_{23},R_{31},R_{32}): \nonumber\\
\qquad\qquad\qquad\qquad\qquad\qquad\qquad\quad\qquad\qquad\qquad\qquad\qquad R_{13} \le I(X_1;Y_3|X_2,X_3),\IEEEeqnarraynumspace\label{eq:mr11}\\
\qquad\qquad\qquad\qquad\qquad\qquad\qquad\quad\qquad\qquad\qquad\qquad\qquad R_{23} \le I(X_2;Y_3|X_1,X_3),\label{eq:mr12}\\
\qquad\qquad\qquad\qquad\qquad\qquad\qquad\quad\qquad\qquad\quad\qquad \hspace{+0.12cm}R_{13} + R_{23} \le I(X_1,X_2;Y_3|X_3),\label{eq:mr13}\\
\qquad\qquad\qquad\qquad\qquad\qquad\qquad\quad\qquad\qquad\qquad\qquad\qquad R_{31} \le I(X_3;X_3 \oplus_q Z_1|V),\label{eq:mr14}\\
\qquad\qquad\qquad\qquad\qquad\qquad\qquad\quad\qquad\qquad\qquad\qquad\qquad R_{32} \le I(V;X_3 \oplus_q Z_1 \oplus_q Z_2)\Big\}.\IEEEeqnarraynumspace\label{eq:mr15}
\end{IEEEeqnarray}
Since \eqref{eq:mr11}-\eqref{eq:mr13} do not depend on $V$ and \eqref{eq:mr14} and \eqref{eq:mr15} do not depend on $(X_1, X_2)$, we have
\begin{IEEEeqnarray}{l}
\mathcal{R}^{\text{MA-DBC}}(P_{X_1,X_2,X_3,V},P_{Y_3|X_1,X_2,X_3},P_{Z_1},P_{Z_2})\nonumber\\
\ =\mathcal{R}^{\text{MA-DBC}}(P_{X_1,X_2|X_3}{\cdot}P_{V,X_3},P_{Y_3|X_1,X_2,X_3},P_{Z_1},P_{Z_2}).\IEEEeqnarraynumspace\label{eq:3userDcomM}
\end{IEEEeqnarray}
To complete the proof, it suffices to show that for every $P_{X_1, X_2|X_3}$ and the corresponding $\tilde{P}_{X_1}\tilde{P}_{X_2}$ (which depends on $P_{X_1, X_2|X_3}$) given by our assumption, satisfies
\begin{IEEEeqnarray}{l}
\mathcal{R}^{\text{MA-DBC}}(P_{X_1,X_2|X_3}P_{V,X_3},P_{Y_3|X_1,X_2,X_3},P_{Z_1},P_{Z_2})\nonumber\\
\ \subseteq {\mathcal{R}^{\text{MA-DBC}}(\tilde{P}_{X_1}{\cdot}\tilde{P}_{X_2}{\cdot}P_{V,X_3}, P_{Y_3|X_1,X_2,X_3},P_{Z_1},P_{Z_2})},\IEEEeqnarraynumspace\label{eq:3Op}
\end{IEEEeqnarray} 
since then we clearly have $\mathcal{C}^{\text{MA-DBC}}_{\text{O}}(P_{Y_3|X_1,X_2,X_3},P_{Z_1},P_{Z_2})\allowbreak\subseteq \mathcal{C}^{\text{MA-DBC}}_{\text{I}}(P_{Y_3|X_1,X_2,X_3},P_{Z_1},P_{Z_2})$. 
To show \eqref{eq:3Op}, consider two input distributions $P_{X_1,X_2,X_3,V}^{(1)}\triangleq P^{(1)}_{X_1,X_2|X_3}\cdot P^{(1)}_{V,X_3}$ and $P_{X_1,X_2,X_3,V}^{(2)}\triangleq \tilde{P}_{X_1}\cdot \tilde{P}_{X_2}\cdot P^{(1)}_{V,X_3}$, where $\tilde{P}_{X_1}\cdot\tilde{P}_{X_2}$ is given by the assumption. 
Then, 
\begin{IEEEeqnarray}{rCl}
I^{(1)}(X_3;X_3 \oplus_q Z_1|V) & =&I^{(2)}(X_3;X_3 \oplus_q Z_1|V)\label{eq:mac1}\\
I^{(1)}(V;X_3 \oplus_q Z_1 \oplus_q Z_2) & =&I^{(2)}(V;X_3 \oplus_q Z_1 \oplus_q Z_2)\label{eq:mac2}\IEEEeqnarraynumspace
\end{IEEEeqnarray}
since $P^{(1)}_{X_1, X_2, X_3, V}$ and $P^{(2)}_{X_1, X_2, X_3, V}$ have the same marginal $P^{(1)}_{V, X_3}$.
Furthermore, 
\begin{IEEEeqnarray}{rCl}
I^{(1)}(X_1;Y_3|X_2,X_3)& =& \sum_{x_3}P^{(1)}_{X_3}(x_3)\cdot I^{(1)}(X_1;Y_3|X_2,X_3=x_3) \nonumber\\
&\le & \sum_{x_3}P^{(1)}_{X_3}(x_3)\cdot I^{(2)}(X_1;Y_3|X_2,X_3=x_3)\nonumber\\
& = & I^{(2)}(X_1;Y_3|X_2,X_3),\nonumber
\end{IEEEeqnarray}
where the inequality follows from \eqref{eq:3include1} and the last equality holds since $P^{(1)}_{X_1, X_2, X_3, V}$ and $P^{(2)}_{X_1, X_2, X_3, V}$ have the same marginal $P^{(1)}_{X_3}$. 
Similarly, we obtain that $I^{(1)}(X_2;Y_3|X_1,X_3)\le \allowbreak I^{(2)}(X_2;Y_3|X_1,X_3)$ and $I^{(1)}(X_1,X_2;Y_3|X_3)\le \allowbreak I^{(2)}(X_1,X_2;Y_3|X_3)$. 
Consequently, \eqref{eq:3Op} holds.
\end{IEEEproof}

\subsection{Proof of Theorem~\ref{thm:exMain2}}\label{Appendix5}
\begin{IEEEproof}
Similar to the proof in Theorem~\ref{thm:exMain}, for any $P_{X_1,X_2|X_3}P_{V,X_3}=P_{X_2|X_3}P_{X_1|X_2, X_3}P_{V, X_3}$, it suffices to show that
\begin{IEEEeqnarray}{l}
\mathcal{R}^{\text{MA-DBC}}(P_{X_1,X_2|X_3}{\cdot}P_{V,X_3},P_{Y_3|X_1,X_2,X_3},P_{Z_1},P_{Z_2})\nonumber\\ 
\ \subseteq {\mathcal{R}^{\text{MA-DBC}}(P^*_{X_1}{\cdot}P_{X_2|X_3}{\cdot}P_{V,X_3}, P_{Y_3|X_1,X_2,X_3},P_{Z_1},P_{Z_2})},\IEEEeqnarraynumspace\label{eq:31p}
\end{IEEEeqnarray} 
where $P^*_{X_1}$ is given by conditions (i). 

For any $P_{X_1,X_2,X_3,V}^{(1)}=P^{(1)}_{X_1, X_2|X_3}\cdot P^{(1)}_{V,X_3}$, let $P_{X_1,X_2,X_3,V}^{(2)}= P^*_{X_1}\cdot P^{(1)}_{X_2}\cdot P^{(1)}_{V,X_3}$, where $P^*_{X_1}$ is given by condition (i) and $P^{(1)}_{X_2}$ denotes the marginal distribution of $X_2$ derived from $P_{X_1,X_2,X_3,V}^{(1)}$. 
For the rate constraints in the DB direction, the same identities as in \eqref{eq:mac1}-\eqref{eq:mac2} can be obtained because $P^{(1)}_{X_1, X_2, X_3, V}$ and $P^{(2)}_{X_1, X_2, X_3, V}$ share a common marginal distribution given by $P^{(1)}_{V, X_3}$. 
For $R_{13}$ in the MA direction, we have
\begin{IEEEeqnarray}{l}
I^{(1)}(X_1;Y_3|X_2,X_3)\ \qquad\qquad\qquad\qquad\qquad\qquad\qquad\qquad\ \nonumber\\
\ = \sum_{x_2, x_3}P^{(1)}_{X_2, X_3}(x_2, x_3)\cdot \scalemath{0.98}{I^{(1)}(X_1;Y_3|X_2=x_2,X_3=x_3)} \nonumber\\
\ = \sum_{x_2, x_3}P^{(1)}_{X_2, X_3}(x_2, x_3)\cdot \mathcal{I}\Big(P^{(1)}_{X_1|X_2=x_2, X_3=x_3}, P_{Y_3|X_1, X_2=x_2, X_3=x_3}\Big) \nonumber\\
\ \le\sum_{x_2, x_3}P^{(1)}_{X_2, X_3}(x_2, x_3)\cdot\Bigg[\max_{P_{X_1|X_2=x_2, X_3=x_3}}\mathcal{I}\Big(P_{X_1|X_2=x_2, X_3=x_3}, P_{Y_3|X_1, X_2=x_2, X_3=x_3}\Big)\Bigg]\nonumber\\
\ =\scalemath{1}{\sum_{x_2, x_3}P^{(1)}_{X_2, X_3}(x_2, x_3)\cdot\mathcal{I}\Big(P^*_{X_1}, P_{Y_3|X_1, X_2=x_2, X_3=x_3}\Big)}\IEEEeqnarraynumspace\label{eq:macnew1} \\
\ =\sum_{x_3}P^{(1)}_{X_3}(x_3)\sum_{x_2}P^{(1)}_{X_2|X_3}(x_2|x_3)\cdot\mathcal{I}\Big(P^*_{X_1}, P_{Y_3|X_1, X_2=x_2, X_3=x_3}\Big)\nonumber\\
\ =\sum_{x_3}P^{(1)}_{X_3}(x_3)\cdot\left(\sum_{x_2}P^{(1)}_{X_2|X_3}(x_2|x_3)\right)\cdot\mathcal{I}\Big(P^*_{X_1}, P_{Y_3|X_1, X_2=x'_2, X_3=x_3}\Big)\IEEEeqnarraynumspace\label{eq:macnew2}\\
\ =\sum_{x'_2}P^{(1)}_{X_2}(x'_2)\sum_{x_3}P^{(1)}_{X_3}(x_3)\cdot\mathcal{I}\Big(P^*_{X_1}, P_{Y_3|X_1, X_2=x'_2, X_3=x_3}\Big) \nonumber\\
\ = I^{(2)}(X_1;Y_3|X_2,X_3),\nonumber
\end{IEEEeqnarray}
where \eqref{eq:macnew1} and \eqref{eq:macnew2} directly follow from condition (i). 

For $R_{23}$, we have
\begin{IEEEeqnarray}{l}
I^{(1)}(X_2;Y_3|X_1,X_3)\nonumber\\
\ = \sum_{x_1,x_3}P^{(1)}_{X_1, X_3}(x_1, x_3)\cdot I^{(1)}(X_2;Y_3|X_1=x_1,X_3=x_3) \nonumber\\
\ = \sum_{x_1,x_3}P^{(1)}_{X_1, X_3}(x_1, x_3)\cdot \mathcal{I}\Big(P^{(1)}_{X_2|X_1=x_1, X_3=x_3}, P_{Y_3|X_1=x_1, X_2, X_3=x_3}\Big) \nonumber\\
\ = \sum_{x_1,x_3}P^{(1)}_{X_1, X_3}(x_1, x_3)\cdot \mathcal{I}\Big(P^{(1)}_{X_2|X_1=x_1, X_3=x_3}, P_{Y_3|X_1=x'_1, X_2, X_3=x'_3}\Big)\label{eq:macnew3}\\
\ \le\mathcal{I}\Bigg(\sum_{x_1,x_3}P^{(1)}_{X_1, X_3}(x_1, x_3)\cdot P^{(1)}_{X_2|X_1, X_3}(x_2|x_1, x_3), P_{Y_3|X_1=x'_1, X_2, X_3=x'_3}\Bigg)\label{eq:macnew4}\\
\ =\mathcal{I}\Big(P^{(1)}_{X_2}, P_{Y_3|X_1=x'_1, X_2, X_3=x'_3}\Big)\nonumber\\
\ =\sum_{x'_1,x'_3}P^{*}_{X_1}(x'_1){\cdot}P^{(1)}_{X_3}(x'_3){\cdot}\mathcal{I}\Big(P^{(1)}_{X_2}, P_{Y_3|X_1=x'_1, X_2, X_3=x'_3}\Big)\IEEEeqnarraynumspace \label{eq:macnew5}\\
\ = I^{(2)}(X_2;Y_3|X_2,X_3),\nonumber
\end{IEEEeqnarray}
where \eqref{eq:macnew3} and \eqref{eq:macnew5} follow from condition (ii) and \eqref{eq:macnew4} is due to convexity of $\mathcal{I}(\cdot, \cdot)$ in its first argument.  

Moreover, for the sum rate $R_{13}+R_{23}$, we have
\begin{IEEEeqnarray}{l}
I^{(1)}(X_1, X_2;Y_3|X_3)\nonumber\\
\ = \sum_{x_3}P^{(1)}_{X_3}(x_3)\cdot I^{(1)}(X_1, X_2;Y_3|X_3=x_3) \nonumber\\
\ = \sum_{x_3}P^{(1)}_{X_3}(x_3)\cdot \mathcal{I}\left(P^{(1)}_{X_1, X_2|X_3=x_3}, P_{Y_3|X_1, X_2, X_3=x_3}\right) \nonumber\\
\ = \sum_{x_3}P^{(1)}_{X_3}(x_3)\cdot \mathcal{I}\left(P^{(1)}_{X_1, X_2|X_3=x_3}, P_{Y_3|X_1, X_2, X_3=x'_3}\right)\label{eq:macnew6}\IEEEeqnarraynumspace\\
\ \le  \mathcal{I}\Bigg(\sum_{x_3}P^{(1)}_{X_3}(x_3)\cdot P^{(1)}_{X_1, X_2|X_3}(x_1, x_2|x_3), P_{Y_3|X_1, X_2, X_3=x'_3}\Bigg)\IEEEeqnarraynumspace\label{eq:macnew7}\\
\ = \mathcal{I}\left(P^{(1)}_{X_1, X_2}, P_{Y_3|X_1, X_2, X_3=x'_3}\right)\nonumber\\
\ \le  \mathcal{I}\left(P^*_{X_1}\cdot P^{(1)}_{X_2}, P_{Y_3|X_1, X_2, X_3=x'_3}\right)\label{eq:macnew8}\\
\ = \sum_{x'_3}P^{(1)}_{X_3}(x'_3)\cdot \mathcal{I}\left(P^*_{X_1}\cdot P^{(1)}_{X_2}, P_{Y_3|X_1, X_2, X_3=x'_3}\right)\nonumber\\
\ = I^{(2)}(X_1, X_2;Y_3|X_3),\nonumber
\end{IEEEeqnarray}
where \eqref{eq:macnew6} and \eqref{eq:macnew8} follow from condition (iii) and \eqref{eq:macnew7} is due to convexity of $\mathcal{I}(\cdot, \cdot)$ in its first argument.
Therefore, \eqref{eq:31p} holds under conditions~(i)-(iii). 
\end{IEEEproof}

\subsection{Proof of Theorem~\ref{thm:exSC}}\label{Appendix2}
It suffices to show that
\begin{IEEEeqnarray}{l}
\mathcal{R}^{\text{MA-DBC}}(P_{X_1,X_2|X_3}{\cdot}P_{V,X_3},P_{Y_3|X_1,X_2,X_3},P_{Z_1},P_{Z_2})\nonumber\\
\ \subseteq {\mathcal{R}^{\text{MA-DBC}}(P^{\text{U}}_{\mathcal{X}_1}{\cdot}P^{\text{U}}_{\mathcal{X}_2}{\cdot}P_{V,X_3}, P_{Y_3|X_1,X_2,X_3},P_{Z_1},P_{Z_2})}\IEEEeqnarraynumspace\label{eq:3uniformOp}
\end{IEEEeqnarray}
for any $P_{X_1,X_2|X_3}P_{V,X_3}$. 
We first give a proof sketch. 
Analogous to Shannon's proof for point-to-point TWCs (see Appendix \ref{appendix:SCproof}), we want to show that for any input distribution $P^{(1)}_{X_1,X_2,X_3, V}=P^{(1)}_{X_1,X_2|X_3}P^{(1)}_{V,X_3}$, if we set  $P^{(2)}_{X_1,X_2,X_3, V}=P^{(2)}_{X_1,X_2|X_3}P^{(1)}_{V,X_3}$ and $P^{(3)}_{X_1,X_2,X_3, V}=P^{(3)}_{X_1,X_2|X_3}P^{(1)}_{V,X_3}$, where  
\begin{align}
P^{(2)}_{X_1,X_2|X_3}(\cdot, \cdot|\cdot) & \triangleq P^{(1)}_{X_1,X_2|X_3}(\tau^{\mathcal{X}_1}_{x_1',x_1''}(\cdot), \cdot|\cdot),\label{eq:mixed}\\
P^{(3)}_{X_1,X_2|X_3}(\cdot, \cdot|\cdot) & \triangleq \frac{1}{2}\left(P^{(1)}_{X_1,X_2|X_3}(\cdot, \cdot|\cdot)+P^{(2)}_{X_1,X_2|X_3}(\cdot, \cdot|\cdot)\right),\label{eq:mixed3}
\end{align}
and $x_1', x''_1 \in \mathcal{X}_1 $, then we have 
\begin{IEEEeqnarray}{l}
\mathcal{R}^{\text{MA-DBC}}(P^{(1)}_{X_1,X_2|X_3}\cdot P^{(1)}_{V,X_3},P_{Y_3|X_1,X_2,X_3},P_{Z_1},P_{Z_2})\nonumber\\
\ = \mathcal{R}^{\text{MA-DBC}}(P^{(2)}_{X_1,X_2|X_3}{\cdot}P^{(1)}_{V,X_3},P_{Y_3|X_1,X_2,X_3},P_{Z_1},P_{Z_2})\IEEEeqnarraynumspace\label{eq:pp2}\\
\ \subseteq \mathcal{R}^{\text{MA-DBC}}(P^{(3)}_{X_1,X_2|X_3}{\cdot}P^{(1)}_{V,X_3},P_{Y_3|X_1,X_2,X_3},P_{Z_1},P_{Z_2}),\IEEEeqnarraynumspace\label{eq:pp3}
\end{IEEEeqnarray}
where the last inclusion is shown using \eqref{eq:md1} and extending Lemma~\ref{lem:Sl2} to the MA/DBC setup. 
Then, we use an induction argument as in the proof of Lemma~\ref{lem:Sl3} to obtain 
\begin{IEEEeqnarray}{l}
\mathcal{R}^{\text{MA-DBC}}(P_{X_1,X_2|X_3}\cdot P_{V,X_3},P_{Y_3|X_1,X_2,X_3},P_{Z_1},P_{Z_2})\nonumber\\
\ \ \subseteq {\mathcal{R}^{\text{MA-DBC}}(P^{\text{U}}_{\mathcal{X}_1}{\cdot}P_{X_2|X_3}P_{V,X_3}, P_{Y_3|X_1,X_2,X_3},P_{Z_1},P_{Z_2})}.\nonumber  \IEEEeqnarraynumspace\label{eq:mcin1}
\end{IEEEeqnarray}
Next, we consider input distributions of the form $P^{(1)}_{X_1, X_2, X_3, V}=P^{\text{U}}_{\mathcal{X}_1}\cdot P^{(1)}_{X_2|X_3}\cdot P^{(1)}_{X_3, V}$ and set $P^{(2)}_{X_1,X_2,X_3, V}=P^{(2)}_{X_1,X_2|X_3}{\cdot}P^{(1)}_{V,X_3}$ and $P^{(3)}_{X_1,X_2,X_3, V}{=}P^{(3)}_{X_1,X_2|X_3}{\cdot}P^{(1)}_{V, X_3}$, where  
\begin{align}
P^{(2)}_{X_1,X_2|X_3}(\cdot, \cdot|\cdot) & \triangleq P^{(1)}_{X_1,X_2|X_3}(\cdot,\tau^{\mathcal{X}_2}_{x_2',x_2''}(\cdot)|\cdot),\nonumber\\
P^{(3)}_{X_1,X_2|X_3}(\cdot, \cdot|\cdot) & \triangleq \frac{1}{2}\left(P^{(1)}_{X_1,X_2|X_3}(\cdot, \cdot|\cdot)+P^{(2)}_{X_1,X_2|X_3}(\cdot, \cdot|\cdot)\right),\nonumber
\end{align}
and $x_2', x''_2 \in \mathcal{X}_2$.
It can be shown via \eqref{eq:md2} that \eqref{eq:pp2}-\eqref{eq:pp3} also hold, and thus applying an induction argument again yields
\begin{IEEEeqnarray}{l}
\mathcal{R}^{\text{MA-DBC}}(P^{\text{U}}_{\mathcal{X}_1}{\cdot}P_{X_2|X_3}{\cdot}P_{V,X_3},P_{Y_3|X_1,X_2,X_3},P_{Z_1},P_{Z_2})\nonumber\\
\ \ \subseteq {\mathcal{R}^{\text{MA-DBC}}(P^{\text{U}}_{\mathcal{X}_1}{\cdot}P^{\text{U}}_{\mathcal{X}_2}{\cdot}P_{V,X_3}, P_{Y_3|X_1,X_2,X_3},P_{Z_1},P_{Z_2})}.\IEEEeqnarraynumspace\label{eq:mcin2}
\end{IEEEeqnarray}
Combining \eqref{eq:mcin1} and \eqref{eq:mcin2} then proves our claim. 
Due to symmetry, we only prove \eqref{eq:mcin1}.

\begin{lemma}\label{lma:mcbase}
For any $P^{(1)}_{X_1,X_2,X_3, V}=P^{(1)}_{X_1,X_2|X_3}\cdot P^{(1)}_{V,X_3}$, let $P^{(2)}_{X_1,X_2,X_3, V}=P^{(2)}_{X_1,X_2|X_3}\cdot P^{(1)}_{V,X_3}$ and $P^{(3)}_{X_1,X_2,X_3, V}=P^{(3)}_{X_1,X_2|X_3}\cdot P^{(1)}_{V,X_3}$, where $P^{(2)}_{X_1,X_2|X_3}$ and $P^{(3)}_{X_1,X_2|X_3}$ are given by \eqref{eq:mixed} and \eqref{eq:mixed3}, respectively. 
Then, \eqref{eq:pp2}-\eqref{eq:pp3} hold. 
\end{lemma}

\begin{IEEEproof}
We have
\begin{IEEEeqnarray}{l}
I^{(2)}(X_1;Y_3|X_2, X_3=x_3) \nonumber\\
\ = \sum_{x_1,x_2,y_3} P^{(2)}_{X_1,X_2|X_3}(x_1,x_2|x_3){\cdot}P_{Y_3|X_1,X_2,X_3}(y_3|x_1,x_2,x_3)\nonumber\\
\qquad\qquad \cdot \log\scalemath{0.93}{\frac{P_{Y_3|X_1,X_2,X_3}(y_3|x_1,x_2,x_3)}{\sum_{\tilde{x}_1}P^{(2)}_{X_1|X_2,X_3}(\tilde{x}_1|x_2,x_3)\cdot P_{Y_3|X_1,X_2,X_3}(y_3|\tilde{x}_1,x_2,x_3)}}\nonumber\\
\ = \sum_{x_1,x_2,y_3} P^{(1)}_{X_1,X_2|X_3}(\tau^{\mathcal{X}_1}_{x_1',x_1''}(x_1),x_2|x_3)\cdot P_{Y_3|X_1,X_2,X_3}(\pi^{\mathcal{Y}_3}[x'_1, x''_1](y_3)|\tau^{\mathcal{X}_1}_{x_1',x_1''}(x_1), x_2, x_3)\nonumber\\
\quad \ \ \cdot \Bigg[\log P_{Y_3|X_1,X_2,X_3}(\pi^{\mathcal{Y}_3}[x'_1, x''_1](y_3)|\tau^{\mathcal{X}_1}_{x_1',x_1''}(x_1), x_2, x_3)\nonumber\\
\ \ -\log\Bigg(\sum_{\tilde{x}_1}P^{(1)}_{X_1|X_2,X_3}(\tau^{\mathcal{X}_1}_{x_1',x_1''}(\tilde{x}_1)|x_2, x_3)\scalemath{1}{\cdot P_{Y_3|X_1,X_2,X_3}(\pi^{\mathcal{Y}_3}[x'_1, x''_1](y_3)|\tau^{\mathcal{X}_1}_{x_1',x_1''}(\tilde{x}_1), x_2, x_3)\Bigg)}\Bigg]\IEEEeqnarraynumspace\label{eq:md11}\\
\ \ = \sum_{x_1,x_2,y_3} P^{(1)}_{X_1,X_2|X_3}(x_1, x_2|x_3){\cdot}P_{Y_3|X_1,X_2,X_3}(y_3|x_1,x_2, x_3)\nonumber\\
\qquad\qquad\ \ \cdot \log\scalemath{0.91}{\frac{P_{Y_3|X_1,X_2,X_3}(y_3|x_1, x_2, x_3)}{\sum_{\tilde{x}_1}P^{(1)}_{X_1|X_2,X_3}(\tilde{x}_1|x_2,x_3){\cdot}P_{Y_3|X_1,X_2,X_3}(y_3|\tilde{x}_1,x_2,x_3)}}\label{eq:md12}\\
\  \ = I^{(1)}(X_1;Y_2|X_2,X_3=x_3),\nonumber
\end{IEEEeqnarray}
where \eqref{eq:md11} follows from \eqref{eq:md1} and \eqref{eq:mixed}, \eqref{eq:md12} holds since $\pi^{\mathcal{Y}_3}[x_1',x_1'']$ and $\tau^{\mathcal{X}_1}_{x_1',x_1''}$ are bijections. 
By a similar argument, we have that $I^{(2)}(X_2;Y_3|X_1,X_3=x_3)=I^{(1)}(X_2;Y_3|X_1,X_3=x_3)$ and that $I^{(2)}(X_1,X_2;Y_3|X_3=x_3)=I^{(1)}(X_1,X_2;Y_3|X_3=x_3)$. 
Next, using the concavity of $I(X_1;Y_3|X_2,X_3=x_3)$, $I(X_2;Y_3|X_1,X_3=x_3)$, and $I(X_1,X_2;Y_3|X_3=x_3)$ in $P_{X_1,X_2|X_3}(\cdot,\cdot|x_3)$\footnote{$I(X_1;Y_3|X_2,X_3=x_3)$ and $I(X_2;Y_3|X_1,X_3=x_3)$ are concave function of $P_{X_1,X_2|X_3}(\cdot,\cdot|x_3)$ since $I(X_1;Y_2|X_2)$ and $I(X_2;Y_1|X_1)$ are both concave in the input distribution $P_{X_1,X_2}$ \cite{Shannon:1961}.} yields that 
\begin{IEEEeqnarray}{rCl}
I^{(3)}(X_1;Y_3|X_2,X_3=x_3)&\ge &\frac{1}{2}\big(I^{(1)}(X_1;Y_3|X_2,X_3=x_3){+}I^{(2)}(X_1;Y_3|X_2,X_3=x_3)\big)\nonumber\\
&=& I^{(1)}(X_1;Y_3|X_2,X_3=x_3),\nonumber\\[+0.2cm]
I^{(3)}(X_2;Y_3|X_1,X_3=x_3)&\ge &\frac{1}{2}\big(I^{(1)}(X_2;Y_3|X_1,X_3=x_3){+}I^{(2)}(X_2;Y_3|X_1,X_3=x_3)\big)\nonumber\\
&=& I^{(1)}(X_2;Y_3|X_1,X_3=x_3),\nonumber\\[+0.2cm]
I^{(3)}(X_1, X_2;Y_3|X_3=x_3)&\ge &\frac{1}{2}\big(I^{(1)}(X_1,X_2;Y_3|X_3=x_3){+}I^{(2)}(X_1,X_2;Y_3|X_3=x_3)\big)\nonumber\\
& =& I^{(1)}(X_1,X_2;Y_3|X_3=x_3),\nonumber
\end{IEEEeqnarray}
and hence
\begin{align*}
I^{(3)}(X_1;Y_3|X_2,X_3)& \ge I^{(1)}(X_1;Y_3|X_2,X_3),\\
I^{(3)}(X_2;Y_3|X_1,X_3)& \ge I^{(1)}(X_2;Y_3|X_1,X_3),\\
I^{(3)}(X_1, X_2;Y_3|X_3)& \ge I^{(1)}(X_1,X_2;Y_3|X_3), 
\end{align*}
since $P^{(1)}_{X_3}=P^{(3)}_{X_3}$. 
Together with the definition of $\mathcal{R}^{\text{MA-DBC}}$ given in Section~\ref{sec2B}, the inclusions in \eqref{eq:pp2}-\eqref{eq:pp3} are proved. 
\end{IEEEproof}

Now, without loss of generality, suppose that $\mathcal{X}_1=\{ 1,2,..., \kappa \}$. 
For $1\le m\le\kappa$, define $\Lambda_m$ as the set of all conditional probability distributions $P_{X_1, X_2|X_3}$ satisfying $P_{X_1, X_2|X_3}(1, x_2|x_3)=P_{X_1, X_2|X_3}(2, x_2|x_3)=\cdots =P_{X_1, X_2|X_3}(m, x_2|x_3)$ for any fixed $x_2\in\mathcal{X}_2$ and $x_3\in\mathcal{X}_3$. 
As in the proof of Lemma~\ref{lem:Sl3}, it can be shown by induction on $m$ that 
\begin{IEEEeqnarray}{l}
\mathcal{R}^{\text{MA-DBC}}(P_{X_1,X_2|X_3}\cdot P_{V,X_3},P_{Y_3|X_1,X_2,X_3},P_{Z_1},P_{Z_2})\nonumber\\
\ \ \subseteq {\mathcal{R}^{\text{MA-DBC}}(\tilde{P}_{X_1, X_2|X_3}\cdot P_{V,X_3}, P_{Y_3|X_1,X_2,X_3},P_{Z_1},P_{Z_2})}\nonumber
\end{IEEEeqnarray}
where $P_{X_1,X_2|X_3}\in\Lambda_{m}$ and $\tilde{P}_{X_1,X_2|X_3}\in\Lambda_{m+1}$ for $1\le m< \kappa$. 
Note that the base case $m=1$  was proved in Lemma~\ref{lma:mcbase}. 
Since $P_{X_1, X_2|X_3}\in\Lambda_{\kappa}$ can be expressed as $P_{X_1, X_2|X_3}=P^{\text{U}}_{\mathcal{X}_1}\cdot P_{X_2|X_3}$, \eqref{eq:mcin1} holds. 
To show \eqref{eq:mcin2}, we consider input probability distributions of the form $P_{X_1, X_2, X_3, V}=P^{\text{U}}_{\mathcal{X}_1}\cdot P_{X_2|X_3}\cdot P_{X_3, V}$. 
By changing the roles of $X_1$ and $X_2$ in the above derivation, the rest of the proof is straightforward. \hfill\IEEEQED

\end{document}